\documentclass[10pt]{article} 
\usepackage[preprint]{tmlr}

\usepackage{microtype}
\usepackage{url}
\usepackage{booktabs}
\usepackage{subfigure}
\usepackage{color}
\usepackage{xcolor}
\usepackage{comment}
\usepackage{rotating}
\usepackage{multirow}
\usepackage{tabularray}
\usepackage{enumitem}
\usepackage{url}
\usepackage{xspace}
\usepackage{subfigure}
\usepackage{amsmath}
\usepackage{tabularx}  
\usepackage{graphicx}
\usepackage{array}
\usepackage{balance}
\usepackage{newtxmath}
\usepackage{listings}
\usepackage[T1]{fontenc}
\usepackage{inconsolata} 
\usepackage{tikz} 
\usetikzlibrary{positioning,shapes,arrows.meta,fit,calc,arrows}
\usepackage{makecell}
\usepackage{tikz}
\usetikzlibrary{positioning, shadows, arrows.meta}
\usepackage{fontawesome5}  

\usepackage{longtable}
\usepackage{array}
\usepackage{ragged2e}
\newcolumntype{L}[1]{>{\raggedright\arraybackslash}p{#1}}
\newcolumntype{C}[1]{>{\centering\arraybackslash}p{#1}}

\usepackage[utf8]{inputenc}

\usepackage{float}   
\usepackage{caption} 

\definecolor{codebg}{RGB}{248,248,248}
\definecolor{indigo}{RGB}{75,0,130}
\definecolor{bittersweet}{rgb}{1.0, 0.44, 0.37}
\usepackage[colorlinks=true, urlcolor=bittersweet]{hyperref}
\usepackage{cleveref}

\usepackage[ruled,vlined,linesnumbered]{algorithm2e}
\SetKwInput{KwIn}{Input}
\SetKwInput{KwOut}{Output}
\SetKwComment{tcp}{//}{}
\usepackage{lineno}

\DeclareMathOperator{\Seq}{Seq}

\definecolor{darkblue}{rgb}{0, 0, 0.5}
\hypersetup{colorlinks=true, citecolor=darkblue, linkcolor=darkblue}
\lstdefinestyle{pythonstyle}{
    language=Python,
    backgroundcolor=\color{codebg},
    basicstyle=\ttfamily\small,
    keywordstyle=\color{blue}\bfseries,
    stringstyle=\color{orange},
    commentstyle=\color{gray},
    showstringspaces=false,
    breaklines=true,
    frame=single,
    rulecolor=\color{black!20},
    tabsize=4,
    numbers=left,
    numberstyle=\tiny\color{gray},
    stepnumber=1,
    numbersep=10pt
}

\usepackage[svgnames]{xcolor}
\usepackage{tikz}
\usetikzlibrary{positioning,fit,shadows.blur}
\usepackage[skins, breakable]{tcolorbox}
\tcbset{enhanced}

\definecolor{CardBG}{RGB}{248,248,248}
\definecolor{CardBorder}{gray}{0.2}
\definecolor{Accent}{gray}{0.20}
\definecolor{Good}{RGB}{0,150,0}

\newcommand{\ufo}{\textsc{UFO}$^3$\xspace}
\newcommand{\cagent}{\textsc{ConstellationAgent}\xspace}
\newcommand{\bench}{\textsc{NebulaBench}\xspace}

\newcommand{\TaskConstellation}{\textsc{TaskConstellation}\xspace}

\newcommand{\TaskStar}{\textsc{TaskStar}\xspace}
\newcommand{\TaskStars}{\textsc{TaskStars}\xspace}
\newcommand{\TaskStarLine}{\textsc{TaskStarLine}\xspace}
\newcommand{\TaskStarLines}{\textsc{TaskStarLines}\xspace}

\newcommand{\eg}{\textit{e.g.},\xspace}
\newcommand{\ie}{\textit{i.e.},\xspace}

\newcommand{\galaxy}{\raisebox{-0.4\height}{\hspace*{-0.2em}\includegraphics[width=2.2em]{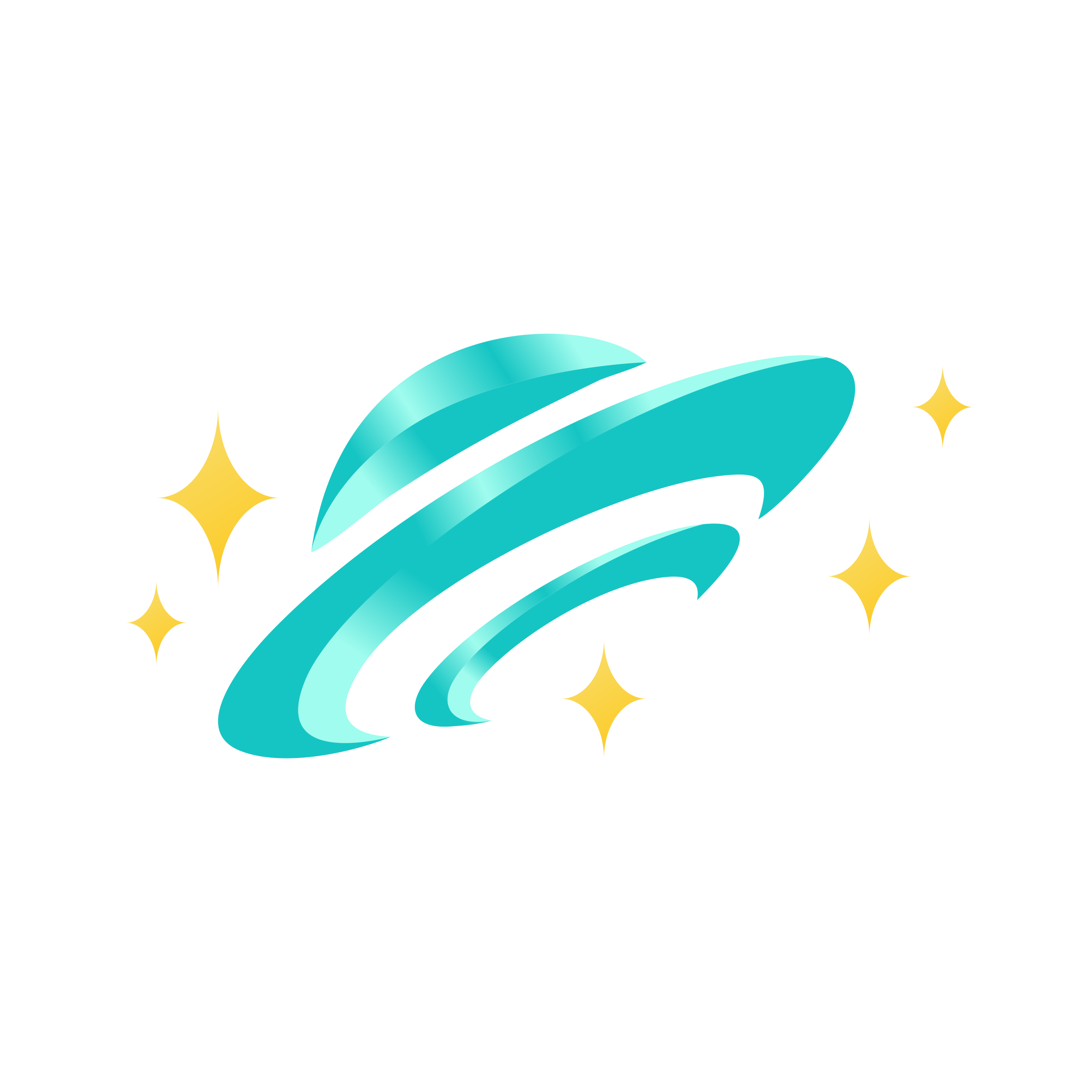}}}

\title{\ufo: Weaving the Digital Agent Galaxy \galaxy}

\author{\name 
Chaoyun Zhang\textsuperscript{1}\thanks{Chaoyun Zhang is the corresponding author: \texttt{chaoyun.zhang@microsoft.com}},
Liqun Li\textsuperscript{1},
He Huang\textsuperscript{1},
Chiming Ni\textsuperscript{2}\thanks{Chiming Ni is with ZJU-UIUC Institute and completed this work while at Microsoft},
Bo Qiao\textsuperscript{1},
Si Qin\textsuperscript{1},
Yu Kang\textsuperscript{1},\\
Minghua Ma\textsuperscript{1},
Qingwei Lin\textsuperscript{1},
Saravan Rajmohan\textsuperscript{1},
Dongmei Zhang\textsuperscript{1}
\\
\addr {\normalsize \textsuperscript{1}Microsoft \quad
\textsuperscript{2}ZJU-UIUC Institute
}
}

\def\month{06}  
\def\year{2025} 
\definecolor{tiffany}{RGB}{10,186,181}

\hypersetup{
    colorlinks=true,
    urlcolor=bittersweet    
}

\begin{document}


\maketitle
\begin{figure*}[h]
  \centering
  \includegraphics[width=\textwidth]{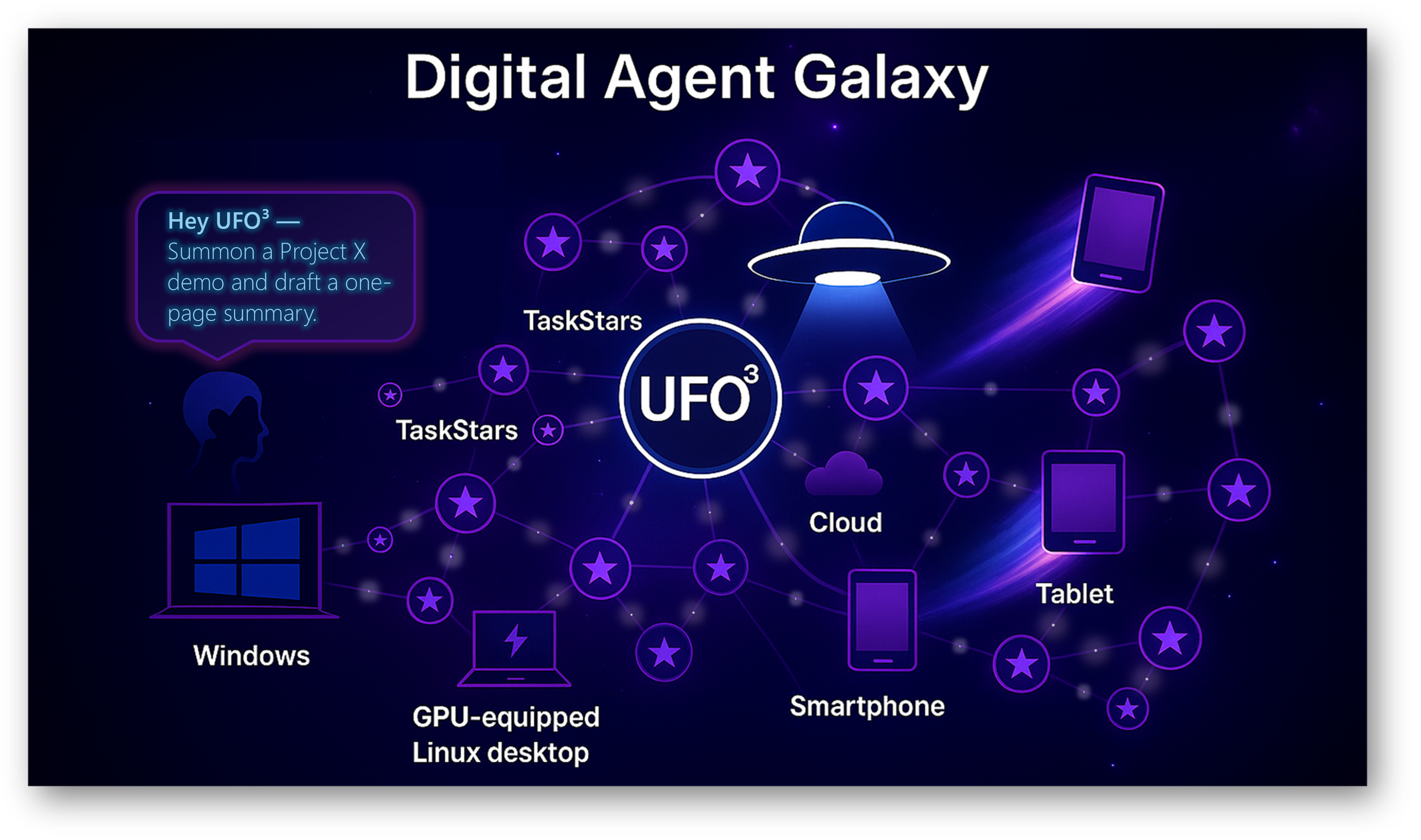}
  \vspace{-1em}
  \caption{\textbf{\ufo: Weaving the Digital Agent Galaxy.}
  A single natural-language intent is decomposed into a dynamically evolved Constellation (DAG) executed across heterogeneous devices. Demo video available at: \url{https://www.youtube.com/watch?v=NGrVWGcJL8o}.}  
  \label{fig:ufo-hero}
\end{figure*}

\begin{abstract}
    Large language model (LLM)-powered agents are transforming digital devices from passive tools into proactive intelligent collaborators. However, most existing frameworks remain confined to a single OS or device, making cross-device workflows brittle and largely manual. We present \textbf{\ufo \galaxy}, a system that unifies heterogeneous endpoints, desktops, servers, mobile devices, and edge, into a single orchestration fabric. \ufo models each user request as a mutable \TaskConstellation: a distributed DAG of atomic subtasks (\TaskStars) with explicit control and data dependencies (\TaskStarLines). The \TaskConstellation continuously evolves as results stream in from distributed devices, enabling asynchronous execution, adaptive recovery, and dynamic optimization. A \emph{Constellation Orchestrator} executes tasks safely and asynchronously while applying dynamic DAG updates, and the \emph{Agent Interaction Protocol (AIP)} provides persistent, low-latency channels for reliable task dispatch and result streaming. These designs dissolve the traditional boundaries between devices and platforms, allowing agents to collaborate seamlessly and amplify their collective intelligence.
    
    We evaluate \ufo on \bench, a benchmark of 55 cross-device tasks across 5 machines and 10 categories. \ufo achieves 83.3\% subtask completion, 70.9\% task success, exposes parallelism with an average width of 1.72, and reduces end-to-end latency by 31\% relative to a sequential baseline. Fault-injection experiments demonstrate graceful degradation and recovery under transient and permanent agent failures. These results show that \ufo achieves accurate, efficient, and resilient task orchestration across heterogeneous devices, uniting isolated agents into a coherent, adaptive computing fabric that extends across the landscape of ubiquitous computing.
    
    We developed \ufo as a fully engineered system with over 73K lines of code, encompassing agent implementations and integrations for Windows, Linux, and Android mobile devices. The entire project is open-sourced at \url{https://github.com/microsoft/UFO/}, accompanied by detailed documentation and tutorials at \url{https://microsoft.github.io/UFO/}
\end{abstract}

\section{Introduction}
The rise of intelligent agents \cite{wang2024survey} marks a new era of human–computer interaction, where large language models (LLMs) \cite{naveed2025comprehensive} are evolving from text-based reasoning engines \cite{qu2025tool} into autonomous digital operators capable of perceiving, acting, and coordinating across tasks \cite{zhang2024large}. Yet despite this progress, most agent frameworks remain confined within a single device or platform, be it a browser tab \cite{ning2025survey, zheng2024gpt}, a desktop environment \cite{zhang2025ufo, zhang2025ufo2}, or a mobile app \cite{zhang2025appagent, wang2024mobile}. This confinement sharply limits their ability to harness the rich, distributed computational ecosystem that modern users inhabit. 

When an agent is trapped within one operating system, it cannot access the complementary strengths of other devices, such as GPU clusters for computation, desktop applications for document editing, or mobile sensors for context capture. The result is a fragmented landscape of intelligent but \emph{siloed} agents: each powerful in isolation yet collectively underutilized. This gap between reasoning ability and real-world actuation leaves vast potential untapped. To truly advance the next frontier of automation and reasoning, agents must operate beyond the boundaries of any single device or OS, forming a coherent digital collective where Windows laptops, Linux servers, mobile devices, and edge nodes collaborate seamlessly \cite{houben2017opportunities, brudy2019cross} to gather and act upon ubiquitous intelligence.

Imagine a future where you could simply say: \emph{``Prepare a production-ready demo of Project X and deliver a one-page executive summary with screenshots and performance numbers.''} Today, this requires tedious, error-prone coordination across devices, checking out code on a laptop, triggering GPU builds on a server, deploying to a cloud instance, recording UI interactions on a phone, and stitching results into a report. Despite recent advances in intelligent agents, most systems remain confined within a single device or platform, leaving vast computational resources underutilized.

To realize this vision of seamless cross-device collaboration, we must overcome three interlocking challenges that go beyond classical workflow engines or single-machine agents. First, \emph{asynchronous parallelism}: many subtasks can and should run concurrently across devices with varying capabilities. Second, \emph{distributed coordination}: agents need reliable, low-latency communication for task dispatch and result streaming despite network variability. Third, \emph{heterogeneous extensibility}: the system should make it easy to develop and integrate new device agents while preserving safety and global consistency.

We present \textbf{\ufo: Weaving the Digital Agent Galaxy}, a cross-device orchestration system that turns isolated devices, desktops, servers, mobile, and edge, into a coherent execution fabric. \ufo models each request as a \TaskConstellation: a dynamic distributed DAG whose nodes (\TaskStars) represent executable subtasks and whose edges (\TaskStarLines) capture data and control dependencies. The Constellation serves as both the logical plan and live runtime substrate: nodes are assigned asynchronously, executed opportunistically, and continuously updated as results stream. Figure~\ref{fig:ufo-hero} illustrates this concept, where one intent decomposed into a distributed DAG and orchestrated across heterogeneous endpoints. 

To realize these capabilities and address the challenges in cross-device agents, \ufo is built around five tightly integrated design principles:
\begin{itemize}
    \item \textbf{Declarative decomposition into a dynamic DAG (\TaskConstellation).} Natural-language or programmatic requests are decomposed by the global \cagent into a structured DAG \cite{bei2025graphs} of \TaskStars and \TaskStarLines that encode workflow logic and dependencies. This declarative structure is amenable to automated scheduling, introspection, and rewriting throughout execution.

    \item \textbf{Continuous, result-driven graph evolution.} The \TaskConstellation is a living data structure. Intermediate outputs, transient failures, and new observations trigger controlled rewrites, adding diagnostic \TaskStars, creating fallbacks, rewiring dependencies, or pruning completed nodes, so the system adapts dynamically instead of aborting on errors \cite{wu2024agentkit}.

    \item \textbf{Heterogeneous, asynchronous, and safe orchestration.} Each \TaskStar is matched to the most suitable device agent via rich AgentProfiles reflecting OS, hardware,  and capabilities. The Constellation Orchestrator executes tasks asynchronously, allowing multiple \TaskStars to progress in parallel. Safe assignment locking, event-driven scheduling, DAG consistency checks, and batched edits collectively ensure correctness and concurrency safety, achieving high efficiency without compromising reliability. These guarantees are further reinforced through \emph{formal verification}.

    \item \textbf{Unified Agent Interaction Protocol (AIP).} Built atop persistent WebSocket channels, we develop AIP, a protocol that provides a unified, secure, and fault-tolerant layer for agent registry, session management, task dispatch, and coordination. It ensures reliability under network fluctuations through automatic reconnection and retry, while exposing a lightweight, extensible interface that allows new agents to integrate seamlessly into the \ufo ecosystem \cite{yang2025survey}.
    
    \item \textbf{Template-driven framework for MCP-empowered device agents.} To democratize agent creation, \ufo provides a lightweight development template and toolkit for rapidly building new device agents. Developers can declare capabilities, bind to local environments, and extend them through one or more Model Context Protocol (MCP) servers \cite{hou2025model} for tool augmentation. This modular design accelerates integration while maintaining consistency across the constellation.
\end{itemize}
Together, these designs enable the system to decompose, schedule, execute, and adapt distributed tasks efficiently while maintaining safety and consistency.

Building on these designs, we implemented the full \ufo system as a \textbf{comprehensive distributed system implementation} with over \textbf{73K lines of Python code}. The implementation integrates all major components, the centralized \cagent, the asynchronous Constellation Orchestrator, the AIP communication layer, and representative device agents for \textit{Windows}, \textit{Linux} and \textit{Android (mobile)}, each designed for containerized deployment and cross-environment compatibility. The system follows a modular, plugin-oriented architecture with type-safe interfaces, persistent telemetry, and built-in tracing, enabling reproducible, large-scale orchestration across heterogeneous devices. We further provide a futuristic WebUI for operator interaction and system visualization.

We evaluated \ufo on \bench, a benchmark of \textbf{55} cross-device tasks spanning \textbf{10} categories across \textbf{5} machines (a Windows 11 desktop, three Ubuntu CPU hosts, and one Ubuntu A100 GPU node). \ufo achieves a \textbf{Subtask Completion Rate (SCR)} of \textbf{83.3\%} and a \textbf{Task Success Rate (TSR)} of \textbf{70.9\%}. It exposes substantial parallelism, with an average execution width of \textbf{1.72} (peaking at $\sim$3.5), and reduces end-to-end latency by \textbf{31\%} compared to a sequential baseline. Fault-injection experiments further demonstrate its robustness: \ufo automatically retries and migrates under transient outages, gracefully degrades under partial failures, and recovers conservatively under global failures.

In essence, \ufo dissolves device boundaries and transforms the digital estate into a single, adaptive collaborator \cite{brudy2018investigating, marks2020multi, zhang2018towards}. It unifies distributed devices into a cohesive digital organism, one that executes user intents safely, asynchronously, and efficiently across heterogeneous environments. At its core, the \textbf{Agent Interaction Protocol (AIP)} serves as the connective tissue of this ecosystem, roviding a unified, fault-tolerant, and extensible communication substrate that allows new agents to join seamlessly and interoperate reliably. Over time, multiple constellations can interconnect through AIP, weaving together agents, devices, and capabilities into a self-organizing \emph{Digital Agent Galaxy}. Through this design, \ufo redefines cross-device automation, elevating it from a brittle engineering challenge to a unified orchestration paradigm, where multi-device workflows become naturally expressive and scalable across the landscape of ubiquitous computing.

\section{Background}
\label{sec:background}

\subsection{Digital Agents}
\label{sec:background:agents}

Leveraging the power of LLMs, modern \emph{digital agents} have emerged as powerful interfaces bridging human intent and the complex software ecosystems people interact with daily \cite{zhang2024large}. These agents can parse natural language requests, interpret screenshots and system state \cite{zheng2025vem, wu2025gui, zhao2025learning}, decompose complex goals into subtasks \cite{huang2024understanding}, and generate scripts or commands to execute tasks using a variety of tools \cite{wang2024survey}. Execution modalities include API calls, operating-system-level commands, GUI interactions via automation or accessibility interfaces, and code generation \cite{zhang2025api, qiao2023taskweaver, wang2024large, zhang2025swe}.

Digital agents operate across a wide spectrum of platforms. Web-based agents can navigate browsers and search the Internet \cite{ning2025survey, zheng2024gpt}, mobile agents are embedded in smartphone applications to automate mobile tasks \cite{zhang2025appagent, wang2024mobile}, and desktop or laptop agents interact with local operating systems and graphical user interfaces \cite{zhang2025ufo, zhang2025ufo2}. Across these platforms, agents observe system state, reason about tasks, and execute actions, enabling applications such as automated customer support flows, desktop productivity macros, and repetitive workflow automation, essentially acting as intelligent digital assistants.

Despite their versatility and growing adoption, existing digital agents are fundamentally limited by their confinement to a single device or environment. In practice, many modern workflows are no longer confined to a single endpoint \cite{brudy2018investigating}: users frequently interact with a combination of desktops, mobile devices, cloud services, and specialized hardware, all as part of a single task \cite{xu2021cross, chen2020multi}. Examples include running data analysis on a GPU cluster, collecting results on a personal laptop, and generating visual summaries on a tablet, or coordinating cross-platform deployments that touch both local workstations and cloud infrastructure. This trend makes the ability to operate seamlessly across devices increasingly urgent and commonplace.

Single-device frameworks face several inherent challenges in meeting this demand:
\begin{itemize}
    \item \textbf{Limited device capabilities.} Each agent is restricted by the hardware and software environment of its host, which limits the scope of tasks it can execute independently.
    \item \textbf{Fragmented personalization and context.} User-specific preferences, personalization, and contextual knowledge captured on one device are difficult to transfer or leverage on another \cite{cemri2025multi}, preventing a coherent multi-device experience.
    \item \textbf{Manual coordination overhead.} Orchestrating multiple single-device agents to accomplish cross-device workflows typically requires extensive manual development and careful sequencing, which is time-consuming, error-prone, and hard to maintain.
\end{itemize}
Together, these limitations highlight the pressing need for a new generation of digital agents that can natively reason about, orchestrate, and adapt across heterogeneous devices.

\subsection{Cross-Device Agent: A New Paradigm}
\label{sec:background:crossdevice}

To overcome the limitations of single-device frameworks, we introduce the concept of \emph{cross-device intelligence}, in which digital agents collaborate seamlessly across multiple heterogeneous endpoints \cite{brudy2019cross}. In this metaphorical \emph{digital galaxy}, each device, desktop, mobile, cloud service, or specialized hardware, acts as a \emph{star}, and coordinated tasks form a \emph{constellation} that collectively fulfills complex user requests. Unlike traditional agents confined to a single host, these cross-device agents can reason about device capabilities, orchestrate subtasks, and execute actions as part of a unified, intelligent ecosystem \cite{cheninternet}

Cross-device intelligence enables a qualitatively new user experience. A single natural-language request, such as ``prepare a production-ready demo, run performance tests, and generate a report'', can trigger an orchestrated constellation of actions: builds executed on developer laptops, computation-heavy tests on GPU clusters, deployments on cloud containers, and visualization or report generation on local or mobile devices. To the user, this appears as one coherent, effortless operation, eliminating the need for manual coordination and device-specific intervention.

This paradigm directly addresses the key limitations of single-device agents. It overcomes the constraints of individual devices by assigning tasks to the endpoints with the necessary capabilities and resources. It allows personalization, context, and user preferences captured on one device to propagate across others, enabling a coherent multi-device experience \cite{giusti2025federation}. And it automates orchestration of complex workflows, removing reliance on brittle glue code or labor-intensive manual integration.


\subsection{Design Challenges in Cross-Device Orchestration}
\label{sec:background:challenges}
However, synchronous control loops, cross-device agent orchestration introduces a fundamentally different class of systems challenges. In a distributed, heterogeneous environment, agents must collaborate across device, network, and platform boundaries, each with distinct runtime contexts and execution semantics \cite{tran2025multi}. Building a reliable and efficient orchestration layer under such conditions requires addressing three core challenges.
\begin{enumerate}
    \item \textbf{Asynchronous Parallelism.}  
    In cross-device workflows, multiple agents may execute concurrently on different endpoints. Unlike linear single-agent plans, task execution must accommodate partial completions, delayed feedback, and dynamic dependency resolution. The orchestrator must therefore reason about concurrency, detect when subtasks can safely proceed in parallel, and continuously adapt its scheduling plan based on evolving runtime states and network latencies \cite{yu2025dyntaskmas}. Failing to do so may lead to wasted computation or stalled progress.
    
    \item \textbf{Distributed Coordination.}  
    Agents in a constellation operate across diverse network and trust domains, implemented in different languages and deployed on various infrastructures. Achieving coherent coordination among them requires standardized mechanisms for registration, capability advertisement, task dispatch, and result collection. These operations must occur through persistent, low-latency channels that can tolerate temporary disconnections and guarantee consistent task states \cite{cheninternet}. In practice, ad-hoc HTTP calls or ephemeral connections are insufficient; a structured communication substrate is essential to sustain long-lived agent interactions.

    \item \textbf{Heterogeneous Extensibility.}  
    A cross-device ecosystem must embrace diversity rather than constrain it. Device agents differ in operating systems, execution environments, and available toolchains. The architecture should therefore allow rapid development, deployment, and integration of new device agents while ensuring consistent semantics in task execution and error propagation \cite{wu2024autogen}. This requires a flexible interface model and a unified protocol to abstract platform-specific complexity, enabling scalable and evolvable orchestration across an open agent federation \cite{yang2025survey}.
\end{enumerate}
These challenges are not merely engineering details, as they delineate the boundary between ad-hoc agent scripts and a principled, distributed orchestration system. To address them, we present \ufo, a reliable and scalable framework for intelligent cross-device agent orchestration. \ufo unifies four essential design elements:
\textit{(i)} a global \textbf{\cagent} that decomposes user intents into dynamic, dependency-aware task DAGs \textbf{(Section~\ref{sec:cagent})};
\textit{(ii)} an event-driven orchestration engine that enables concurrent, asynchronous, and adaptive execution \textbf{(Section~\ref{sec:orchestrator})};
\textit{(iii)} a standardized \textbf{Agent Interaction Protocol (AIP)} that supports persistent, low-latency, and extensible communication across heterogeneous agents \textbf{(Section~\ref{sec:aip})}; and
\textit{(iv)} an easy-to-use development interface that allows developers to quickly build new device agents and seamlessly integrate them into the ecosystem \textbf{(Section~\ref{sec:device-agent})}.
Together, these mechanisms transform isolated device agents into a coherent, cooperative constellation capable of executing complex, distributed tasks efficiently, safely, and at scale.

\section{\ufo: Design Principles and Overview}
\label{sec:architecture}

\begin{figure}[t]
    \centering
    \includegraphics[width=0.65\textwidth]{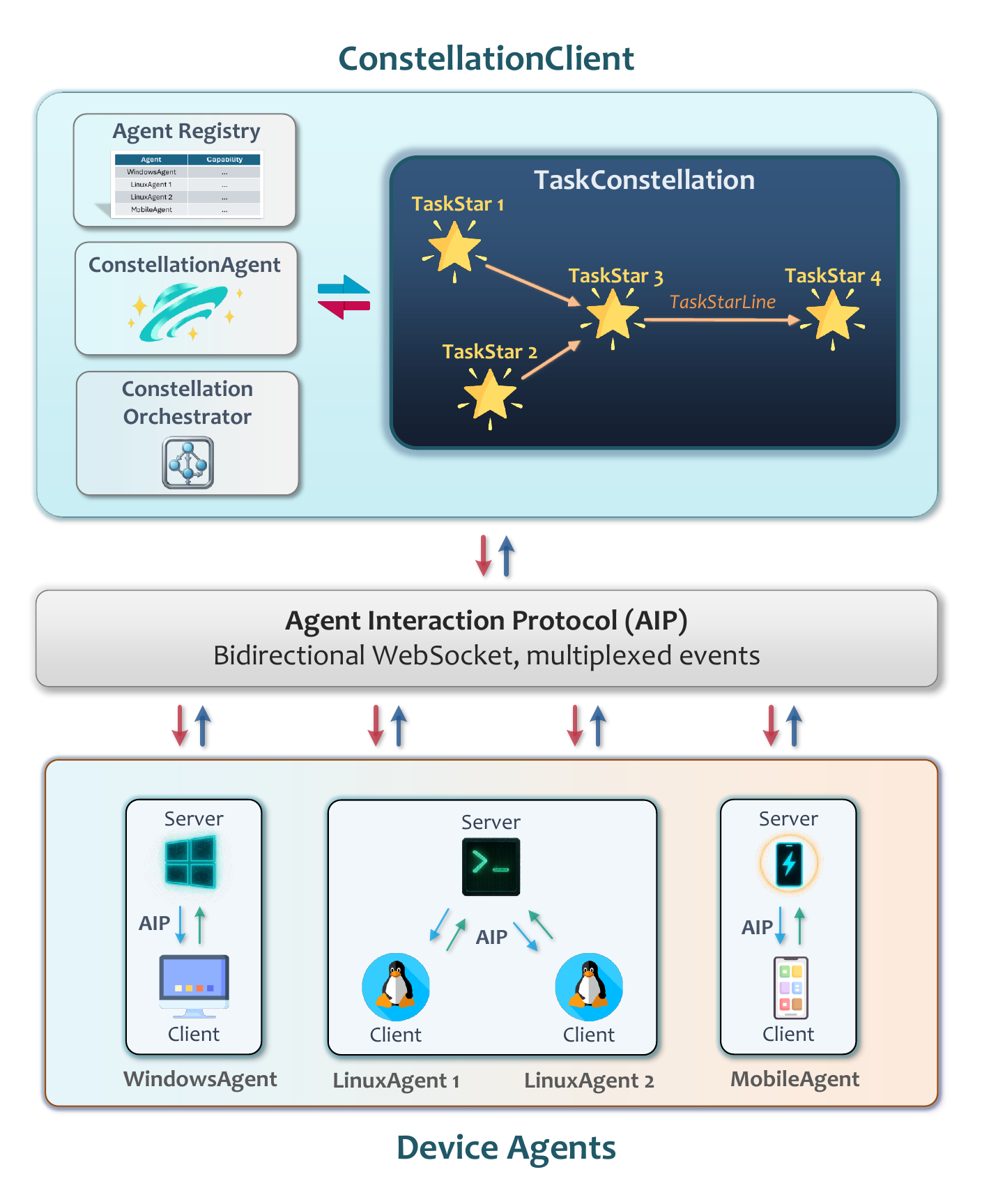}
    \vspace{-1.5em}
    \caption{Layered architecture of \ufo.}
    \label{fig:architecture}
\end{figure}

We now present an overview of the \ufo architecture, which transforms a collection of heterogeneous device agents into a unified, fault-tolerant execution fabric. Figure~\ref{fig:architecture} depicts its layered design. At a high level, \ufo follows a \emph{hierarchical orchestration model} that separates global coordination from local execution. This separation enables scalable cross-device orchestration while maintaining consistent control and responsiveness across diverse operating systems and network environments.

\subsection{Hierarchical Control Plane}
At the top of the hierarchy, the \textbf{ConstellationClient} serves as the global control plane. It maintains a live registry of all connected \textbf{device agents}, including their capabilities, system specifications, and runtime health metrics. This registry allows the orchestrator to place tasks on devices that can satisfy their resource requirements, avoiding mismatches between task demands and device capacity.

Each device hosts a \textbf{device agent server} that manages local orchestration. The server maintains a persistent WebSocket session with the ConstellationClient and oversees execution contexts on the host. A lightweight \textbf{device client} on each host provides a unified interface to underlying tool environments, exposed via MCP servers, enabling task execution, telemetry streaming, and resource monitoring. This layered control plane cleanly decouples global orchestration policies from device-specific heterogeneity, providing a consistent abstraction across endpoints that may differ in OS, hardware, or network conditions.

\subsection{Orchestration Flow}
When a high-level user request arrives, the ConstellationClient invokes the \cagent to construct a \TaskConstellation: a dynamic directed acyclic graph (DAG) that encodes task decomposition, dependencies, and candidate device mappings. Each node, or \TaskStar, represents an atomic execution unit assigned to a suitable device agent according to its capability profile and current system load.

The \textbf{Constellation Orchestrator} executes the DAG asynchronously and in an event-driven manner. Task completions trigger dependent nodes, while failures prompt retry, migration, or partial DAG rewrites. This design allows workflows to adapt to real-time system dynamics, such as device churn, network variability, or incremental task updates, while preserving overall progress and global consistency. The result is an execution model that is both highly parallel and resilient, sustaining workflow completion even as subsets of devices fail or reconnect.

\subsection{Cross-Agent Communication}
All cross-agent interactions, including agent registration, capability synchronization, task dispatch, progress reporting, and result aggregation, are handled by the \textbf{Agent Interaction Protocol (AIP)}. Built on persistent WebSocket channels, AIP provides a lightweight, bidirectional, and multiplexed substrate for structured event messages. Its design ensures low-latency propagation of control signals and consistent global state, even in the presence of intermittent connectivity or asynchronous updates.

Together, these design elements allow \ufo to orchestrate large-scale, heterogeneous, and adaptive workflows, forming a cohesive foundation for building a resilient, multi-device execution fabric.

\section{Formal Constellation Model}
\label{sec:constellation-model}

We now formalize the concept of a \TaskConstellation, the central abstraction that captures the concurrent and asynchronous structure of distributed task execution and dependencies. This model provides the theoretical foundation for reasoning about task dependencies, execution order, and fault-tolerant orchestration across heterogeneous devices. At its core, a \TaskConstellation represents a decomposed view of a complex user request in a directed acyclic graph (DAG) representation: a set of interdependent subtasks connected through explicit dependency edges. This formalism not only enables consistent scheduling and recovery but also supports runtime dynamism, allowing new tasks or dependencies to be introduced as the workflow evolves.

This representation provides clear advantages. Task ordering and dependencies are explicitly captured, ensuring correctness across distributed execution. The DAG topology naturally exposes parallelism and asynchronous execution, enabling efficient concurrency across heterogeneous devices. Moreover, nodes and edges can be dynamically added, removed, or rewired based on predecessor completion, allowing adaptive execution without compromising consistency. These properties make DAGs a natural and effective abstraction for modeling complex, cross-device workflows in heterogeneous environments.

\subsection{\TaskStar: Atomic Execution Unit}
\label{subsec:taskstar}

A \TaskStar denotes the atomic unit of computation in the \ufo framework, the smallest indivisible task scheduled on a device agent. Each \TaskStar encapsulates the complete context necessary for autonomous execution, including its semantic description, assigned device, execution state, and dependency relationships.

\begin{itemize}
    \item \textbf{Description:} a natural-language specification of the task, sent to the target device agent;
    \item \textbf{Tips:} A list of natural-language guidance designed to help the device agent successfully complete the task;
    \item \textbf{Device:} the identifier of the device agent responsible for execution;
    \item \textbf{Status:} the current execution state (\eg \textit{pending}, \textit{running}, \textit{completed});
    \item \textbf{Dependencies:} references to prerequisite tasks that must complete before execution.
\end{itemize}

Formally, a \TaskStar $t_i$ is defined as:
\[
t_i = (\text{name}_i, \text{description}_i, \text{device}_i, \text{tips}_i, \text{status}_i, \text{dependencies}_i)
\]

Intuitively, a \TaskStar ``knows'' what it should do, where it should run, how far it has progressed, and which other tasks it depends on. This self-contained representation enables fine-grained monitoring and decentralized scheduling across devices.

\subsection{\TaskStarLines: Dependency Edge}
\label{subsec:taskstarline}

A \TaskStarLine represents a dependency relation between two \TaskStars, forming a directed edge in the task graph. Let $t_i$ and $t_j$ denote two \TaskStars. A \TaskStarLines $e_{i \rightarrow j}$ specifies that $t_j$ cannot begin until certain conditions on $t_i$ are satisfied:
\[
e_{i \rightarrow j} = (\text{from\_task}_i, \text{to\_task}_j, \text{type}, \text{description})
\]

The \textit{type} parameter determines the nature of the dependency:
\begin{itemize}
    \item \textbf{Unconditional:} $t_j$ always waits for $t_i$ to complete;
    \item \textbf{Success-only:} $t_j$ proceeds only if $t_i$ succeeds;
    \item \textbf{Conditional:} $t_j$ proceeds based on a user-defined or runtime condition.
\end{itemize}
These dependency edges enforce causal consistency within the constellation, ensuring that concurrent execution respects logical task ordering while maximizing parallelism.

\subsection{TaskConstellation: Directed Acyclic Task Graph}
\label{subsec:taskconstellation}

A complete \TaskConstellation is a DAG that encodes the structure of a distributed workflow:
\[
\mathcal{C} = (\mathcal{T}, \mathcal{E})
\]
where $\mathcal{T}$ is the set of all \TaskStars and $\mathcal{E}$ is the set of \TaskStarLines. Each \TaskConstellation provides a compact yet expressive representation of how a complex task decomposes into independently executable units with explicit dependencies.

\begin{figure}[t]
    \centering
    \includegraphics[width=0.8\textwidth]{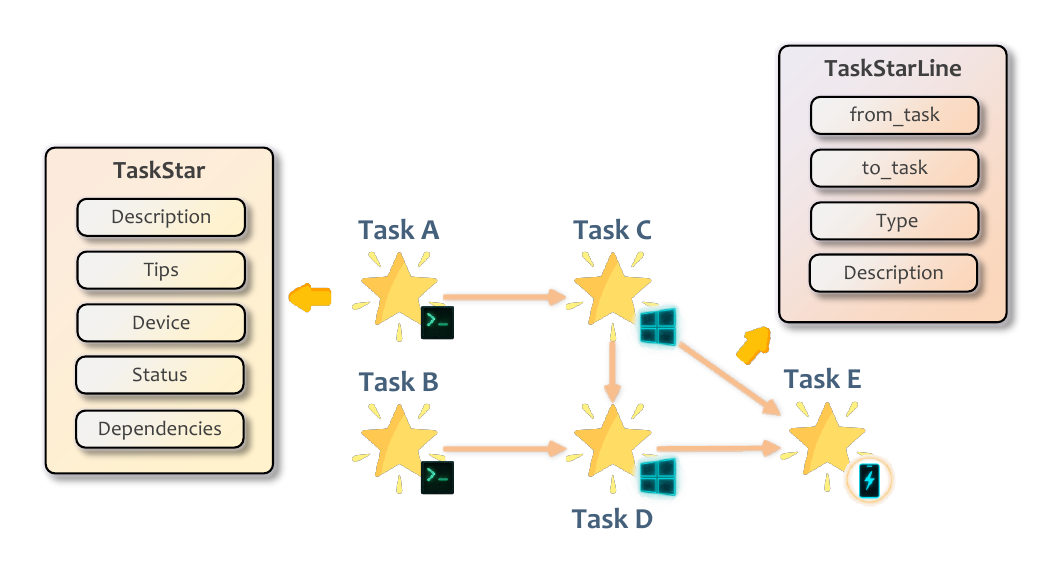}
    \vspace{-1.5em}
    \caption{Example of a \TaskConstellation illustrating both sequential and parallel dependencies.}
    \label{fig:taskconstellation-example}
\end{figure}

Figure~\ref{fig:taskconstellation-example} illustrates a simple \TaskConstellation spanning multiple devices. Task~A (LinuxAgent) must finish before Task~C starts; both Task~B (LinuxAgent) and Task~C (WindowsAgent) must complete before Task~D (WindowsAgent) begins; and Task~E (MobileAgent) depends on the successful completion of both Task~C and Task~D. This example demonstrates how the model naturally captures both sequential and parallel dependencies within a unified structure.

\subsection{Runtime Dynamism and Adaptivity}
Unlike static DAG schedulers \cite{Islam2012Oozie, di2017nextflow, argo}, \ufo treats TaskConstellations as \emph{mutable} objects. Tasks and dependency edges can be inserted, removed, or modified at runtime. This design enables \ufo to dynamically react to evolving execution contexts, such as new user inputs, intermediate results, or device failures, without restarting the entire workflow. Such adaptivity is key to maintaining progress and efficiency in long-running, cross-device task orchestration.

In summary, the \TaskConstellation formalism serves as the conceptual backbone of \ufo. It provides a precise and extensible representation of distributed workflows, enabling both rigorous reasoning and practical orchestration under asynchronous and failure-prone environments.

\section{\cagent: The Centralized Constellation Weaver}
\label{sec:cagent}

\begin{figure}[t]
    \centering
    \includegraphics[width=0.9\textwidth]{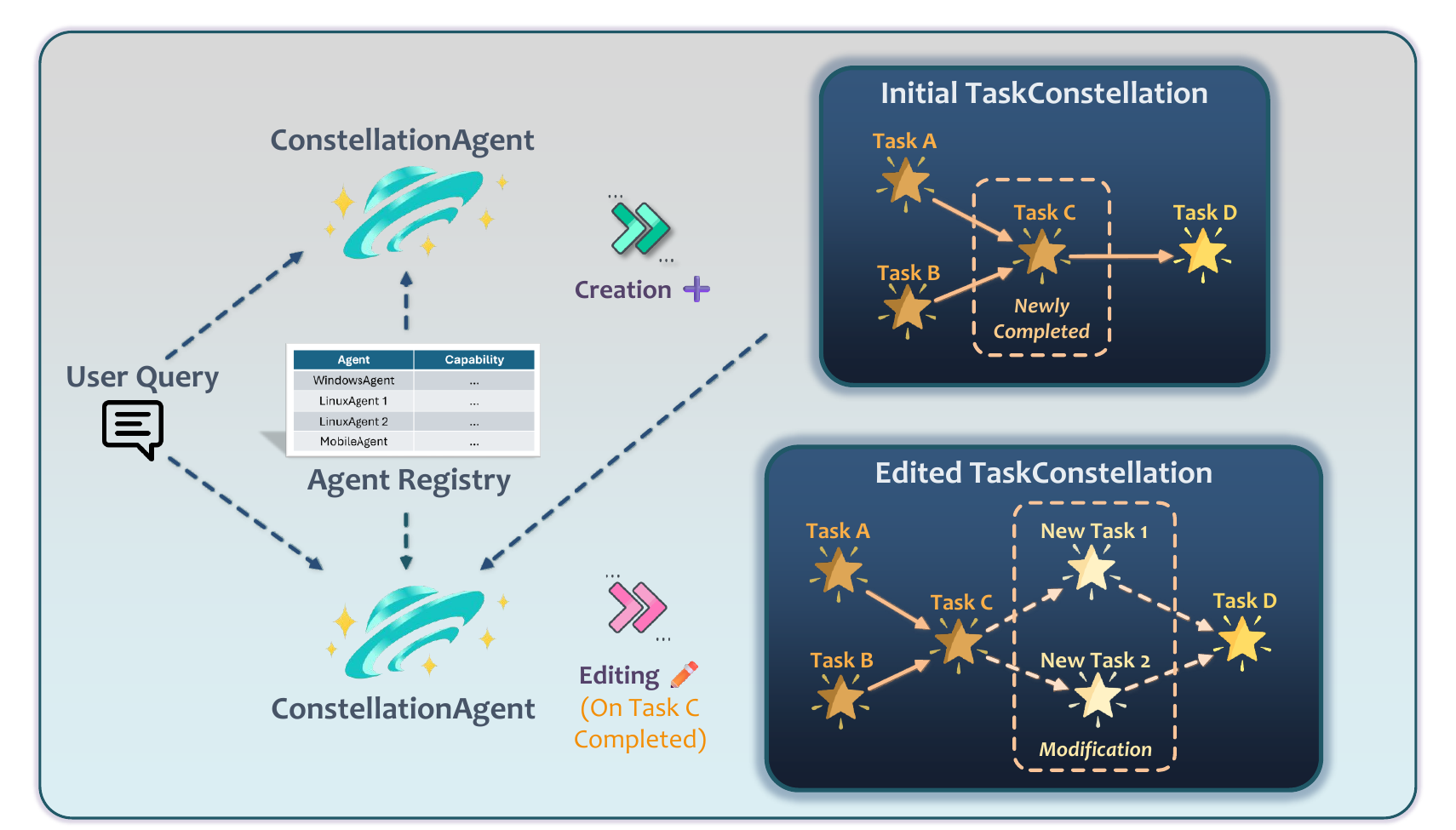}
    \vspace{-.5em}
    \caption{An overview of the \cagent.}
    \label{fig:cagent}
\end{figure}

While the previous section formalized the \TaskConstellation as an abstract model of distributed task structure, we now turn to its realization in the runtime control plane. The \cagent is the central intelligence of \ufo, responsible for interpreting user intent, constructing executable constellations, and steering their evolution across heterogeneous devices. Residing within the \textbf{ConstellationClient}, \cagent bridges the gap between high-level natural-language goals and concrete multi-agent execution through a unified orchestration interface.

As illustrated in Figure~\ref{fig:cagent}, \cagent acts as both a \textit{planner} and a \textit{replanner}: it first synthesizes a \TaskConstellation from user instructions, then incrementally refines this constellation as feedback arrives from distributed agents. Internally, \cagent is implemented as an LLM-driven ReAct agent \cite{yao2022react} governed by a finite-state machine (FSM) \cite{schneider2005state}. This FSM alternates between two complementary operating modes, namely \textit{creation} and \textit{editing}, forming a closed-loop control cycle that continuously updates the global execution graph in response to runtime conditions.

This design achieves a tight coupling between symbolic reasoning and distributed execution: declarative goals are grounded into concrete task graphs, and dynamic feedback drives continuous graph mutation. Through this feedback-driven control loop, \cagent maintains global consistency, ensures forward progress, and adapts seamlessly to changing device conditions. Overall, the design provides three core benefits:
\begin{enumerate}
    \item \textbf{Unified reasoning and control:} High-level task synthesis and low-level execution coordination are decoupled yet remain tightly synchronized via the \TaskConstellation abstraction.
    \item \textbf{Dynamic adaptability:} The editable \TaskConstellation enables recovery, reallocation, and opportunistic task generation in the face of partial failures or evolving goals.
    \item \textbf{End-to-end observability:} \cagent maintains a complete lineage of task states and dependencies, enabling introspection, debugging, and verifiable traceability.
\end{enumerate}

\subsection{Responsibilities and I/O Interface}
\label{subsec:responsibilities}

\cagent serves as the reasoning core of the \ufo control plane, orchestrating a structured feedback loop that alternates between \textit{creation} and \textit{editing} phases. Each phase defines explicit inputs, outputs, and operational responsibilities, ensuring that cross-device execution remains both consistent and adaptive.

\paragraph{Creation Mode.}
In the creation phase, \cagent receives three primary inputs:
\emph{(i)} a user-issued goal, expressed in natural or structured language;  
\emph{(ii)} the \textbf{AgentProfile} registry, describing each available device agent's capabilities, environment, and metadata; and  
\emph{(iii)} demonstration examples to support in-context learning (ICL) \cite{dong2022survey, jiang2024xpert}.

Leveraging the LLM's semantic reasoning capabilities, \cagent decomposes the user goal into a structured execution graph, the \TaskConstellation. Each node (\TaskStar) is annotated with explicit dependencies, resource constraints, and a target device assignment. This graph constitutes the initial execution plan, which is then handed off to the Constellation Orchestrator for distributed scheduling.

The \TaskConstellation is generated in a structured, machine-readable JSON format. Beyond the raw graph, \cagent produces a detailed reasoning trace: it first formulates an \texttt{Observation} of the input and AgentProfile, followed by a structured \texttt{Thought} capturing its analysis and decision-making rationale \cite{wei2022chain, ding2024everything}. It then outputs the next \texttt{State} in the control loop and the overall \texttt{Result} for user-request or the generated \TaskConstellation. This ensures that the \TaskConstellation is derived through a transparent, deliberate reasoning process.

\paragraph{Editing Mode.}
During distributed execution, \cagent enters the editing phase of its adaptive control loop. In this mode, it continuously consumes:  
\emph{(i)} the original user request;  
\emph{(ii)} the current AgentProfile registry;  
\emph{(iii)} a serialized snapshot of the \TaskConstellation; and  
\emph{(iv)} demonstration examples for ICL.

Upon receiving completion or failure events, \cagent evaluates whether modifications are necessary, such as adding follow-up subtasks, removing redundant ones, or refining dependency edges. Edits are applied only to non-terminal \TaskStars (\ie not in \textsc{Running}, \textsc{Completed}, or \textsc{Failed}), ensuring runtime correctness while allowing the \TaskConstellation to evolve dynamically.

Editing actions are invoked through one or more of the tools defined in Section~\ref{subsec:mcp-server} to facilitate parsing and execution. Concurrently, \cagent produces a structured reasoning trace, including a \texttt{Thought} narrative that explains the current constellation state, justifies any modifications (or lack thereof), identifies the next \texttt{State} in the control loop, and summarizes the overall user-request \texttt{Result}. This transparent, step-by-step reasoning guarantees that adaptations are both explainable and consistent.

This dual-mode control pattern realizes a balance between \textbf{global consistency} and \textbf{local adaptivity}. By structuring feedback integration through explicit modes and invariants, \cagent achieves stable yet flexible orchestration, avoiding common pitfalls of uncontrolled LLM self-modification. The resulting system embodies a feedback-driven orchestration loop that remains both reactive and verifiable.

\subsection{Finite-State Machine and Lifecycle}
\label{subsec:fsm-lifecycle}

\begin{figure}[t]
    \centering
    \includegraphics[width=0.65\textwidth]{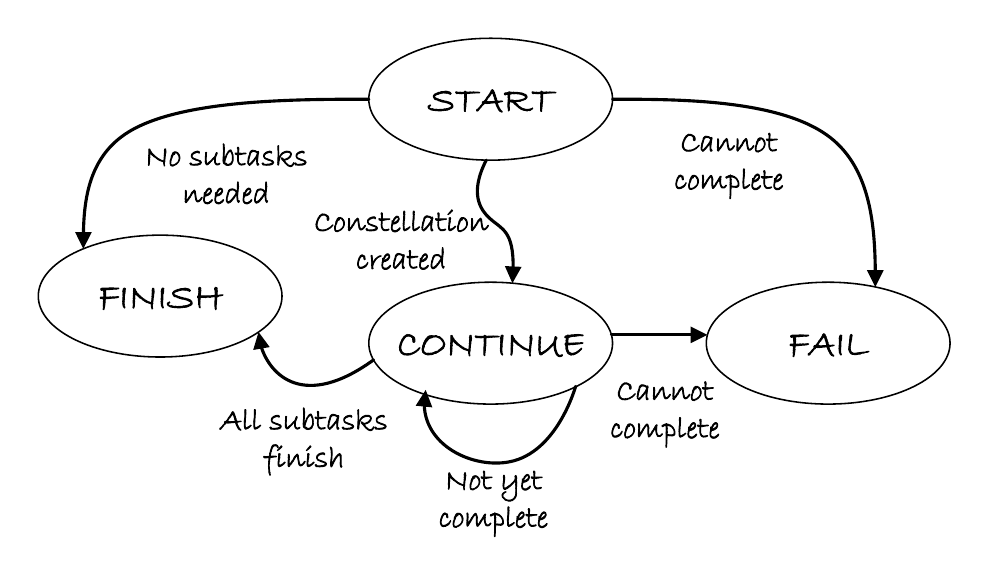}
    \vspace{-1.5em}
    \caption{Lifecycle state transitions of the \cagent.}
    \label{fig:cagent-fsm}
\end{figure}

The internal logic of \cagent is expressed as a \textbf{finite-state machine (FSM)}, providing a clear, enforceable structure for task lifecycle management. We show its state transition in Figure~\ref{fig:cagent-fsm}. The FSM governs how \cagent transitions across four primary operational states:

\begin{itemize}
    \item \texttt{START:} Initialization state where \cagent receives the user goal and constructs the initial \TaskConstellation (\textit{creation mode}).
    \item \texttt{CONTINUE:} The steady-state loop, where task progress is monitored and incremental edits are applied based on runtime feedback (\textit{editing mode}).
    \item \texttt{FINISH:} The successful termination state, triggered when all subtasks are completed or no further edits are required. Results are aggregated and reported to the user.
    \item \texttt{FAIL:} The terminal error state, entered upon irrecoverable failures or unreachable goals, prompting abort and logging for user recovery.
\end{itemize}

This FSM-based design ensures deterministic task transitions, consistent global state evolution, and predictable recovery paths. It also provides a natural boundary between LLM reasoning and deterministic control logic, which improves safety and debuggability in complex, cross-device workflows.

\subsection{Constellation MCP Server: Structured Task Management}
\label{subsec:mcp-server}

\begin{table}[t]
\centering
\caption{Core tools exposed by the Constellation MCP Server for managing tasks and dependencies.}
\begin{tabular}{l p{3.5cm} p{2.5cm} p{3cm}}
\toprule
\textbf{Tool Name} & \textbf{Purpose} & \textbf{Input} & \textbf{Output} \\
\midrule
\texttt{add\_task} & Add a new atomic task (\TaskStar) & Task ID, name, description, target device, tips & Updated \TaskConstellation  \\\midrule
\texttt{remove\_task} & Remove a task and all associated dependencies & Task ID & Updated \TaskConstellation  \\\midrule
\texttt{update\_task} & Modify task fields (name, description, device, tips) & Task ID + updated fields & Updated \TaskConstellation  \\\midrule
\texttt{add\_dependency} & Establish a dependency between two tasks & Dependency ID, from task, to task, condition & Updated \TaskConstellation  \\\midrule
\texttt{remove\_dependency} & Remove a dependency line & Dependency ID & Updated \TaskConstellation  \\\midrule
\texttt{update\_dependency} & Update the condition or description of a dependency & Dependency ID, condition description & Updated \TaskConstellation  \\\midrule
\texttt{build\_constellation} & Batch-create tasks and dependencies from structured input & Configuration dictionary, clear flag & Built \TaskConstellation  \\
\bottomrule
\end{tabular}
\label{tab:mcp-tools}
\end{table}

To operationalize dynamic graph construction, \cagent interacts with a lightweight \textbf{Constellation MCP Server} \cite{hou2025model}, a modular component exposing a standardized set of task and dependency management primitives. This server serves as the structured manipulation layer that bridges LLM-level reasoning and concrete execution state. Each operation encapsulates a single, idempotent transformation on the \TaskConstellation, ensuring reproducibility and easy rollback.

The MCP Server supports both fine-grained (single task or edge) and bulk (batch graph) operations, all of which return a serialized, globally consistent constellation snapshot. Table~\ref{tab:mcp-tools} summarizes the core toolset. Through this uniform interface, \cagent can safely evolve task graphs at runtime while preserving key invariants such as DAG validity and single assignment.

By decoupling reasoning (handled by \cagent) from structured mutation (handled by MCP), \ufo achieves a clean separation of concerns: \cagent focuses on semantic decision-making, while the MCP Server enforces syntactic and structural integrity. This design not only enhances robustness and auditability but also facilitates future extensibility, for instance, incorporating new agent types, validation hooks, or task optimization heuristics without modifying the orchestration core.

\paragraph{Summary.}
In summary, \cagent serves as the ``central weaver'' of distributed intelligence within \ufo. By combining LLM-driven reasoning, a finite-state control backbone, and a structured task manipulation interface, it transforms abstract user goals into live, evolving constellations, maintaining both rigor and adaptability across the lifecycle of multi-device orchestration.

\section{Asynchronous Dynamic Constellation Orchestrator}
\label{sec:orchestrator}

\begin{figure}[t]
    \centering
    \includegraphics[width=\textwidth]{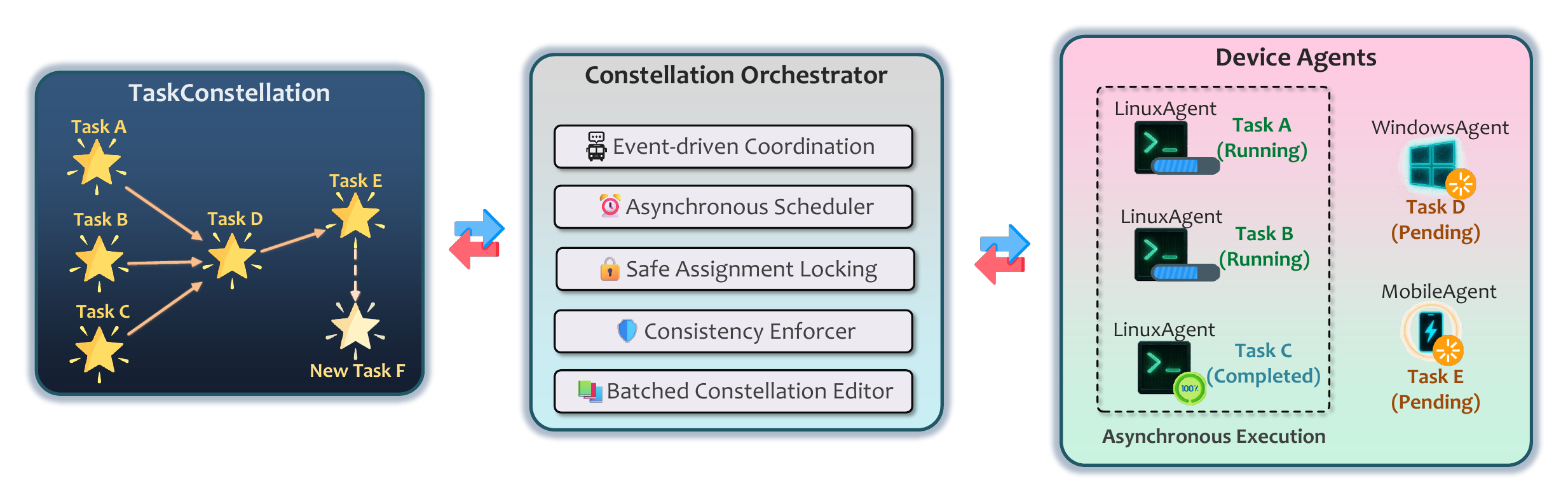}
    \vspace{-2em}
    \caption{The Constellation Orchestrator bridges \TaskConstellation  and execution, enabling asynchronous, adaptive task orchestration across devices.}
    \label{fig:orchestrator}
\end{figure}

With the \cagent governs the reasoning and evolution of a \TaskConstellation, the \emph{Constellation Orchestrator} brings that plan to life by executing, monitoring, and adapting interdependent tasks across heterogeneous devices \cite{li2022dag}, as shown in Figure~\ref{fig:orchestrator}. Conceptually, it transforms a static DAG into a \emph{living execution fabric}, where tasks evolve concurrently, react to runtime signals, and adapt to new decisions generated by the reasoning agent. 

Unlike traditional serial or static agent workflows, \ufo's orchestration must satisfy three often conflicting goals: (\emph{i}) asynchronous parallelism to leverage device heterogeneity, (\emph{ii}) safety and consistency under concurrent DAG updates, and (\emph{iii}) adaptivity to runtime feedback from both devices and LLM reasoning. Achieving these goals poses several technical challenges. First, subtasks must be assigned and executed asynchronously without violating data dependencies. Second, task execution may overlap with live \TaskConstellation edits, requiring strict consistency control. Third, edited graphs must remain valid and acyclic. Finally, frequent updates should not degrade performance or introduce synchronization bottlenecks.

To address these challenges, the Constellation Orchestrator is built around five design pillars: 
\textbf{(1) event-driven coordination}, 
\textbf{(2) asynchronous scheduling}, 
\textbf{(3) safe assignment locking}, 
\textbf{(4) consistency enforcement}, and 
\textbf{(5) batched constellation editing}. 
Together, these principles enable scalable, adaptive orchestration over evolving task graphs while preserving correctness and efficiency.

\subsection{Event-Driven Coordination}
Traditional DAG schedulers rely on polling or global checkpoints to detect task completion, introducing latency and synchronization overhead. In contrast, the Constellation Orchestrator operates as a fully \textbf{event-driven} system built on an internal event bus and an observer design pattern.

Two key components manage this process: the \texttt{ConstellationProgressObserver}, which tracks task execution and orchestrates DAG edits, and the \texttt{ConstellationModificationSynchronizer}, which ensures global consistency after each modification. The orchestrator emits and reacts to four primary event types:
\begin{enumerate}
    \item \texttt{TASK\_STARTED}: triggered when a \TaskStar is assigned to a device and begins execution.
    \item \texttt{TASK\_COMPLETED}: emitted upon successful task completion, prompting potential DAG updates.
    \item \texttt{TASK\_FAILED}: emitted on failure, triggering re-planning or fallback logic.
    \item \texttt{CONSTELLATION\_MODIFIED}: emitted once DAG edits are committed and synchronized across agents.
\end{enumerate}
These events collectively capture the lifecycle of each task and the evolution of the constellation. All handlers operate asynchronously, ensuring immediate, fine-grained reactions to runtime signals without centralized coordination delays. This event-driven design provides high responsiveness and forms the foundation of adaptive orchestration in \ufo.

\subsection{Asynchronous Scheduling}
\label{sec:async}
\begin{figure}[t]
  \centering
  \includegraphics[width=\textwidth]{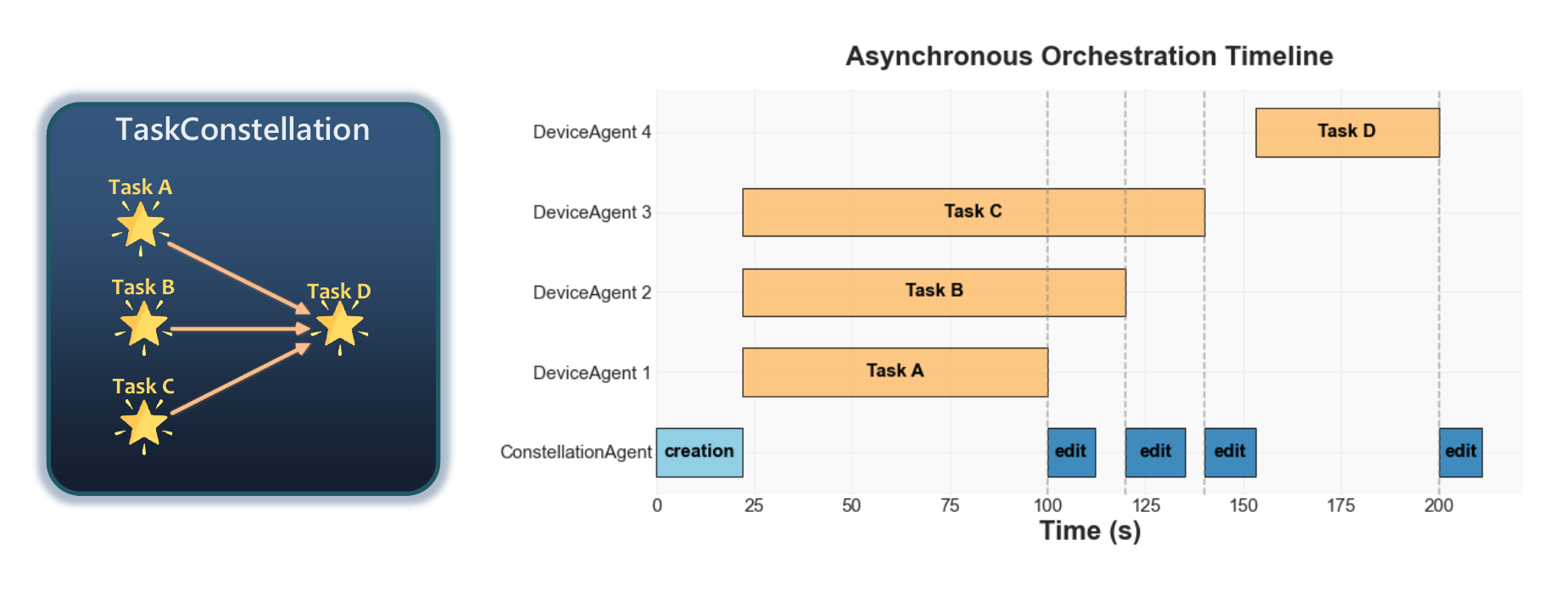}
  \vspace{-2.5em}
  \caption{Illustration of asynchronous scheduling and concurrent \TaskConstellation editing.}
  \label{fig:async_timeline}
\end{figure}

At the core of the orchestrator lies a fully \textbf{asynchronous scheduling loop}. Unlike traditional schedulers that alternate between discrete planning and execution phases, the orchestrator continuously monitors the evolving DAG to identify \emph{ready} \TaskStar, those whose dependencies are satisfied, and dispatches them concurrently to available devices. Each \TaskStar runs within an \texttt{asyncio} coroutine that encapsulates its full lifecycle: execution, result collection, and event publication. When a task starts, a \texttt{TASK\_STARTED} event is emitted; upon completion or failure, corresponding \texttt{TASK\_COMPLETED} or \texttt{TASK\_FAILED} events are immediately published to trigger downstream orchestration updates.

Notably, \TaskStar execution and \TaskConstellation editing can proceed concurrently. As shown in Figure~\ref{fig:async_timeline}, when Task A completes and triggers an edit, the edit operation executes in parallel with the ongoing Tasks B and C. Similarly, edits triggered by Task B overlap with subsequent executions. This concurrency further reduces end-to-end latency by overlapping computation and orchestration, allowing the system to adapt in real time as results stream in.

This asynchronous design maximizes device utilization and eliminates idle waiting, enabling parallel progress across heterogeneous devices. It is essential for scaling cross-device workflows, where independent subtasks (\eg log collection, file aggregation, or model execution) can execute concurrently while higher-level orchestration continuously adapts to dynamic task states and outcomes.

\subsection{Safe Assignment Locking and Synchronization}
\begin{algorithm}[t]
\caption{Safe Assignment Locking and Asynchronous Rescheduling Protocol}
\label{alg:safe-locking}
\KwIn{Event stream $\mathcal{E}$, current \TaskConstellation $\mathcal{C}$}
\KwOut{Consistent and updated $\mathcal{C}$ with newly scheduled ready tasks}
\While{system is running}{
    \ForEach{event $e \in \mathcal{E}$}{
        \If{$e$ is \texttt{TASK\_COMPLETED} or \texttt{TASK\_FAILED}}{
            \textbf{async} enqueue($e$) \tcp*{record completion/failure for processing asynchronously}
        }
    }

    acquire(\texttt{assign\_lock}) \tcp*{suspend new assignments}
    
    \While{queue not empty}{
        $e \gets$ dequeue() \tcp*{get next event for processing}
        
        $\Delta$ = invoke(\cagent, edit($\mathcal{C}$, $e$)) \tcp*{propose DAG edits}
        $\mathcal{C} \gets$ apply($\mathcal{C}$, $\Delta$) \tcp*{update the constellation structure}
        validate($\mathcal{C}$) \tcp*{ensure acyclicity and invariants (I1–I3)}
        publish(\texttt{CONSTELLATION\_MODIFIED}, $t$) \tcp*{notify DAG update}
        $\mathcal{C} \gets$ synchronize($\mathcal{C}$, $\mathcal{T}_{C}$) \tcp*{merge newly completed \TaskStars} 
    }
    
    release(\texttt{assign\_lock}) \tcp*{resume orchestration after all queued events are processed}            

    \textbf{// Rescheduling Phase (outside lock)}
    
    $\mathcal{T}_{R} \gets$ get\_ready\_tasks($\mathcal{C}$) \tcp*{collect newly ready \TaskStars} 
    \ForEach{$t \in \mathcal{T}_{R}$}{
        \textbf{async} dispatch($t$) \tcp*{send to available device agent asynchronously}
        \textbf{async} publish(\texttt{TASK\_STARTED}, $t$) \tcp*{notify asynchronously}
    }  
}
\end{algorithm}

\begin{figure}[t]
    \centering
    \includegraphics[width=\textwidth]{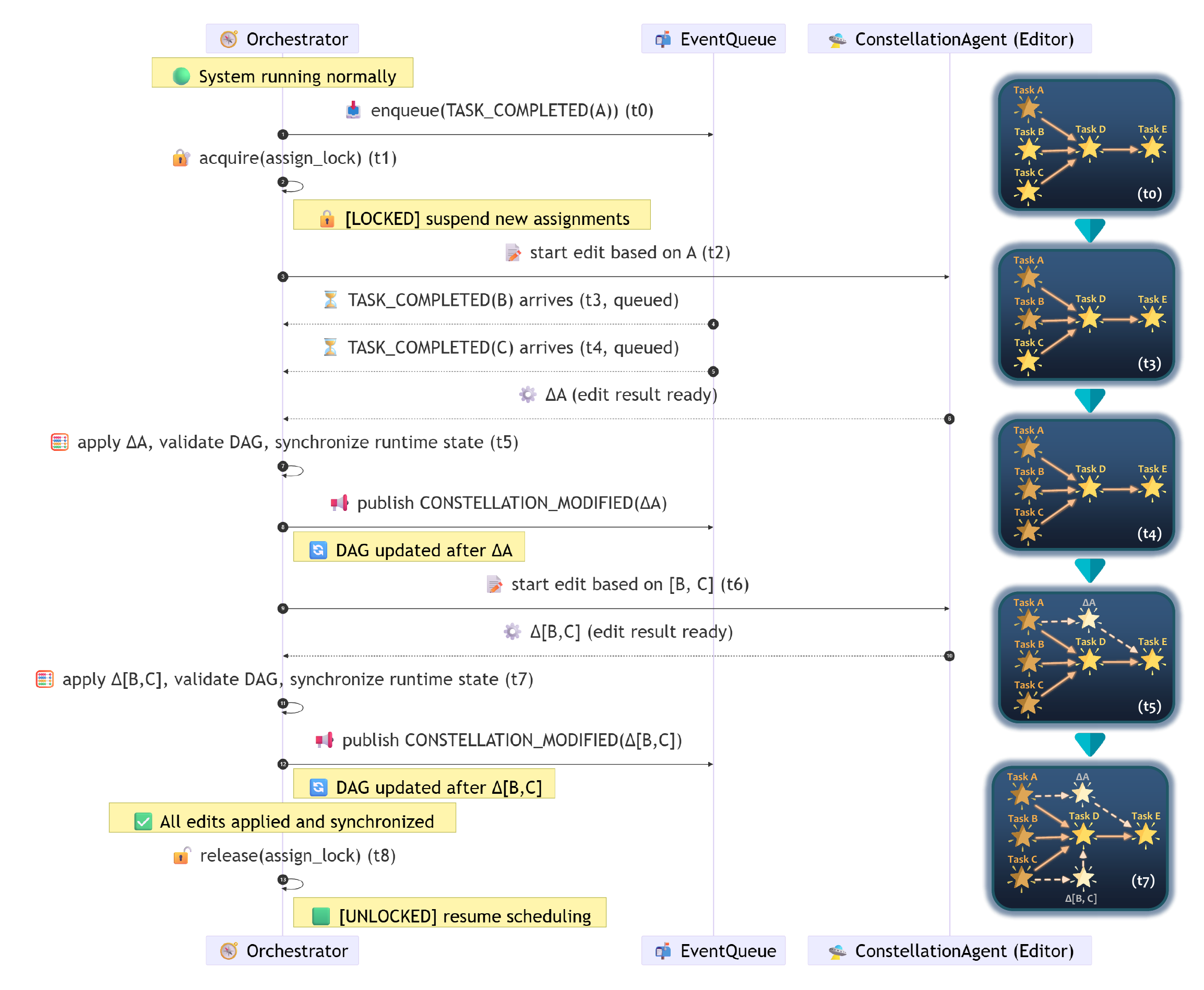}
    \caption{An example of the safe assignment locking and event synchronization workflow.}
    \label{fig:lock}
\end{figure}

While asynchrony improves efficiency, it introduces correctness challenges when task execution overlaps with DAG updates. Specifically, the orchestrator must prevent race conditions when the \cagent dynamically adds, removes, or rewires \TaskStar during execution. Without safeguards, a task could be dispatched based on a stale DAG, leading to duplicated or invalid execution.

To ensure atomicity, the orchestrator employs a \textbf{safe assignment lock}. When an edit cycle begins, bounded by a \texttt{TASK\_COMPLETED}/\texttt{TASK\_FAILED} event and its corresponding \texttt{CONSTELLATION\_MODIFIED} event, the scheduler suspends new task assignments to prevent dispatching based on stale DAG states. During this period, incoming events are queued, and the \texttt{ConstellationModificationSynchronizer} guarantees that all edits are applied atomically. A key aspect of this process is \textbf{synchronization}: once the \cagent finishes editing, the orchestrator merges its structural changes with runtime updates from concurrently running tasks (\ie new completions or failures that occurred during the editing window). This ensures that the final \TaskConstellation reflects a globally consistent view of both reasoning-time modifications and execution-time progress. 

The complete protocol is summarized in Algorithm~\ref{alg:safe-locking}, which demonstrates how locking, validation, synchronization, and rescheduling jointly ensure correctness and consistency under concurrent task updates. Figure~\ref{fig:lock} further illustrates this process. When multiple \texttt{TASK\_COMPLETED} events arrive simultaneously, the orchestrator acquires a global lock to prevent inconsistent scheduling decisions. During this locked phase, the \cagent performs DAG modifications (\eg $\Delta$A, $\Delta$[B, C]), which are atomically merged, validated, and synchronized with the live constellation state. Once synchronization completes, the lock is released, and normal scheduling resumes based on the updated DAG. Each edit cycle is linearized (publish--release) and validated before scheduling resumes; the linearization argument and TLA$^{+}$ model are provided in Appendix~\ref{app:formal} and~\ref{app:tla}.

This mechanism guarantees that task assignments remain immutable during edits, preventing conflicts between execution and modification. By maintaining atomicity and synchronization without blocking overall progress, it preserves both safety and consistency for concurrent, LLM-driven orchestration at scale.

\subsection{Consistency and Safety Guarantees}
\label{subsec:orchestrator-consistency}
Since the DAG may be dynamically rewritten by an LLM, the orchestrator enforces runtime invariants to preserve correctness even under partial or invalid updates:
\begin{itemize}
  \item \textbf{I1 (Single Assignment):} Each \TaskStar has at most one active device assignment at any time.  
  \item \textbf{I2 (Acyclic Consistency):} Edits must preserve DAG acyclicity; the orchestrator performs local cycle detection before committing modifications.  
  \item \textbf{I3 (Valid Update):} Only \texttt{PENDING} tasks and their dependent nodes may be modified; \texttt{RUNNING}, \texttt{COMPLETED}, and \texttt{FAILED} nodes are immutable. 
\end{itemize}
Together, these invariants ensure that even as the constellation evolves, new stars form, old ones fade, the overall structure remains stable and semantically valid. We enforce three runtime invariants (I1--I3) under a lock-bounded editing regime; see Appendix~\ref{app:formal} for the formal state model and proof sketch.

\subsection{Batched Constellation Editing}
Frequent LLM-driven edits can introduce significant overhead if processed individually. To balance responsiveness with efficiency, the orchestrator supports \textbf{batched constellation editing}. During a reasoning round, multiple \texttt{TASK\_COMPLETED} or \texttt{TASK\_FAILED} events may accumulate; instead of invoking the \cagent after each event, the orchestrator aggregates them and applies the resulting modifications atomically once reasoning is complete. As illustrated in Figure~\ref{fig:lock}, when \texttt{task\_A} completes ($t_0$), the \cagent starts an edit cycle. During this process, new completion events from \texttt{task\_B} and \texttt{task\_C} arrive ($t_3$, $t_4$) and are temporarily queued. After the first edit result~$\Delta_A$ is validated and synchronized ($t_5$), the orchestrator batches the pending updates from \texttt{B} and \texttt{C} into a single reasoning round ($t_6$–$t_7$).

This batching mechanism amortizes LLM invocation and synchronization overhead while preserving atomicity and consistency. It enables the orchestrator to remain both efficient and adaptive, reacting swiftly to meaningful state transitions without incurring excessive micro-edits or redundant reasoning calls. We prove an edit--sync confluence lemma showing that folding runtime events commutes with lock-bounded edits within the same window; see Appendix~\ref{app:formal} (Edit--Sync Confluence).

\paragraph{Design Summary.}
In contrast to static DAG or synchronous schedulers, the Constellation Orchestrator treats task execution as an \textbf{open-world process}, continuously evolving, reacting, and converging toward user intent. Its event-driven backbone ensures responsiveness; asynchronous scheduling maximizes concurrency; locking and batching ensure safety; and DAG validity checks preserve correctness under dynamic reasoning.  

Together, these components realize a new form of orchestration, \emph{asynchronous, adaptive, and reasoning-aware}, that bridges declarative intent and distributed execution across a heterogeneous universe of intelligent agents.

\section{Agent Interaction Protocol (AIP)} 
\label{sec:aip}
The orchestration model described in Section~\ref{sec:orchestrator} requires a communication substrate that remains \emph{correct under continuous DAG evolution}, \emph{dynamic agent participation}, and \emph{fine-grained event propagation}. Legacy HTTP-based coordination approaches (\eg A2A~\cite{duan2025agent}, ACP~\cite{ehtesham2025survey}) assume short-lived, stateless interactions, incurring handshake overhead, stale capability views, and fragile recovery when partial failures occur mid-task. These assumptions make them unsuitable for the continuously evolving workflows and long-running reasoning loops characteristic of \ufo.

\subsection{Design Overview}
\label{sec:aip:overview}
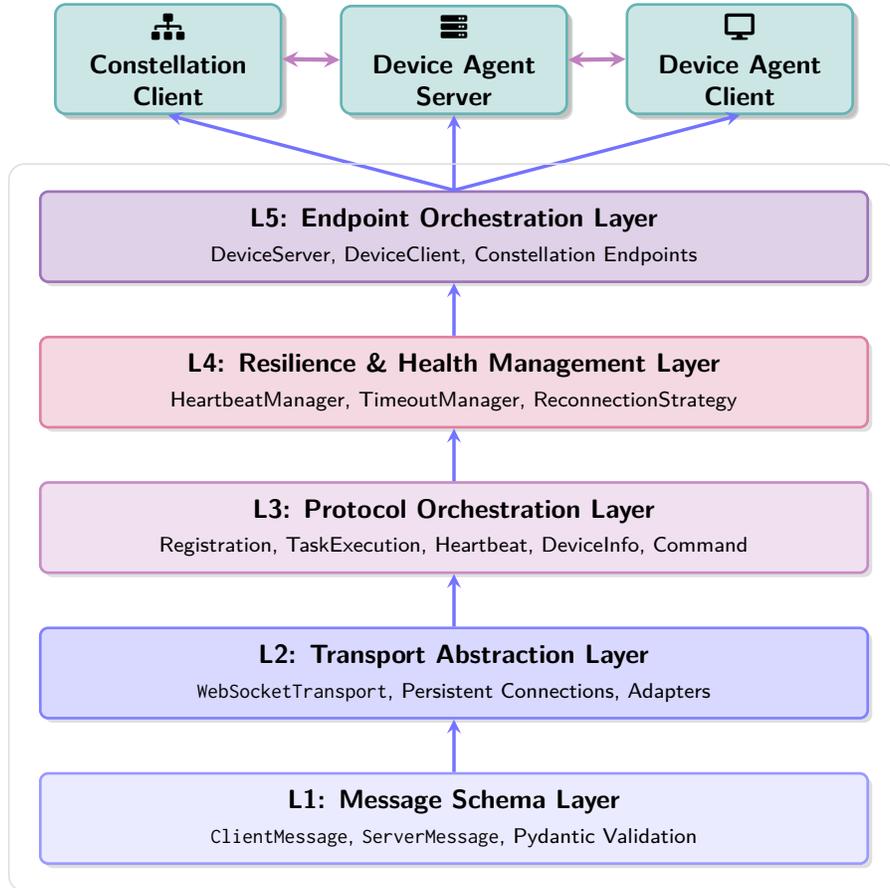
\begin{figure}[t]
    \centering
    \begin{tikzpicture}[
        node distance=0.7cm,
        box/.style={
            rectangle, 
            draw=black!30, 
            line width=1pt,
            minimum width=11cm, 
            minimum height=1.2cm, 
            align=center, 
            font=\small\sffamily,
            rounded corners=3pt,
            drop shadow={shadow xshift=0.6mm, shadow yshift=-0.6mm, opacity=0.25}
        },
        layer1/.style={box, fill=blue!8, draw=blue!40},
        layer2/.style={box, fill=blue!15, draw=blue!50},
        layer3/.style={box, fill=violet!12, draw=violet!45},
        layer4/.style={box, fill=purple!15, draw=purple!50},
        layer5/.style={box, fill=indigo!18, draw=indigo!55},
        endpoint/.style={
            box, 
            fill=teal!20, 
            draw=teal!60,
            minimum width=3cm,
            minimum height=1.05cm,
            font=\normalsize\sffamily\bfseries,
            rounded corners=4pt,
            drop shadow={shadow xshift=0.8mm, shadow yshift=-0.8mm, opacity=0.3}
        },
        arrow/.style={
            ->, 
            >=stealth, 
            line width=1.3pt,
            draw=blue!55
        },
        bidir/.style={
            <->, 
            >=stealth, 
            line width=1.6pt, 
            draw=violet!50
        }
    ]
        \node[layer1] (L1) at (0,0) {
            \textbf{\normalsize L1: Message Schema Layer}\\[2pt]
            {\footnotesize\texttt{ClientMessage}, \texttt{ServerMessage}, Pydantic Validation}
        };
        
        \node[layer2] (L2) [above=of L1] {
            \textbf{\normalsize L2: Transport Abstraction Layer}\\[2pt]
            {\footnotesize\texttt{WebSocketTransport}, Persistent Connections, Adapters}
        };
        
        \node[layer3] (L3) [above=of L2] {
            \textbf{\normalsize L3: Protocol Orchestration Layer}\\[2pt]
            {\footnotesize Registration, TaskExecution, Heartbeat, DeviceInfo, Command}
        };
        
        \node[layer4] (L4) [above=of L3] {
            \textbf{\normalsize L4: Resilience \& Health Management Layer}\\[2pt]
            {\footnotesize HeartbeatManager, TimeoutManager, ReconnectionStrategy}
        };
        
        \node[layer5] (L5) [above=of L4] {
            \textbf{\normalsize L5: Endpoint Orchestration Layer}\\[2pt]
            {\footnotesize DeviceServer, DeviceClient, Constellation Endpoints}
        };
        
        \node[endpoint] (constellation) [above=1cm of L5, xshift=-3.8cm] {
            \faSitemap\\[2pt]
            Constellation\\Client
        };
        \node[endpoint] (server) [above=1cm of L5] {
            \faServer\\[2pt]
            Device Agent \\ Server
        };
        \node[endpoint] (client) [above=1cm of L5, xshift=3.8cm] {
            \faDesktop\\[2pt]
            Device Agent \\Client
        };
        
        \draw[arrow] (L1) -- (L2);
        \draw[arrow] (L2) -- (L3);
        \draw[arrow] (L3) -- (L4);
        \draw[arrow] (L4) -- (L5);
        
        \draw[arrow] (L5.north) -- (constellation.south);
        \draw[arrow] (L5.north) -- (server.south);
        \draw[arrow] (L5.north) -- (client.south);
        
        \draw[bidir] (constellation.east) -- (server.west);
        \draw[bidir] (server.east) -- (client.west);
        
        \draw[draw=gray!25, line width=0.6pt, rounded corners=6pt]
            ([xshift=-0.4cm, yshift=-0.35cm]L1.south west) rectangle 
            ([xshift=0.4cm, yshift=0.35cm]L5.north east);
        
    \end{tikzpicture}
    \caption{AIP Architecture: Five-layer protocol stack enabling persistent, resilient multi-agent orchestration.}
    \label{fig:aip-overview}
\end{figure}

AIP serves as the \textbf{nervous system} of \ufo, connecting the ConstellationClient, device agent services, and device clients under a unified, event-driven control plane, as shown in Figure~\ref{fig:aip-overview}.
It is designed as a lightweight yet evolution-tolerant protocol to satisfy six goals: 
\begin{enumerate}[label=\textit{(G\arabic*})]
    \item Maintain persistent bidirectional sessions to eliminate per-request overhead;
    \item Unify heterogeneous capability discovery via multi-source profiling;
    \item Ensure fine-grained reliability through heartbeats and timeout managers for disconnection and failure detection;
    \item Preserve deterministic command ordering within sessions;
    \item Support composable extensibility for new message types and resilience strategies;
    \item Provide transparent reconnection and task continuity under transient failures.
\end{enumerate}

To meet these requirements, AIP adopts a persistent, bidirectional WebSocket transport and decomposes the
orchestration substrate into \textbf{five} logical strata (Figure~\ref{fig:aip-overview}), each responsible for a distinct aspect of reliability and adaptability:
\begin{itemize}[leftmargin=2em]
    \item \textbf{L1: Message Schema Layer} — Defines strongly-typed, Pydantic-validated contracts (\texttt{ClientMessage}, \texttt{ServerMessage}) for message direction, purpose, and task transitions. Structured metadata (system info, capabilities) supports unified capability discovery (\textit{G2}) and deterministic ordering via explicit ID correlation (\textit{G4}).

    \item \textbf{L2: Transport Abstraction Layer} — Provides a protocol-agnostic \texttt{Transport} interface with a production-grade WebSocket implementation supporting configurable pings, timeouts, and large payloads. Decoupled transport logic ensures low-latency persistent sessions (\textit{G1}) and future extensibility (\textit{G5}).
    
    \item \textbf{L3: Protocol Orchestration Layer} — Implements modular handlers for registration, task execution, heartbeat, and command dispatch (see Appendix~\ref{app:aip-messages}), each extending a common \texttt{AIPProtocol} base with middleware hooks (logging, metrics, auth). This design ensures ordered state transitions (\textit{G4}) and composable extensibility (\textit{G5}).
    
    \item \textbf{L4: Resilience and Health Management Layer} — Encapsulates \texttt{HeartbeatManager}, \texttt{TimeoutManager}, and \texttt{ReconnectionStrategy} with exponential backoff and automatic session recovery. It guarantees reliability (\textit{G3}) and seamless task continuity under transient disconnections (\textit{G6}).
    
    \item \textbf{L5: Endpoint Orchestration Layer} — Provides role-specific facades: \texttt{DeviceServerEndpoint}, \texttt{DeviceClientEndpoint}, and \texttt{ConstellationEndpoint}, integrating lower layers into deployable components. These endpoints unify connection lifecycle, task routing, and health monitoring across roles, reinforcing \textit{G1}–\textit{G6}.
\end{itemize}

Together, these layers form a vertically integrated stack where \textbf{L1} establishes semantic contracts, \textbf{L2} provides transport flexibility, \textbf{L3} implements protocol logic, \textbf{L4} ensures operational resilience, and \textbf{L5} delivers deployment-ready orchestration primitives. This design enables \ufo to maintain correctness and availability under DAG evolution (\textit{G4}, \textit{G5}), agent churn (\textit{G3}, \textit{G6}), and heterogeneous execution environments (\textit{G1}, \textit{G2}).

\subsection{Agent Registration and Profiling}
\label{subsec:aip-registration}

Agent registration in AIP corresponds to the entry point of the orchestration pipeline, anchoring the capability discovery and topology formation processes outlined in \textit{(G2)} and implemented primarily through the \textbf{L1–L3 layers}.
\begin{figure}[t]
  \centering
  \includegraphics[width=\textwidth]{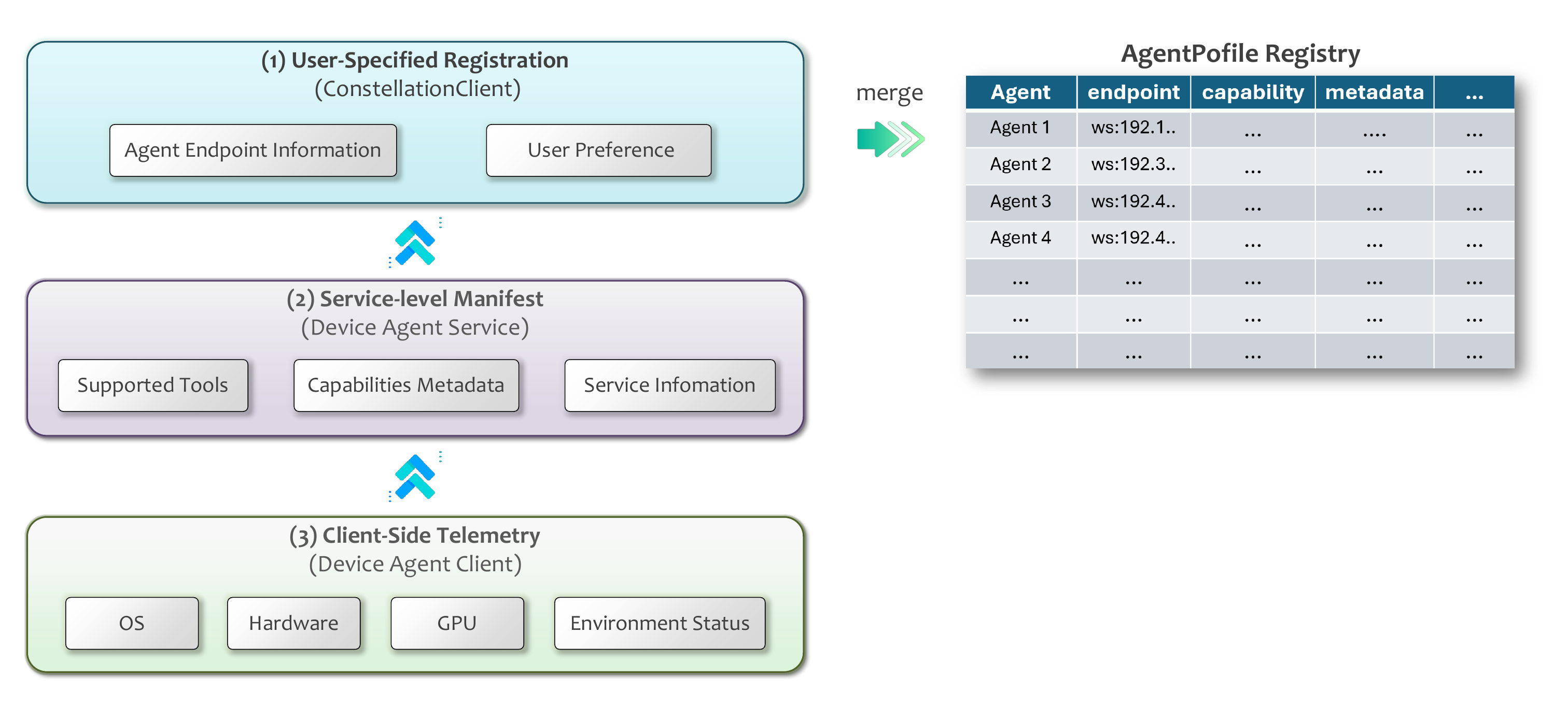}
  \vspace{-2em}
  \caption{Agent registration flow: multi-source AgentProfile construction and registration.}
  \label{fig:agent-registration}
\end{figure}

As illustrated in Figure~\ref{fig:agent-registration}, the registration pipeline consists of three complementary stages that together establish a unified and continuously refreshed view of the constellation's capabilities:
\begin{enumerate}
    \item \textbf{User-specified registration (ConstellationClient).} Administrators provide endpoint identities and specify user preferences. These initial configurations define the logical boundaries of the constellation and seed the connection parameters required for session establishment.
    
    \item \textbf{Service-level manifest (device agent service).} Each device agent advertises its supported tools, environment variables, and operational metadata through a \texttt{REGISTER} message. These descriptors are validated and normalized by the \textbf{Message Schema Layer (L1)} and merged through the \textbf{Protocol Orchestration Layer (L3)}, ensuring semantic consistency and structured capability discovery.
    
    \item \textbf{Client-side telemetry (device agent client).} Local clients continuously report runtime metrics, such as OS version, hardware status, GPU utilization, and software environment, to the device agent service. This telemetry stream keeps the global registry up to date and enables adaptive re-scheduling under resource drift or device churn.
\end{enumerate}

\begin{figure}[t]
\centering
\scalebox{0.85}{
\begin{tcolorbox}[
  enhanced,
  width=0.92\linewidth,
  colback=CardBG,
  colframe=CardBorder,
  boxrule=0.4pt,
  arc=2mm,
  sharp corners=south,
  drop shadow=black!10
]
\begin{tikzpicture}[font=\sffamily]
\node (title) at (0,0) [anchor=west, text=Accent] {\Large \textbf{AgentProfile:} \texttt{gpu\_agent}};
\node (status) [right=1cm of title, anchor=west, draw=CardBorder,
  rounded corners=2pt, inner xsep=6pt, inner ysep=2pt] 
  {\textbf{Status:} \textcolor{Good}{idle}};
\end{tikzpicture}

\vspace{4pt}\hrule height0.4pt \color{CardBorder}\vspace{6pt}

\begin{minipage}[t]{0.48\linewidth}
\textbf{System}\\
OS: \texttt{linux} \quad(\texttt{Ubuntu 24.04})\\[2pt]
Capabilities: \texttt{cli}, \texttt{file\_system}, \texttt{data\_processing}, \texttt{model\_training}\\[6pt]

\textbf{Paths}\\
Dataset: \texttt{\textasciitilde/dataset}\\
Training: \texttt{\textasciitilde/script}\\
Code: \texttt{\textasciitilde/code}\\
\end{minipage}\hfill
\begin{minipage}[t]{0.48\linewidth}
\textbf{Performance}\\
GPU: \texttt{4 $\times$ NVIDIA A100 80G}\\
CPU: \texttt{96 cores}\\
Memory: \texttt{866.1 GB}\\[6pt]

\textbf{Network \& Host (optional)}\\
Host: \texttt{gui-model-a100}\\
IP: \texttt{172.19.0.4}
\end{minipage}

\vspace{6pt}\hrule height0.4pt \color{CardBorder}\vspace{4pt}
{\footnotesize Last heartbeat: 2025-10-28 07:46:24 UTC\quad | \quad Server: \texttt{ws://localhost:5005/ws}}
\end{tcolorbox}
}
\caption{An example \textbf{AgentProfile} of a GPU agent with Linux system.}
\label{fig:agentprofile-card}
\end{figure}

The aggregated results from these three stages are merged by the ConstellationClient into a unified \textit{AgentProfile} that represents each agent's real-time operational state. An example \textit{AgentProfile} for a GPU-enabled device is shown in Fig.~\ref{fig:agentprofile-card}, demonstrating how multi-level profiling captures hardware, software, and dynamic runtime descriptors within a single schema.

Through this registration and profiling pipeline, AIP achieves continuous capability discovery, evolution-tolerant orchestration, and consistent topology awareness across heterogeneous devices. The process directly fulfills \textit{(G1–G3)} by maintaining persistent sessions, ensuring reliable metadata propagation, and enabling transparent adaptation as the constellation evolves.

\subsection{Task Dispatch and Result Delivery}
\label{subsec:aip-dispatch}
Task dispatch operationalizes the event-driven execution model envisioned in \textit{(G1)} and \textit{(G4)} through tightly managed sessions that persist across multiple task rounds. When the ConstellationClient assigns a \TaskStar to a device, the Transport Abstraction Layer guarantees low-latency delivery of the serialized \texttt{TASK} message to the target agent service. The Protocol Orchestration Layer coordinates message routing, while the Resilience and Health Management Layer monitors the session heartbeat to ensure reliability.

Each task follows a deterministic life cycle, from \texttt{TASK} to \texttt{TASK\_END}, with strict ordering guarantees enforced by the session-level sequence manager. Intermediate logs and evaluator outputs are streamed back incrementally to the ConstellationClient, which updates the global \TaskConstellation state and triggers potential DAG adjustments.

This continuous, feedback-driven execution loop transforms AIP from a mere transport protocol into a temporal coordination substrate, harmonizing asynchronous reasoning, scheduling, and execution across distributed devices.

\subsection{Command Execution}
\label{subsec:aip-commands}
At a finer operational granularity, AIP implements a unified \textbf{command execution model} that directly fulfills \textit{(G4)} and \textit{(G5)} by ensuring deterministic, extensible control within persistent sessions.

Each \texttt{COMMAND} message specifies a unique identifier, target function, and a typed argument list. The Message Schema Layerenforces structure and validation, while the Protocol Orchestration Layer executes commands sequentially within the session context to maintain determinism. To optimize multi-action workflows, multiple commands may be batched in a single message, reducing round-trip overhead. Execution results are returned as structured \texttt{Result} objects, containing status codes, return values, and error metadata. Failures or timeouts are propagated through the same channel, enabling the orchestrator to apply adaptive recovery or task reassignment strategies via \textbf{L4}'s resilience mechanisms.

Unified with \ufo's MCP tool-calling interface, this model bridges system-level orchestration with model-generated actions, ensuring that high-level reasoning and low-level execution operate under a consistent and evolvable protocol surface.

\subsection{Resilient Connection Protocol}
\label{subsec:aip-resilience}

The distributed and volatile nature of device environments necessitates a dedicated resilience layer to uphold \textit{(G3)} and \textit{(G6)}. AIP's \textbf{Resilient Connection Protocol}, implemented primarily within \textbf{L4}, guarantees synchronized fault handling and seamless recovery across client–server boundaries.

When a \textit{Device Agent} disconnects unexpectedly, the orchestrator immediately marks it as \texttt{DISCONNECTED}, removes it from the active scheduling pool, and triggers background reconnection attempts using exponential backoff. Upon recovery, the agent re-registers automatically, restoring its prior session state and resuming task participation without manual intervention.
If disconnection occurs during task execution, all affected tasks are transitioned to \texttt{TASK\_FAILED}, and corresponding updates are propagated to the \cagent for DAG revision, ensuring that the orchestration view remains globally consistent.

Symmetrically, when the \textit{ConstellationClient} itself disconnects, the corresponding \textit{Device Agent Server} receives a termination signal and \emph{proactively aborts} all ongoing tasks associated with that client. This bidirectional fault-handling policy prevents resource leakage, avoids orphaned execution states, and guarantees that both endpoints maintain a consistent global view of task progress.

Together, these mechanisms realize an end-to-end resilient orchestration substrate that preserves correctness, availability, and synchronization under transient network failures or partial system outages,closing the reliability loop envisioned in AIP's layered design.

\paragraph{Summary.}
AIP consolidates registration, task dispatch, command execution, and resilience into a coherent, evolution-tolerant communication fabric that embodies the six goals outlined in Section~\ref{sec:aip:overview}. Functionally, AIP forms the \textbf{nervous system of \ufo}, enabling reasoning, execution, and recovery to operate seamlessly within an evolving constellation of intelligent agents. Its minimal yet extensible design ensures that as workflows, models, and environments evolve, the underlying communication protocol remains stable, adaptive, and correct by construction.

\section{Design and Development of Device Agents}
\label{sec:device-agent}
With the structured reasoning of the \cagent, the dynamic execution model of the Constellation Orchestrator, and the low-latency, persistent communication enabled by AIP, the next challenge is clear: how do we design a device agent that can be quickly onboarded into \ufo, adapt to a heterogeneous platform, and seamlessly participate in evolving task constellations? Our solution provides a standardized template for device agent development, minimizing engineering effort while maximizing system scalability and reliability.

\begin{figure}[t]
  \centering
  \includegraphics[width=0.7\textwidth]{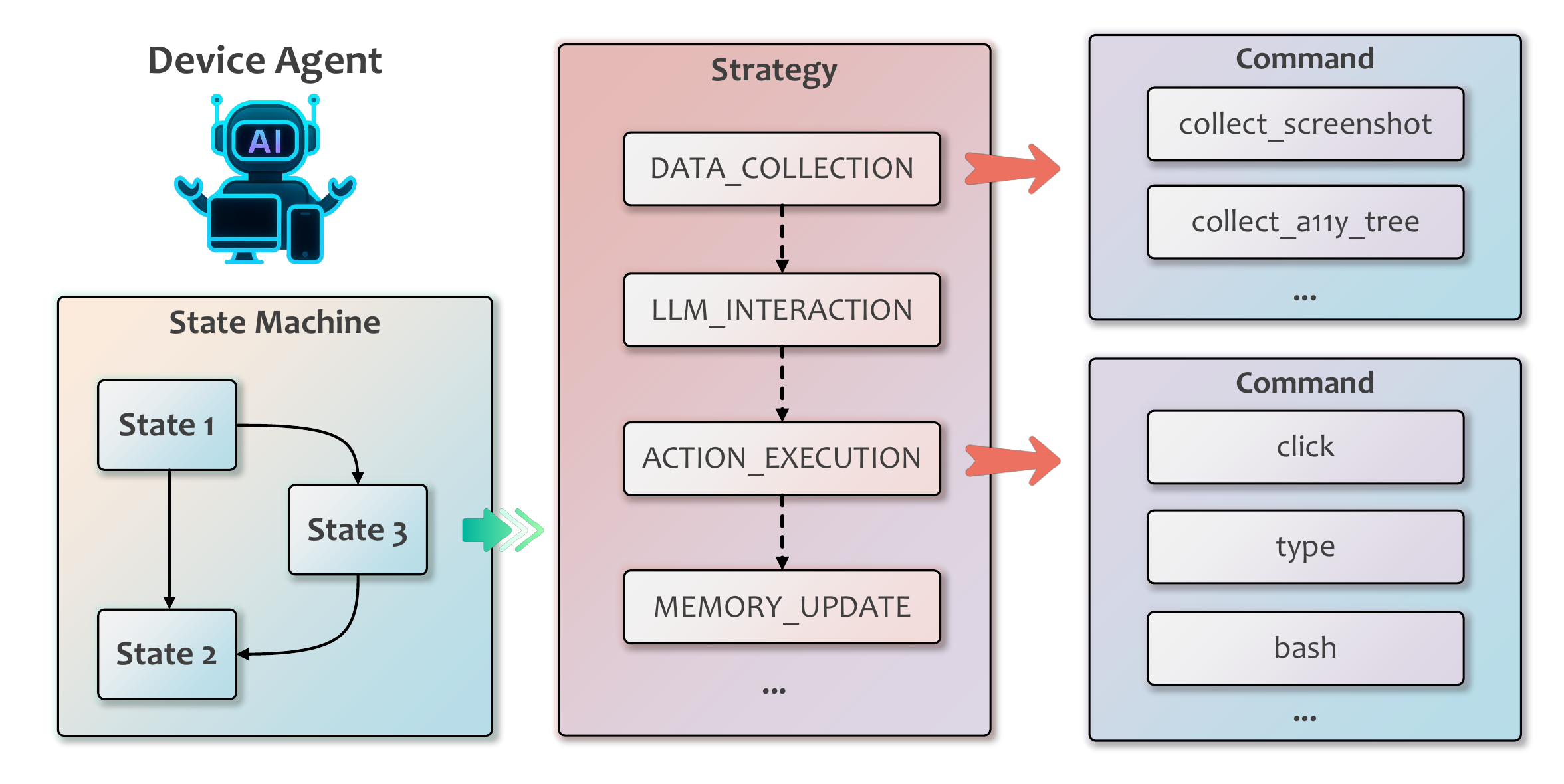}
  \vspace{-1em}
  \caption{The three-layer framework of a device agent.}
  \label{fig:device-agent}
\end{figure}

\subsection{Architecture Overview}  
A device agent serves as the execution endpoint of \ufo, translating high-level task directives into concrete actions on a target device. To support rapid development and flexible deployment, each agent is structured as a \textbf{three-layer, state-machine-driven framework}, illustrated in Figure~\ref{fig:device-agent}. To enable safe and scalable execution across heterogeneous devices, each agent is further partitioned into a \textbf{server} that manages orchestration and FSM logic, and a \textbf{client} that executes low-level commands locally. This server-client separation, combined with the layered FSM design, balances \textbf{modularity, extensibility, and runtime robustness}, and supports both single-agent and multi-agent deployment scenarios depending on platform requirements.

Specifically, the architecture decomposes agent behavior into three hierarchical levels:  
\begin{enumerate}
    \item \textbf{Level-1: State (Finite-State Machine Layer).}  
    Each agent maintains an internal state machine that governs its behavior at each execution step. A state encapsulates a \emph{processor}, the next state to transition to, and optionally the next agent to invoke in multi-agent setups. The collection of states defines a finite-state machine that ensures predictable, controllable lifecycle progression.

    \item \textbf{Level-2: Strategy (Execution Logic Layer).}  
    Within each state, the processor manages a sequence of \emph{strategies} that implement the step-level workflow. Strategies handle tasks such as data collection, environment inspection, prompt construction, action planning, or tool invocation. This separation allows the agent to \textbf{compose complex behaviors from modular, reusable strategies}, while maintaining clear boundaries between decision logic and execution mechanics.

    \item \textbf{Level-3: Command (System Interface Layer).}  
    Each strategy can invoke a set of \emph{commands} from a configured MCP server, which provides standardized operations for perceiving system state, executing tools, or interacting with device resources. Commands are executed deterministically and report structured outcomes, allowing higher layers to react and adapt without managing low-level device specifics.
\end{enumerate}
By instantiating an agent with concrete \emph{State}, \emph{Strategy}, and \emph{Command} definitions, a fully functional device agent can be realized. This hierarchical, layered approach offers several advantages. First, the same framework can accommodate a wide variety of devices, platforms, and execution environments.  Second, developers can onboard new devices by defining only the relevant states, strategies, and commands, without rewriting orchestration or communication logic. Finally, the layered FSM structure allows the agent to respond to dynamic task edits, partial failures, or concurrent executions while maintaining correctness.  

Together, this architecture positions device agents as \textbf{plug-and-play execution units}, seamlessly bridging the high-level reasoning of the \cagent, the dynamic orchestration of the Constellation Orchestrator, and the event-driven communication of AIP. It forms the final, essential layer that enables \ufo to operate as a cohesive, scalable, and resilient multi-device system.

\subsection{Server-Client Architecture}
\begin{figure}[t]
  \centering
  \includegraphics[width=0.75\textwidth]{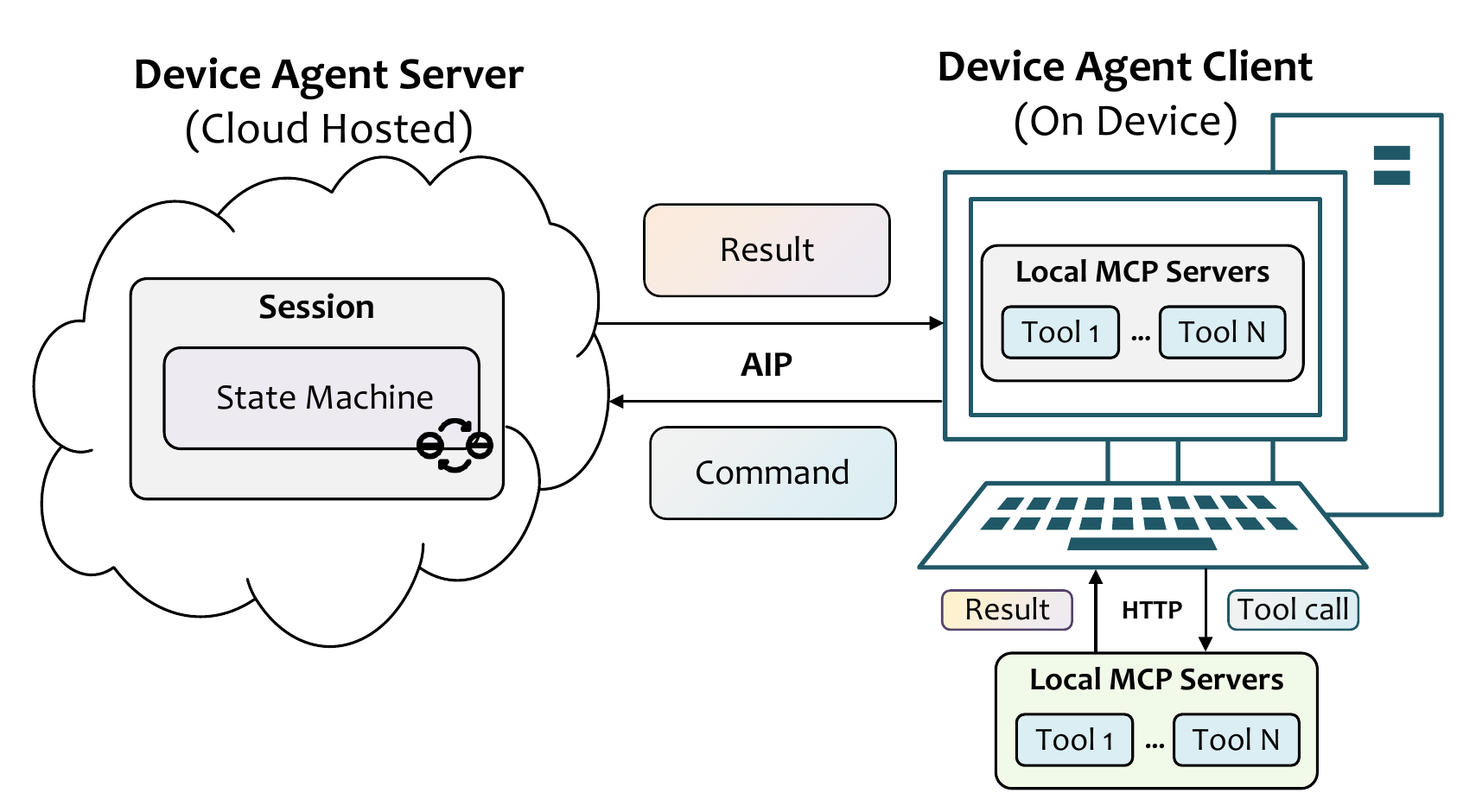}
  \vspace{-0.5em}
  \caption{The server-client architecture of a device agent.}
  \label{fig:server-client}
\end{figure}

To support safe, scalable, and flexible execution across heterogeneous devices, each device agent is partitioned into a server and a client, as illustrated in Figure~\ref{fig:server-client}. This separation of responsibilities aligns naturally with the layered FSM architecture and leverages AIP for reliable, low-latency communication.

\paragraph{Server: Orchestration and State Management.}  
The \emph{agent server} is responsible for managing the agent's state machine lifecycle, executing high-level strategies, and interacting with the \cagent or the orchestrator. It handles task decomposition, prompt construction, decision-making, and command sequencing. Crucially, the server maintains full control over the agent's workflow logic, enabling updates to decision strategies without impacting low-level execution on the device. 

Each server instance exposes its AgentProfile, a structured description of its capabilities, configurations, and runtime status. This metadata allows the orchestrator to dynamically select suitable agents for specific subtasks, improving task distribution efficiency. A single server can manage multiple agent clients concurrently, maintaining isolation across devices while supporting centralized supervision and coordination.

\paragraph{Client: Commands Execution and Resource Access.}  
The \emph{agent client} runs on the target device and manages a collection of MCP servers or tool interfaces. These MCP servers can operate locally (via direct invocation) or remotely (through HTTP requests), and each client may register multiple MCP servers to access diverse tool sources. Upon receiving commands from the agent server, such as collecting telemetry, invoking system utilities, or interacting with hardware components, the client translates them into MCP tool calls, executes them deterministically, aggregates the results, and returns structured outputs via AIP. The client remains stateless with respect to reasoning: it faithfully executes directives without engaging in high-level decision-making.

During initialization, each client connects to the agent server through the AIP endpoint, performs self-checks (\eg disk, CPU, memory, GPU, and network configuration), and registers its hardware–software profile. This profile is integrated into the server's \textit{AgentProfile}, giving the orchestrator complete visibility into system topology and resource availability for informed task assignment and scheduling.

\paragraph{Server-Client Communication.}  
All communication between the server and client is routed through the \emph{AIP}, leveraging persistent WebSocket connections. This allows bidirectional, low-latency messaging that supports both synchronous command execution and asynchronous event reporting. By decoupling high-level reasoning from low-level execution, the system can safely update server logic or client MCP tools independently, without disrupting ongoing workflows.

\paragraph{Design Consideration.}  
This server–client architecture provides strong modularity and scalability. Device clients can be rapidly deployed with minimal setup, immediately joining \ufo as execution endpoints. The server focuses on high-level reasoning and orchestration, while clients ensure deterministic command execution, preventing cross-layer interference and simplifying maintenance. Persistent sessions and structured AIP event semantics enhance robustness under intermittent connectivity and dynamic task updates. The design also scales efficiently across devices: a single server can orchestrate multiple clients, and extensibility is achieved by adding new tools or interfaces at the client side or new reasoning strategies at the server without mutual dependencies.

\subsection{State: Finite-State Machine Layer}

\begin{figure}[t]
\centering
\begin{minipage}{\columnwidth}
\begin{lstlisting}[style=pythonstyle, label={lst:state-interface}]
class AgentState(ABC):
    """Abstract interface for a device agent state."""

    @abstractmethod
    async def handle(self, agent, context=None):
        """Execute the logic for the current state."""
        pass

    @abstractmethod
    def next_state(self, agent) -> "AgentState":
        """Return the next state in the FSM."""
        pass

    @abstractmethod
    def is_round_end(self) -> bool:
        """Determine whether the current task round has ended."""
        pass
\end{lstlisting}
\end{minipage}
\caption{Simplified interface of a device agent state, highlighting the core methods for execution, state transition, and termination checks.}
\vspace{-1em}
\label{fig:state-interface}
\end{figure}
The top-level lifecycle of each device agent is governed by a \textbf{finite state machine (FSM)}, which provides a structured and predictable execution framework. Each state encapsulates the logic for handling a specific step of the agent's workflow, including invoking the corresponding processor, determining the next state to transition to, selecting the next agent in multi-agent setups, and deciding whether the current task has reached completion. Figure~\ref{fig:state-interface} illustrates the interface exposed by a state.

At runtime, the agent invokes the state's \texttt{handle} function, which executes the state-specific behavior (\eg a \texttt{processor}) and returns control decisions to the FSM. Transitions between states can be determined dynamically by the agent based on LLM reasoning, or triggered by rule-based logic in response to errors, timeouts, or external events. This flexibility allows the agent to react promptly to runtime conditions, while maintaining a predictable execution path for normal workflows.

The FSM-based design offers several advantages. First, it enforces a clear separation of concerns: state transitions govern workflow progression, processors implement step-level strategies, and commands handle low-level system interactions. Second, it simplifies reasoning about agent behavior and facilitates debugging, since each state represents an isolated, testable unit. Finally, the FSM enables device agent to safely manage dynamic agent behaviors and failures, and concurrent executions.

In essence, the \emph{State} layer provides a robust backbone for device agent execution, allowing high-level orchestration from the \cagent and Constellation Orchestrator to be reliably translated into stepwise, adaptive actions across heterogeneous devices.

\subsection{Strategy: Composable Execution Logic}
Each agent state delegates step-level workflow management to a \texttt{processor}, which orchestrates a sequence of \emph{strategies}. Strategies encapsulate modular execution logic, enabling fine-grained control over the agent's behavior while maintaining a clear separation between decision-making (State layer) and concrete actions (Command layer).

In a typical processor, we define four core strategy types and execute sequentially:
\begin{itemize}
    \item \texttt{DATA\_COLLECTION:} Gather necessary context from the device, such as screenshots, accessibility information, system status, or user input.
    \item \texttt{LLM\_INTERACTION:} Construct prompts using the collected data and interact with the LLM to obtain actionable instructions or decisions.
    \item \texttt{ACTION\_EXECUTION:} Perform the commands returned by the LLM or pre-defined toolkits, applying them to the device environment deterministically.
    \item \texttt{MEMORY\_UPDATE:} Update the agent's short-term or long-term memory to reflect task progress and provide context for subsequent steps.
\end{itemize}

This strategy-based design offers several advantages. First, it allows \textbf{flexible customization} for different device types and task requirements; strategies can be reordered, added, or replaced without modifying the core state machine. Second, it provide a template that \textbf{modularizes execution}, making the workflow easier to test, debug, and extend. Finally, by clearly delineating data collection, reasoning, action, and memory update, the processor ensures that each step in the agent's lifecycle is \textbf{composable, observable, and adaptable} to dynamic runtime conditions.

Overall, the Strategy layer acts as the operational bridge between the high-level reasoning of the State layer and the low-level Command execution, enabling device agents to carry out complex, multi-step tasks reliably and efficiently.

\subsection{Command: Atomic Execution Units}
Commands represent the \textbf{atomic execution units} within a device agent. Each command encapsulates a specific operation, defined by a \texttt{function} and its corresponding \texttt{arguments}, which maps directly to a tool call on the MCP server co-located with the device client. By treating commands as self-contained units, the agent can systematically decompose complex workflows into discrete, testable actions.

Each Strategy invokes commands to realize its operational intent, for example, a \texttt{DATA\_COLLECTION} strategy might request a screenshot or accessibility tree, while an \texttt{ACTION\_EXECUTION} strategy triggers a system command or UI interaction. Commands are transmitted via the AIP to the client, which performs the actual tool execution on the device and returns structured results back to the server. This separation of command logic from device-level execution ensures \textbf{deterministic, reliable, and auditable operations} across heterogeneous platforms. At runtime, the agent can query the client for available commands and their usage metadata, enabling dynamic selection by the LLM and adaptive workflows. This design supports extensibility: new tools or device capabilities can be integrated by simply registering additional commands at the client layer, without modifying the server-side logic or State/Strategy definitions.

In essence, the Command layer completes the device agent's execution pipeline: it bridges high-level reasoning (State), step-wise workflow orchestration (Strategy), and concrete device interaction, providing a robust and flexible foundation for reliable, multi-step automation across diverse environments.

\subsection{Example Device Agents: LinuxAgent and WindowsAgent}
To demonstrate how the layered device agent architecture and server-client design can be instantiated in practice, we present two representative case studies: \textbf{LinuxAgent} and \textbf{WindowsAgent}. These examples illustrate how \ufo's templates enable rapid development, integration, and execution of device agents across different platforms.

The \textbf{LinuxAgent} showcases a \emph{single-agent deployment}, highlighting how a standalone agent can leverage the FSM-based State, Strategy, and Command layers to interact with system tools, collect telemetry, and execute workflow tasks. In contrast, the \textbf{WindowsAgent} demonstrates a \emph{multi-agent deployment}, where multiple agent instances coordinate via a server-client setup to manage complex workflows involving local MCP servers, UI automation, and dynamic LLM-driven task decomposition. Together, these two case studies provide concrete examples of how \ufo can accommodate diverse execution environments, and serve as the foundational agents for the experimental evaluations in Section~\ref{sec:experiments}.

\subsubsection{LinuxAgent: Single-Agent CLI Execution}
The \textbf{LinuxAgent} is designed as a lightweight, single-agent instance capable of executing command-line instructions to fulfill user requests. It demonstrates how a standalone device agent can leverage the layered FSM architecture and server-client design to perform intelligent, iterative task execution on a CLI-based environment.

\paragraph{State.}  

\begin{figure}[t]
  \centering
  \includegraphics[width=0.6\textwidth]{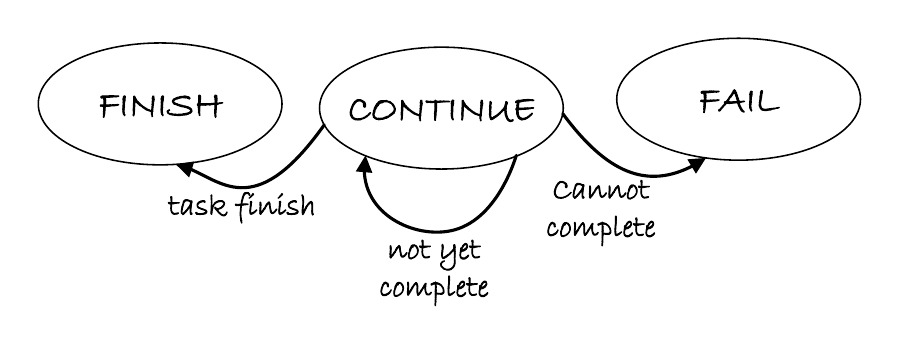}
  \vspace{-1.5em}
  \caption{Lifecycle state transitions of the LinuxAgent.}
  \label{fig:linuxagent-state}
\end{figure}
The LinuxAgent's lifecycle is governed by a minimal set of states, capturing the essential execution progression while maintaining simplicity and predictability:  
\begin{itemize}
    \item \textbf{CONTINUE:} The task is ongoing and requires further CLI command execution to progress toward completion.  
    \item \textbf{FAIL:} The task cannot proceed under current system conditions or resources, signaling an unrecoverable error.  
    \item \textbf{FINISH:} The task has successfully completed all required operations.  
\end{itemize}  
At each step, the agent evaluates execution outcomes and determines the appropriate state transition, allowing the FSM to drive both normal progress and error handling in a structured, deterministic manner. 

\paragraph{Strategy.}  
When in the \textbf{CONTINUE} state, the LinuxAgent executes a processor that orchestrates a small, modular set of strategies:  
\begin{itemize}
    \item \textbf{LLM\_INTERACTION:} Construct prompts using prior execution results and predefined templates to request next-step commands from the LLM.  
    \item \textbf{ACTION\_EXECUTION:} Execute the CLI commands returned by the LLM, ensuring results are captured and structured for downstream processing.  
    \item \textbf{MEMORY\_UPDATE:} Persist execution results and issued commands into the agent's memory for future reference, enabling iterative refinement and error recovery.  
\end{itemize}  
This layered strategy design separates decision-making, such as prompt construction, from execution and state updates, enhancing modularity, reproducibility, and extensibility for future workflow modifications. In particular, unlike traditional polling-based or externally triggered data collection, the agent can \emph{proactively} obtain system and environment information by invoking CLI commands on demand, eliminating unnecessary overhead and increasing responsiveness.

\paragraph{Command.}  
The LinuxAgent interacts with the MCP server via two primary commands:  
\begin{itemize}
    \item \textbf{EXEC\_CLI:} Execute arbitrary shell commands, capturing \texttt{stdout} and \texttt{stderr} for structured feedback.  
    \item \textbf{SYS\_INFO:} Collect system-level information, such as memory usage, disk space, and hardware configuration, to inform decision logic or precondition checks.  
\end{itemize}  
These commands provide the atomic building blocks for the agent's strategies, isolating system-specific operations within the client layer while enabling the server layer to focus on workflow orchestration and LLM-guided reasoning.

The LinuxAgent illustrates the minimal viable design for a single-agent system that integrates FSM control, strategy orchestration, and command execution. By maintaining a small, deterministic state set, modular strategies, and well-defined commands, the agent achieves robust, flexible, and traceable CLI task execution.

\subsubsection{WindowsAgent: Multi-Agent Coordination on GUI Systems}
\begin{figure}[t]
  \centering
  \includegraphics[width=0.7\textwidth]{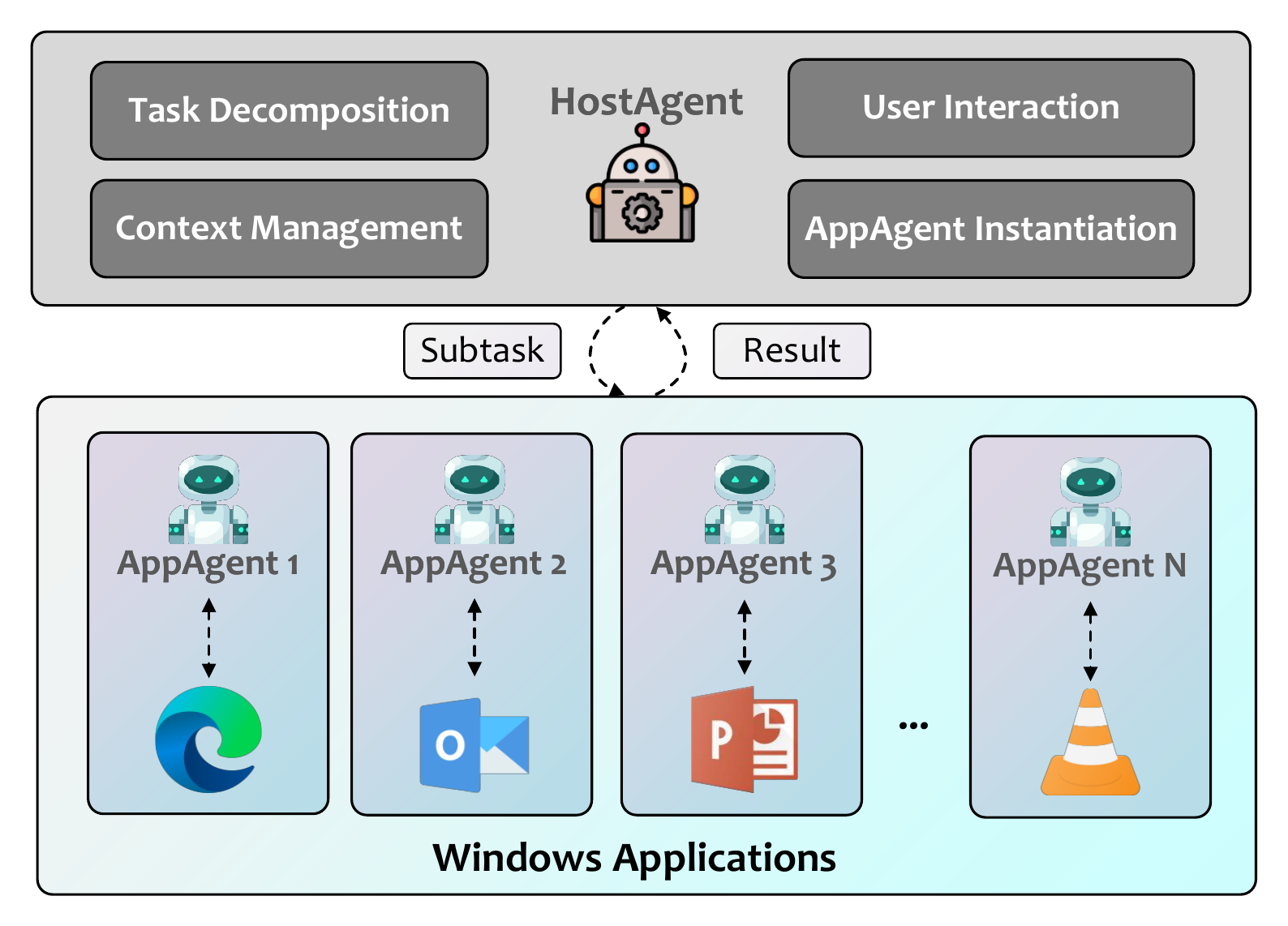}
  \vspace{-.5em}
  \caption{Overall architecture of the WindowsAgent built upon UFO\textsuperscript{2}. Figure adapted from the original paper.}
  \label{fig:windowsAgent}
\end{figure}
While the LinuxAgent represents a lightweight, single-agent model for command-line environments, the WindowsAgent embodies a more sophisticated multi-agent framework tailored for GUI-based systems. We leverage the Desktop AgentOS \textsc{UFO}\textsuperscript{2} as the implementation of the WindowsAgent, which follows the same architectural principles introduced above but extends them to support multi-application coordination. Specifically, \textsc{UFO}\textsuperscript{2} consists of a \textit{HostAgent} that decomposes a user request into multiple subtasks and assigns each subtask to an \textit{AppAgent}, which executes the assigned subtask within an individual application. Below, we highlight the core design ideas and the major differences from LinuxAgent, and refer readers to the \textsc{UFO}\textsuperscript{2} paper for full details.

\paragraph{State.} 
The HostAgent adopts a state machine similar to that of the LinuxAgent, but introduces an additional \textbf{ASSIGN} state responsible for delegating subtasks to AppAgents. Once assigned, control transitions to the selected AppAgent for execution. Importantly, when an AppAgent reaches the \textbf{FINISH} or \textbf{FAIL} state, the control does not terminate but instead returns to the HostAgent, enabling it to decide whether to retry, re-plan, or advance to the next subtask. This hierarchical state transition mechanism naturally supports cooperative task completion across multiple applications.

\paragraph{Strategy.} 
Unlike LinuxAgent, where system states can be dynamically queried through CLI commands, GUI-based environments require explicit perception of the screen and interface hierarchy. Both HostAgent and AppAgent therefore begin each round with a \texttt{DATA\_COLLECTION} strategy that captures screenshots and accessibility (a11y) metadata to construct a structured view of the GUI environment. These inputs are crucial for grounding subsequent reasoning and action decisions. The remaining strategies, \texttt{LLM\_INTERACTION}, \texttt{ACTION\_EXECUTION}, and \texttt{MEMORY\_UPDATE}, follow the same modular workflow as in the LinuxAgent, and are omitted here for brevity. This design unifies the overall agent logic across heterogeneous platforms while allowing for system-specific customization.

\paragraph{Command.} 
The WindowsAgent exposes a significantly richer command set compared to the LinuxAgent. The HostAgent can invoke MCP tools to launch or select applications, while the AppAgent operates through a dedicated GUI MCP server capable of simulating user interactions such as mouse clicks and keyboard inputs. Furthermore, each AppAgent can integrate application-specific MCP servers that bridge to internal APIs, enabling faster and more reliable automation than purely vision-based approaches. This hybrid interaction model, combining GUI manipulation with API-level control strikes a practical balance between generality and robustness.

\smallskip
Overall, LinuxAgent and WindowsAgent jointly demonstrate how the proposed architecture can be adapted for both single-agent and multi-agent environments, from lightweight CLI systems to complex desktop ecosystems. The same design paradigm can be extended to other platforms, such as mobile or in-vehicle infotainment systems, ensuring scalability and reusability across diverse device types.

\section{Implementation and Engineering}
\label{sec:implementation}

We implemented \ufo as a large-scale system consisting of approximately core \textbf{73K lines of Python code}, integrating the centralized \cagent, the asynchronous Constellation Orchestrator, the AIP communication layer, and 3 representative device agents, \ie, \textit{LinuxAgent}, \textit{WindowsAgent} and \textit{MobileAgent (Android)}. An additional 6.1K lines of code were developed for the frontend web UI. The implementation is further accompanied by \textbf{over 77K lines of user documentation}, detailing module interfaces, orchestration protocols, and configuration schemas. This extensive documentation ensures maintainability, reproducibility, and smooth multi-team integration, reflecting substantial engineering effort and demonstrating \ufo's readiness for real-world deployment.

We leverage Python's \texttt{asyncio} framework for concurrent event handling, allowing dynamic DAG updates, agent registration, and task execution to proceed asynchronously without blocking the orchestrator's control loop. The system runs seamlessly across Linux and Windows environments, enabling dynamic workflow orchestration, persistent cross-device communication, and plugin-based extensibility.

\subsection{Interactive \ufo WebUI}
\label{sec:webui}

\begin{figure}[t]
\centering
\includegraphics[width=\textwidth]{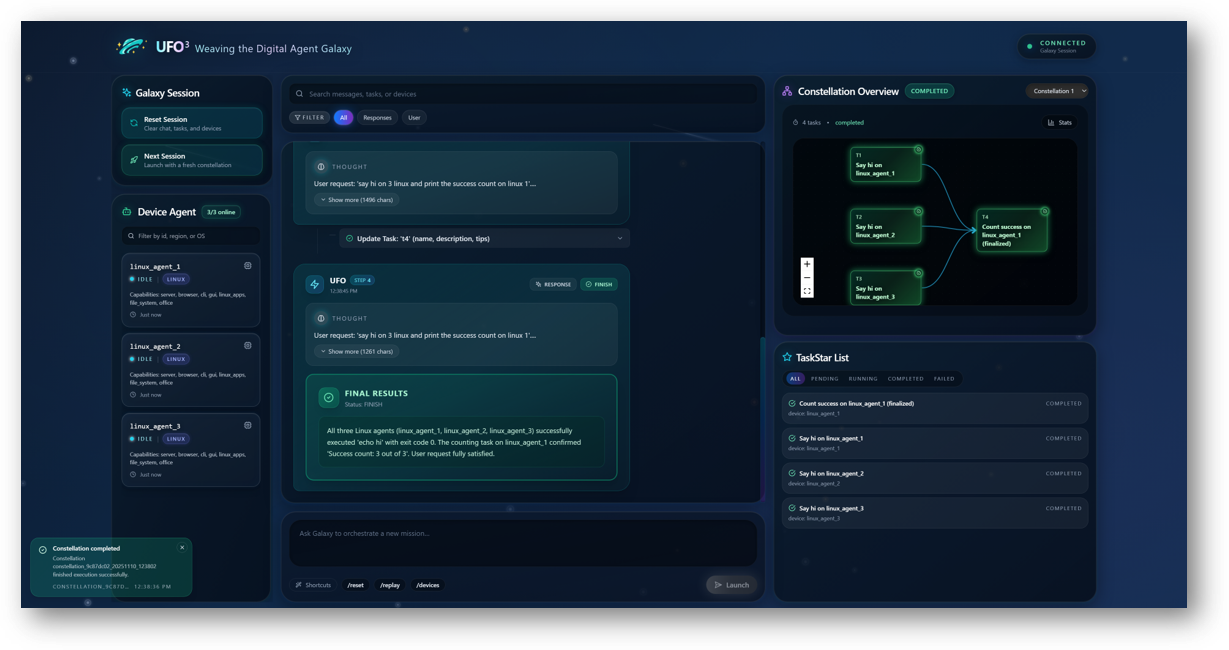}
\vspace{-2em}
\caption{Snapshot of the \ufo WebUI. The interface integrates natural-language interaction, \TaskConstellation visualization, and device agent management in real time.}
\label{fig:webui}
\end{figure}

We build a modern, futuristic WebUI that serves as the operator-facing control surface of \ufo, integrating chat interaction, real-time task monitoring, and device management within a single view (Figure~\ref{fig:webui}). The chatbox at the center allows users to issue natural-language requests and observe agent reasoning and replies in real time. The right panel visualizes the evolving \TaskConstellation as a DAG and lists all \TaskStars with status indicators and detailed logs accessible via expansion. The lower-left registry displays all connected device agents with their capabilities, and states, enabling operators to quickly inspect, reconnect, or migrate tasks as needed. Execution events stream continuously through an event-driven backend built on FastAPI and WebSocket, ensuring sub-millisecond update latency and seamless synchronization between orchestrator and visualization.

This design provides high observability and transparency across the multi-agent workflow. Users can trace each decision from natural-language reasoning to execution outcomes, diagnose failures via per-\TaskStar logs, and visualize dependency satisfaction in real time. By unifying conversation, orchestration, and monitoring, the WebUI bridges human intent and agentic execution, allowing rapid debugging, safe intervention, and fine-grained control without disrupting asynchronous task execution. Overall, the WebUI turns \ufo from a background automation engine into an \textbf{interactive, transparent, and trustworthy orchestration environment}.

\subsection{Plugin and Extension Framework}
Both the \cagent and device agents expose configurable MCP servers interfaces implemented using the \texttt{FastMCP} package. Upon startup, each agent launches an embedded MCP server whose toolset is dynamically registered according to its role (\eg Linux CLI tools, Windows GUI automation, or system telemetry collectors). This plugin mechanism allows developers to add new capabilities, such as a novel GUI driver or API connector, by implementing a lightweight interface, without altering any orchestration or protocol logic. This design significantly improves maintainability, reduces coupling, and enables rapid onboarding of new device types into the constellation.

\subsection{Prompt and Model Integration}
Prompts are modularly defined via a hierarchical configuration. Each agent maintains a core system prompt template augmented by a collection of in-context exemplars for few-shot adaptation. The \cagent centrally manages the LLM backend through a unified API layer compatible with OpenAI-style interfaces, enabling model-agnostic deployment across GPT-based, Claude-based, or local open-weight models. This separation between orchestration and inference ensures the entire system remains robust against future LLM model changes.

\begin{figure}[t]
  \centering
  \includegraphics[width=0.9\linewidth]{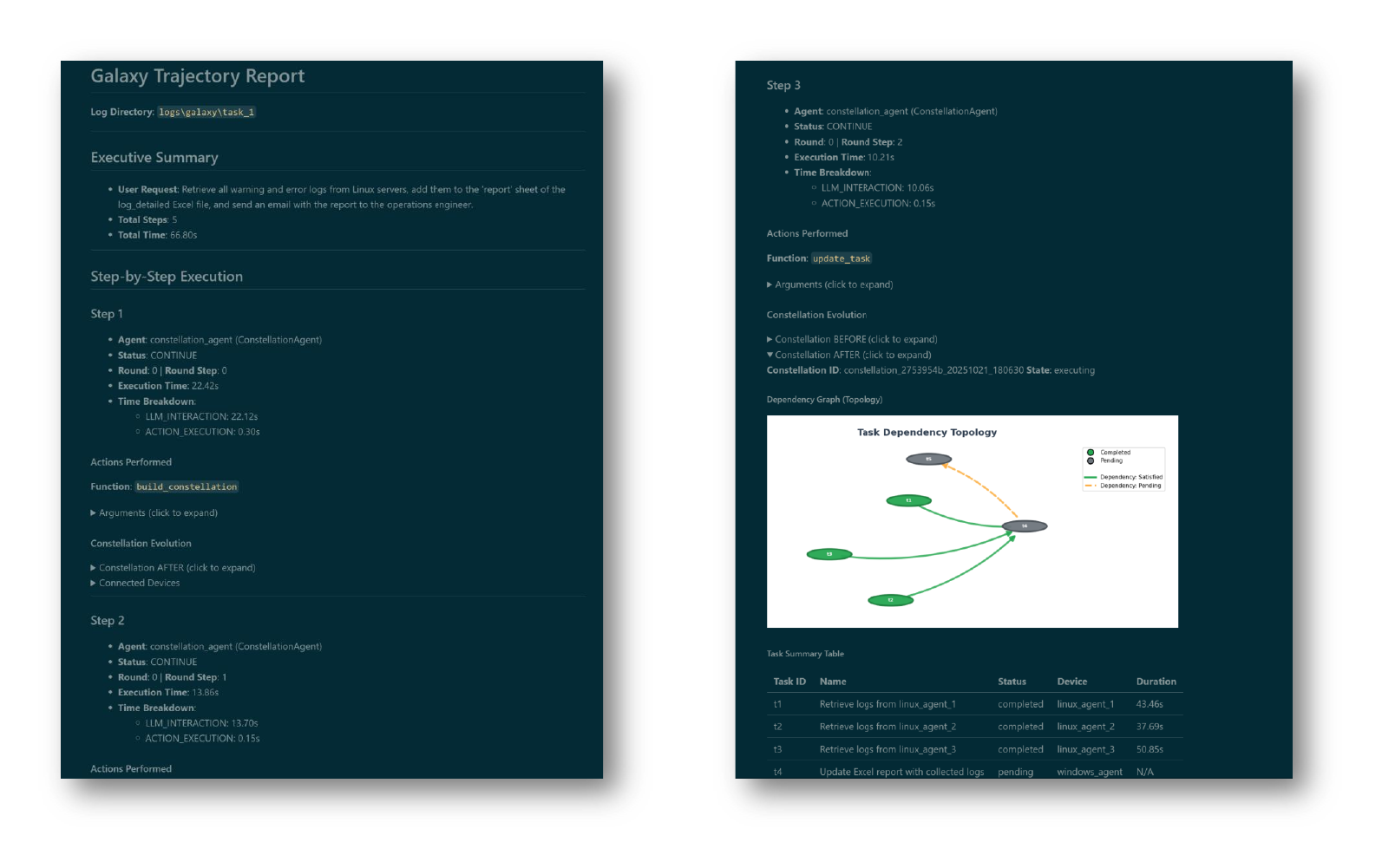}
  \vspace{-1.5em}
  \caption{Example Markdown log generated by \ufo, showing agent actions, reasoning, \TaskConstellation DAG (before and after edits).}
  \label{fig:task_logging}
\end{figure}

\subsection{Automated Task Logging.} 
To facilitate debugging and system introspection, \ufo automatically generates detailed logs in Markdown format after each task execution. These logs capture the complete trace of agent actions, reasoning steps, and intermediate outputs, including \textit{thoughts}, invoked commands, and returned results. They also include visualizations of the \TaskConstellation DAG, showing both the initial and modified topologies, as well as task execution timelines (\TaskStar and \TaskStarLine). Figure~\ref{fig:task_logging} illustrates an example log output, demonstrating how these comprehensive records provide a clear, structured view of task execution for analysis and debugging.

\section{Experimental Evaluation}
\label{sec:experiments}

Following the system implementation, we evaluate the \ufo framework to understand its effectiveness, robustness, and coordination efficiency in realistic, heterogeneous environments.  Existing agent benchmarks predominantly focus on single-device or single-OS settings (\eg text-based API workflows or GUI automation) \cite{mu2025gui}, which fail to capture the cross-device, cross-platform orchestration that \ufo is designed for. Moreover, to the best of our knowledge, there exists \emph{no prior agent system} capable of orchestrating multi-device tasks that span both Linux and Windows environments with unified control and shared context. 
This makes direct comparison against existing systems infeasible and potentially misleading.

Therefore, instead of benchmarking against prior single-agent or single-platform baselines, we focus on a comprehensive internal evaluation that characterizes \ufo's performance across multiple dimensions, covering its planning accuracy, execution reliability, coordination efficiency, and fault tolerance. This approach allows us to isolate the impact of \ufo's architectural innovations and evaluate how well it scales to realistic multi-device orchestration scenarios.

Our evaluation aims to answer the following research questions:

\begin{itemize}
    \item \textbf{RQ1 (Task Completion):} Can \ufo successfully complete diverse, multi-agent tasks across heterogeneous devices and platforms?
    \item \textbf{RQ2 (Orchestration and Adaptation):} How effectively does \ufo orchestrate a user query into a structured \TaskConstellation DAG, and how does it adapt the DAG dynamically in response to intermediate subtask results during execution?
    \item \textbf{RQ3 (Parallelism Exploitation):} To what extent can \ufo identify and exploit parallelism across independent subtasks to accelerate overall execution without compromising correctness?
    \item \textbf{RQ4 (Performance and Scalability):} How efficient is \ufo in task planning, scheduling, and end-to-end execution latency under different network conditions and system scales?
    \item \textbf{RQ5 (Robustness):} How does \ufo handle partial failures, network delays, or unavailable agents during distributed execution?
\end{itemize}

\subsection{Experiment Setup}
\label{sec:setup}

We deploy \ufo across \textbf{five physical machines} in a controlled environment:
\begin{itemize}
    \item \textbf{1× Windows 11 desktop}, running the \textit{WindowsAgent};
    \item \textbf{3× Ubuntu 22.04 workstations (CPU-only)}, running \textit{LinuxAgent};
    \item \textbf{1× Ubuntu 24.04 GPU node} equipped with four NVIDIA A100 GPUs, running \textit{LinuxAgent}.
\end{itemize}

Agents communicate via a simulated local network with 1--10\,ms latency and a wide-area link (50--100\,ms latency) for the GPU node using the AIP. All components are implemented in Python~3.10 and powered by the same large language model (\textsc{GPT-5-Chat-20251003}) \cite{openai2025gpt5systemcard} for both the \cagent and device agents. The orchestrator and controller run on a dedicated management node that coordinates agent discovery, task scheduling, and monitoring. This setup emulates a realistic hybrid enterprise environment that combines cloud servers, local desktops, and GPU compute nodes.

\subsection{\bench: A Crossed-Device Evaluation Benchmark}
\label{sec:scenarios}

\begin{table}[t]
\centering
\caption{Overview of the 10 task categories used in evaluation. Each category includes 4--10 representative cases.}
\label{tab:task_categories}
\vspace{3pt}
\begin{tabular}{p{0.21\linewidth} | p{0.65\linewidth} | p{0.06\linewidth} }
\toprule
\textbf{Category} & \textbf{Description} & \textbf{Count} \\ 
\midrule
Logs \& Monitoring & Log retrieval, aggregation, and report generation across devices. & 6 \\\midrule
System State \& Configuration & Managing environment variables, users, permissions, and disk information. & 5 \\\midrule
Processes \& Services & Starting, stopping, and monitoring services and scheduled tasks. & 5  \\\midrule
Data Wrangling \& Scripting & Parsing CSV/text files, performing statistical summaries, and executing scripts. & 4\\  \midrule
DevOps \& Containers & Managing Git repositories, CI/CD pipelines, and container operations. & 10\\\midrule
Networking \& Connectivity & Diagnosing connectivity issues, pinging hosts, and verifying endpoints. & 5\\\midrule
Browsing Tasks & Web-assisted cross-device operations that require interacting with web resources via a browser and then applying or verifying results on remote Linux hosts. & 5\\\midrule
Cross-Device Orchestration & Coordinating multi-agent workflows involving data transfer and dependency handling. & 5\\\midrule
GPU \& ML Workloads & Launching and monitoring GPU-based ML training and inference jobs. & 5\\\midrule
Negative Tasks & Infeasible or invalid tasks used to test failure detection and safe termination. & 5\\\midrule
\bottomrule
\end{tabular}
\end{table}

To assess \ufo's generality and real-world applicability, we construct \bench, a benchmark of \textbf{55 representative multi-agent tasks} spanning typical productivity and system administration workflows. Five volunteers with diverse Linux and Windows experience proposed realistic queries they would naturally issue to a digital assistant in daily use. The resulting tasks span ten functional categories, summarized in Table~\ref{tab:task_categories}. A detailed list of queries in \bench is shown in Appendix~\ref{sec:bench_details} and Table~\ref{tab:galaxy_results}.

\begin{figure}[t]
  \centering
  \includegraphics[width=0.9\textwidth]{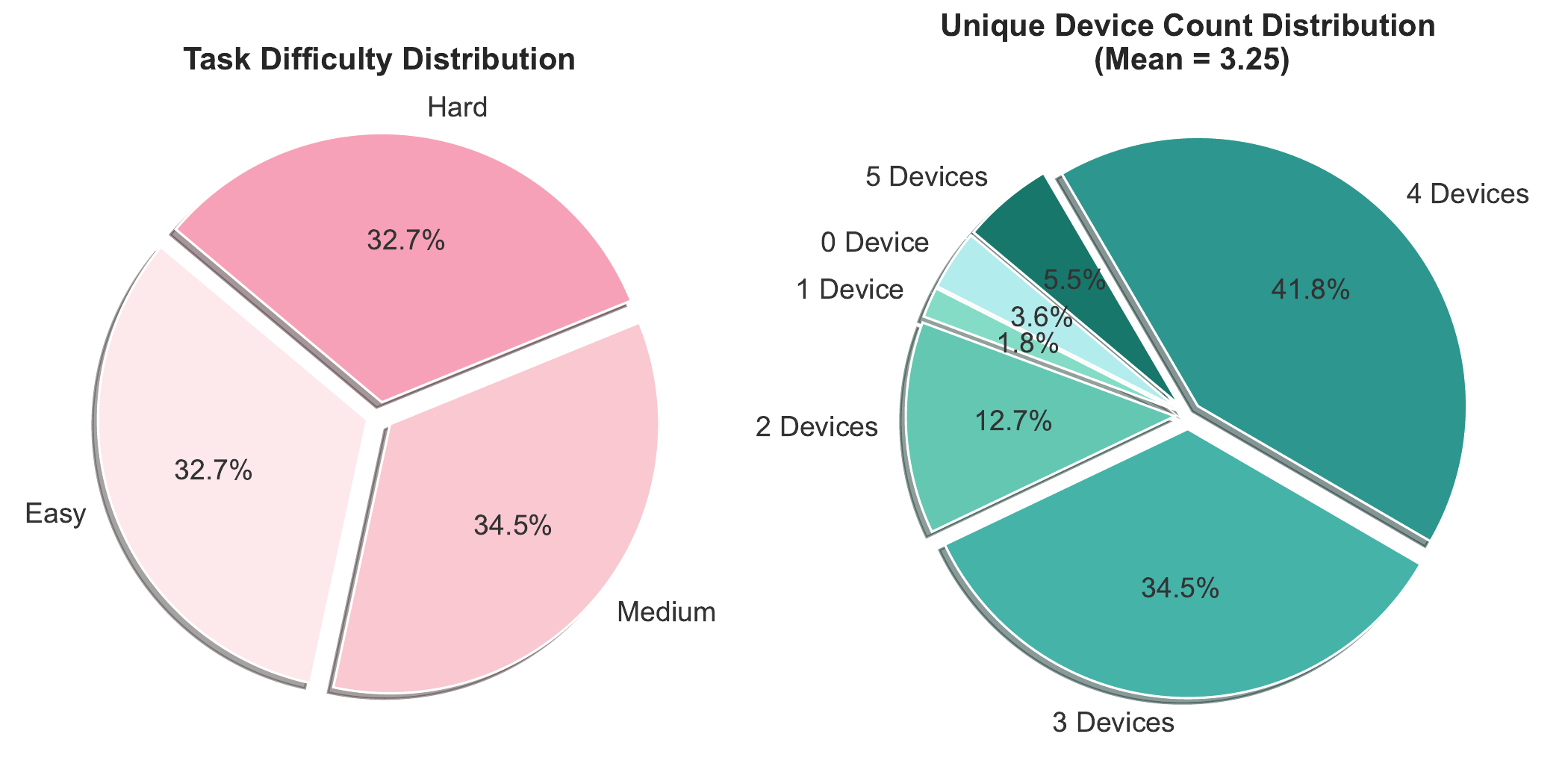}
  \vspace{-1.5em}
  \caption{Distribution of \bench tasks. Left: proportion of tasks by difficulty. Right: number of devices involved per task.}
  \label{fig:task_distribution}
\end{figure}

Each scenario is labeled with one of three difficulty levels, namely \textit{Easy}, \textit{Medium}, and \textit{Hard}, reflecting the degree of orchestration and reasoning required:
\begin{itemize}
    \item \textbf{Easy:} Single-host operations such as log inspection, one-off checks, service control, small file transfers, or short summarization. These tasks require minimal coordination and succeed with standard administrative privileges.
    \item \textbf{Medium:} Moderately orchestrated tasks involving multiple machines or light cross-platform activity (\eg container run-and-verify, metric aggregation, spreadsheet updates). They often require conditional logic, transient failure handling, or simple parsing and aggregation.
    \item \textbf{Hard:} Complex multi-step workflows requiring cross-platform orchestration, CI/build pipelines, container image management, distributed data processing, or live patching. These tasks involve high privilege levels, non-trivial verification, and greater exposure to race conditions or dependency issues.
\end{itemize}
Figure~\ref{fig:task_distribution} visualizes the composition of \bench. The left pie chart shows the distribution of task difficulty, which is roughly balanced across Easy (33\%), Medium (35\%), and Hard (33\%) tasks. The right chart depicts the number of devices required per task, ranging from 0 to 5. Tasks requiring no devices correspond to \textit{negative tasks} that are inherently infeasible; the system is expected to detect and fail these safely without executing any agent actions. Most tasks, however, involve 3–5 devices, highlighting \bench's focus on multi-agent orchestration scenarios.

This stratification enables a controlled evaluation of \ufo's reasoning, coordination, and execution robustness under increasing complexity. Each task is executed end-to-end, from natural language request to final result through the orchestrator–agent hierarchy, with all required data, scripts, and code pre-deployed. By including negative tasks, we also validate \ufo's safety mechanisms, ensuring that infeasible goals are correctly identified and aborted without unintended side effects.

\subsection{Metrics and Methodology}
To rigorously assess the effectiveness and robustness of \ufo, we design a comprehensive, multi-dimensional evaluation methodology covering correctness, adaptivity, efficiency, and fault tolerance. Our goal is not only to measure whether tasks are completed, but also to understand how well \ufo plans, adapts, and sustains performance under dynamic and heterogeneous conditions. We organize our analysis around four research questions (RQ1–RQ4), each corresponding to a key aspect of \ufo's system behavior.

\paragraph{RQ1: Task Completion.}
We measure the overall \textit{Task Success Rate (TSR)} and \textit{Subtask Completion Rate (SCR)} to evaluate \ufo's execution reliability. 
TSR captures the fraction of user queries that are successfully completed end-to-end, as judged by the query authors based on final outcomes. 
SCR measures the success rate of individual subtasks executed by device agents, annotated automatically by an LLM evaluator that inspects the task trajectories and completion logs. Together, TSR and SCR provide a top-down and bottom-up view of system reliability across the orchestration hierarchy. Note that for negative test cases, we mark a task or subtask as successful when the agent correctly detects and reports the intended failure, rather than attempting to complete it erroneously.

\paragraph{RQ2: Orchestration and Adaptation.}  
We evaluate \ufo's reasoning and dynamic re-planning by analyzing the initial \TaskConstellation DAG and its evolution during execution. To quantify adaptability, we track DAG modification metrics, including edits, node insertions and deletions, and dependency changes, which reflect how the system adjusts the task constellation in response to runtime results. Comparing the initial and final DAGs provides a holistic measure of \ufo's ability to adapt its plan as execution unfolds.

\paragraph{RQ3: Parallelism and Execution Efficiency.} 
To evaluate \ufo's ability to exploit concurrency and optimize execution, we measure several metrics derived from the \TaskConstellation DAG: 
\begin{itemize}
    \item \textbf{Maximum Parallel Width:} the largest number of subtasks that can be executed concurrently at any point in the DAG.
    \item \textbf{Critical Path Length (L):} the execution time of the longest serial dependency chain in the DAG.
    \item \textbf{Total Task Execution Time (W):} the sum of execution times of all tasks, representing the cumulative workload.
    \item \textbf{Parallelism Ratio (P):} $P = W / L$, capturing the degree to which \ufo leverages parallel execution across devices.    
\end{itemize}
These metrics provide insight into \ufo's efficiency in orchestrating multi-agent tasks and its effectiveness in parallelizing independent subtasks to reduce overall task latency.

\paragraph{RQ4: Latency and Scalability.}
We evaluate \textit{End-to-End Latency}, \textit{Orchestration Time} (including \TaskConstellation creation and edits), and \textit{Total Task Execution Time} to characterize system efficiency. These metrics capture the balance between reasoning overhead, arising from LLM-based planning, scheduling, and coordination, and the execution gains achieved through distributed parallelism. Together, they provide a holistic view of \ufo's scalability across heterogeneous agents and devices under realistic workload conditions.

\paragraph{RQ5: Robustness and Fault Handling.}
We assess robustness through a case study on a single request executed under three controlled failure modes: \textsf{(i)} a single device agent fails and recovers via dynamic DAG reassignment, \textsf{(ii)} a single agent fails without recovery, and \textsf{(iii)} all agents fail simultaneously. These scenarios capture \ufo's fault tolerance boundary and illustrate its adaptive recovery mechanisms under partial and global disruptions.

Overall, this evaluation framework enables a holistic examination of \ufo's behavior under diverse operational conditions, revealing how design principles such as event-driven scheduling, runtime DAG modification, and agent autonomy translate into measurable system benefits.

\subsection{RQ1: Task Completion Analysis}
\label{subsec:rq1}


\begin{figure}[t]
  \centering
  \includegraphics[width=\textwidth]{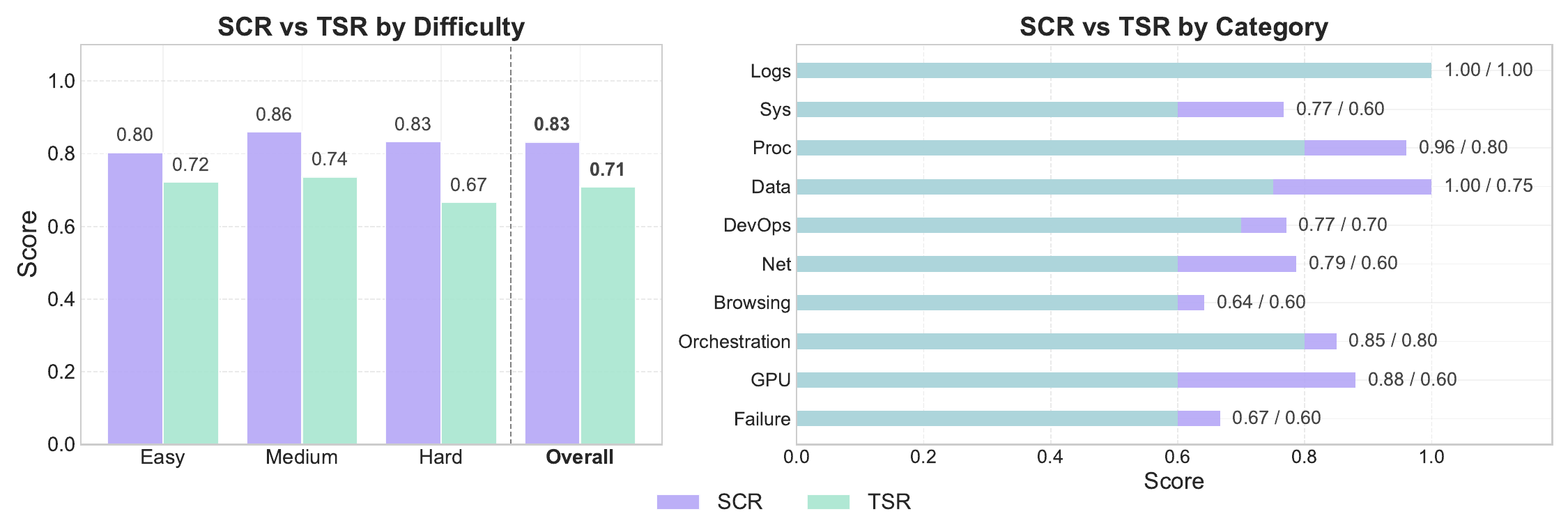}
  \vspace{-.5em}
  \caption{Task and subtask success rates by difficulty level (left) and by query category (right) of \ufo.}
  \label{fig:success_rate}
\end{figure}

To evaluate \ufo's effectiveness in executing distributed tasks, we first examine the primary metrics of interest: Subtask Completion Rate (SCR) and Task Success Rate (TSR). We analyze performance both by task difficulty and by functional category, as summarized in Figure~\ref{fig:success_rate} \footnote{A complete list of \ufo's performance on all queries is provided in Appendix~\ref{sec:bench_details} and Table~\ref{tab:galaxy_results}.}. Overall, \ufo demonstrates strong reliability across the benchmark, achieving an SCR of 83.3\% and a TSR of 70.9\%. These high rates indicate that the orchestrator effectively coordinates subtasks across devices, while individual device agents reliably execute their assigned operations, validating the robustness of \ufo's design. Breaking down by difficulty, easy and medium tasks exhibit comparable TSRs (72.2\% and 73.7\%, respectively), showing that \ufo handles routine and moderately interdependent tasks effectively. For hard tasks, TSR declines to 66.7\%, yet even in these complex multi-host scenarios \ufo successfully completes the majority of requests, highlighting its capability under challenging orchestration conditions.

By functional category, performance trends align with task characteristics. Structured, deterministic tasks such as \texttt{Data} (SCR 100\%) and \texttt{Proc} (SCR 96\%) achieve the highest reliability. Tasks involving user-like interactions or ambiguity, including \texttt{Browsing} (SCR 64.2\%) and \texttt{Negative} scenarios (SCR 66.7\%), show lower success rates. Intermediate categories requiring cross-device coordination, such as \texttt{Orchestration} (SCR 85\%) and \texttt{DevOps} (SCR 77.1\%), demonstrate solid subtask reliability, evidencing \ufo's capacity to manage dependencies across heterogeneous devices.

Across both analyses, SCR generally exceeds TSR. This gap arises because a single subtask failure among multiple dependencies can cause an overall task failure, even when most subtasks succeed. The pattern underscores the value of \ufo's dynamic DAG management and reasoning, which mitigate partial failures and maintain high system reliability in complex, distributed workflows.

\paragraph{Error Analysis.}
To better understand the remaining failures, we examined common error patterns. First, tasks requiring file transfers across devices occasionally fail because AIP currently supports only textual communication; device agents may not have direct network access to each other. Future work includes maintaining a shared memory and extending AIP to support file read/write operations. Second, some failures arise when agents attempt to complete tasks regardless of preconditions. For example, the request ``Start \ufo service on all Linux devices'' triggered the LinuxAgent to create the service even when it did not exist, rather than reporting a failure. Enhancing agent self-awareness and enforcing conditional execution will mitigate such issues. 
Third, tasks executed via WindowsAgent show relatively lower success due to GUI dependencies and interface variability; improving robustness of GUI-based agents remains an important direction. 

Despite these errors, \ufo's generated \TaskConstellation are largely correct, indicating that the \cagent effectively decomposes complex requests, identifies dependencies, and exposes parallelism, validating its reasoning design.

\subsection{RQ2: Orchestration and Adaptation}
\label{subsec:rq2}

\begin{table}[t]
\centering
\caption{Average modifications per edit and per request by type.}
\begin{tabular}{lccccccc}
\toprule
 & \makecell{Added\\Tasks} & \makecell{Removed\\Tasks} & \makecell{Modified\\Tasks} & \makecell{Added\\Dependencies} & \makecell{Removed\\Dependencies} & \makecell{Modified\\Dependencies} & Total \\
\midrule
Per Edit & 0.05 & 0.00 & 0.99 & 0.04 & 0.01 & 0.00 & 1.09 \\
Per Request & 0.41 & 0.00 & 5.20 & 0.24 & 0.07 & 0.00 & 5.91 \\
\bottomrule
\end{tabular}
\label{tab:modifications}
\end{table}

\begin{figure}[t]
  \centering
  \includegraphics[width=\textwidth]{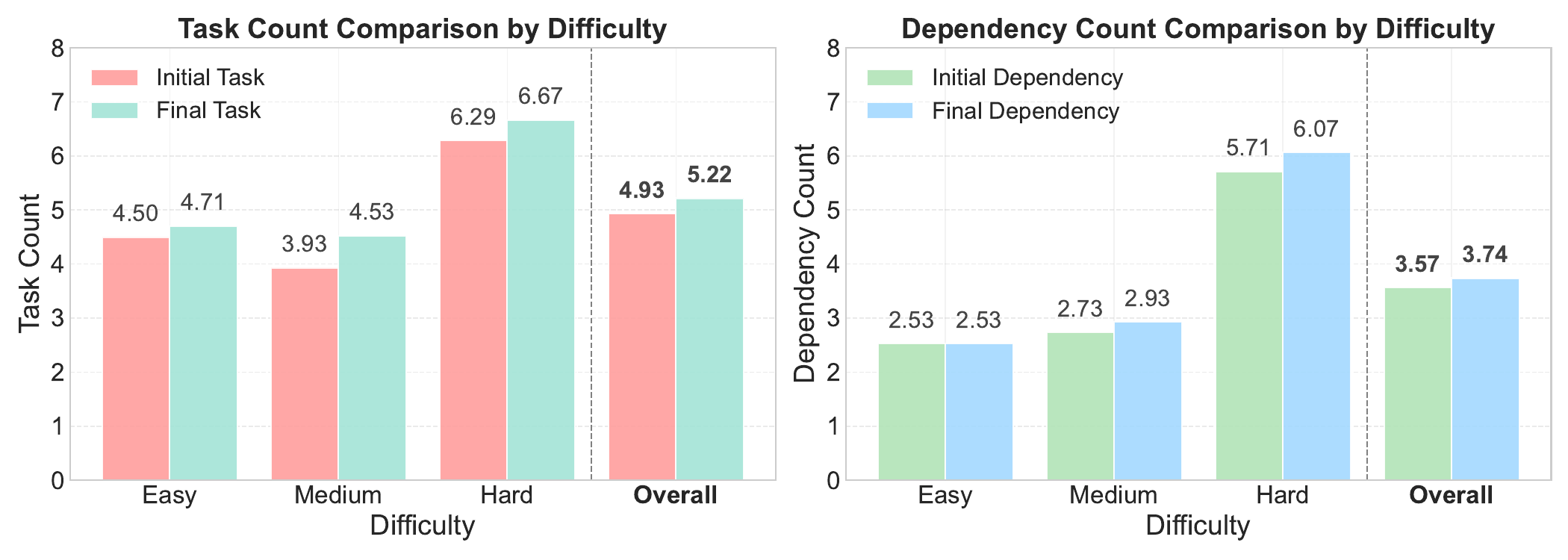}
  \vspace{-1.5em}
  \caption{Comparison of the initial and final task (left) and dependency (right) counts in the \TaskConstellation of \ufo.}
  \label{fig:dag_comparison}
\end{figure}

We next examine \ufo's ability to orchestrate a user query into an executable \TaskConstellation DAG and adapt it dynamically as execution unfolds. To quantify this, we track modifications to the DAG across all requests, including node insertions, deletions, and dependency updates in Figure~\ref{fig:dag_comparison}, and compare the initial and final task and dependency counts in Table~\ref{tab:modifications}.

Overall, \ufo demonstrates a highly active editing behavior, with an average of 1.09 modifications per edit and 5.91 modifications per request. Most changes occur in the \textbf{Modified Tasks} category, reflecting the system's strategy of enriching downstream tasks based on the results of preceding subtasks. For example, logs collected by earlier tasks are incorporated into subsequent document-writing or analysis subtasks, ensuring that later steps have complete context and accurate information. In contrast, the number of added or removed tasks and dependencies is minimal, indicating that \cagent generally produces well-structured DAGs from the outset and only fine-tunes tasks during execution.

Breaking down by difficulty (Figure~\ref{fig:dag_comparison}), we observe that \ufo maintains a robust initial decomposition across Easy, Medium, and Hard tasks, with only modest increases in both task and dependency counts by the end of execution. This demonstrates that while the orchestrator adapts to runtime results, it does so without fundamentally restructuring the workflow, preserving overall plan stability. Harder tasks show slightly more edits, consistent with their longer multi-step workflows and richer context propagation.

Taken together, these observations indicate that \ufo exhibits strong adaptive orchestration capabilities: \textit{(i)} it actively enriches downstream tasks based on prior subtask outputs, ensuring context-aware execution; \textit{(ii)} it generates high-quality initial DAGs, requiring minimal structural modifications; and \textit{(iii)} it balances stability with runtime flexibility, applying refinements without disrupting the overall task plan. These features highlight \ufo's ability to both plan effectively and adjust dynamically across distributed devices.

\subsection{RQ3: Parallelism and Execution Efficiency}
\label{subsec:rq3}

\begin{table}[t]
\centering
\caption{Parallelism characteristics of task constellations by difficulty.}
\begin{tabular}{lcccc}
\toprule
\textbf{Difficulty} & 
\makecell{\textbf{Max}\\\textbf{Parallel Width}} & 
\makecell{\textbf{Critical Path}\\\textbf{Length (L)}} & 
\makecell{\textbf{Total Exec.}\\\textbf{Time (W)}} & 
\makecell{\textbf{Parallelism}\\\textbf{Ratio (P = W/L)}} \\
\midrule
Easy & 3.53 & 144.10 & 315.28 & 1.86 \\
Medium & 2.73 & 166.94 & 212.54 & 1.51 \\
Hard & 3.21 & 232.12 & 339.65 & 1.77 \\
\midrule
Overall & 3.17 & 178.34 & 289.20 & 1.72 \\
\bottomrule
\end{tabular}
\label{tab:parallelism}
\end{table}

We next investigate \ufo's ability to exploit concurrency and optimize distributed task execution.  
Table~\ref{tab:parallelism} reports four DAG-derived metrics that capture the structural and runtime efficiency of each task constellation.  
Across all scenarios, \ufo achieves an average \textbf{parallelism ratio} of 1.72, with up to \textbf{3.5 concurrent subtasks} executing at peak, demonstrating that \ufo can effectively uncover and schedule independent subtasks across devices.

We observe several trends.   First, the \textit{maximum parallel width} remains consistently high (around 3 tasks) even as task difficulty increases, indicating that \ufo's orchestrator maintains concurrency opportunities even in complex, multi-host settings.  Second, while the \textit{critical path length} naturally grows with task complexity (144s $\to$ 232s), the increase in total execution time (315s $\to$ 340s) is modest, showing that \ufo successfully overlaps execution through concurrent scheduling.  Third, medium-difficulty tasks exhibit slightly lower parallelism ($P=1.51$), as many involve short verification or data aggregation steps with limited parallel components, while both easy and hard tasks benefit more from concurrent flows.

These results reveal that \ufo's DAG-based orchestration is not only structurally well-parallelized but also runtime-efficient. By modeling tasks as DAGs, \ufo naturally identifies and executes independent subtasks asynchronously across heterogeneous devices, keeping the system utilization high. Together, these findings demonstrate that \ufo delivers strong execution efficiency through fine-grained parallel orchestration, validating the effectiveness of its DAG-based design.

\label{subsec:rq4}
\begin{figure}[t]
  \centering
  \includegraphics[width=\textwidth]{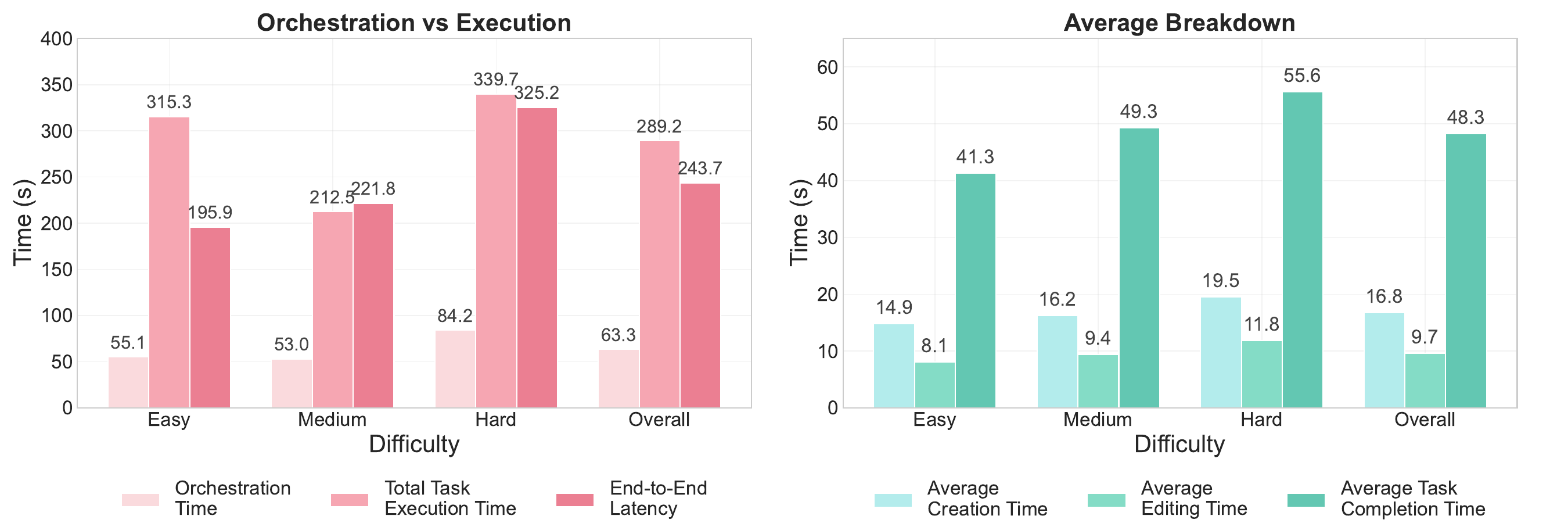}
  \vspace{-.5em}
  \caption{Task timing and orchestration breakdown by difficulty.}
  \label{fig:time_comparison}
\end{figure}

\begin{figure}[t]
  \centering
  \vspace{-1.em}
  \includegraphics[width=\textwidth]{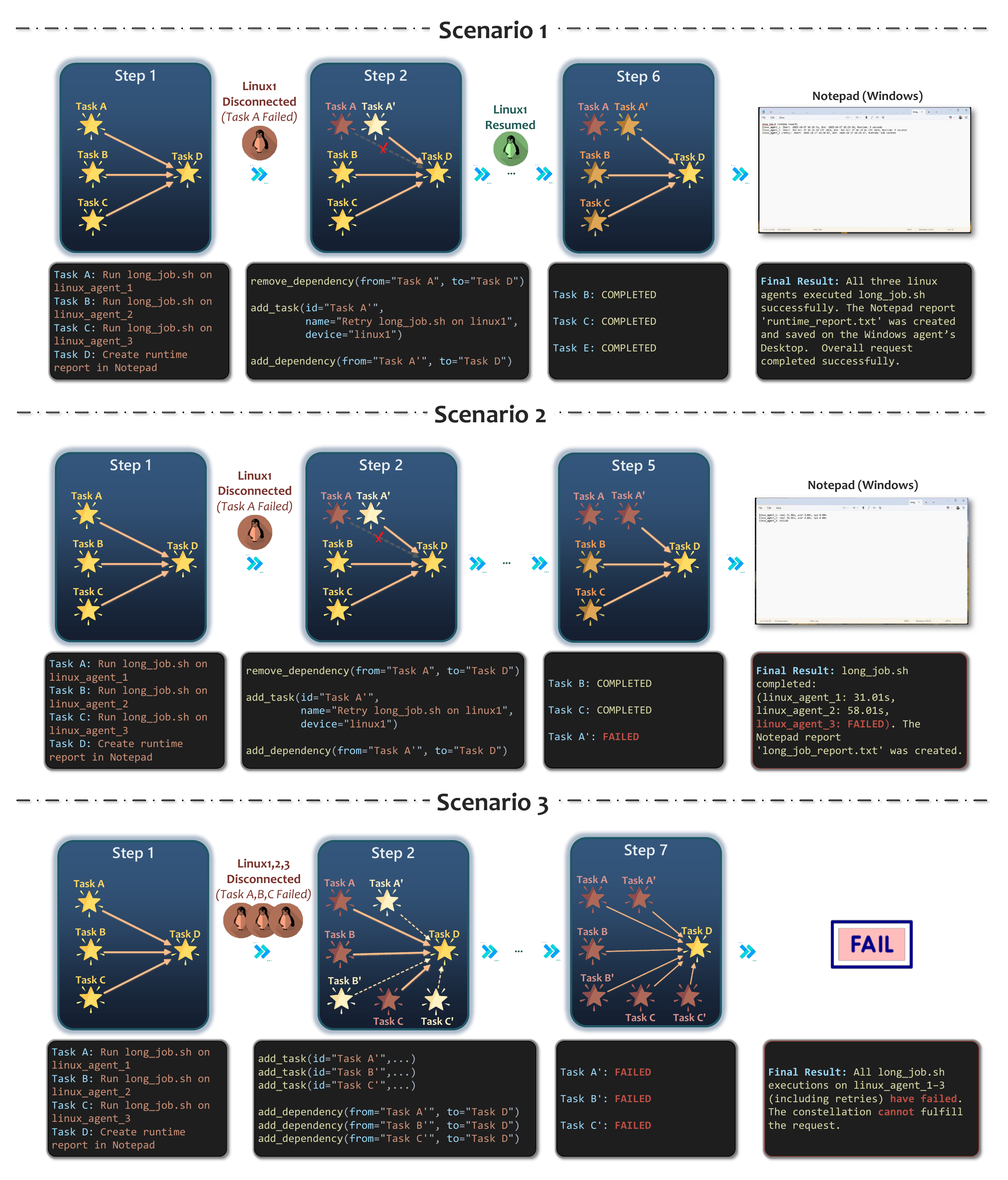}
  \vspace{-1.5em}
  \caption{\textbf{Fault injection scenarios for robustness evaluation.} 
  \ufo's adaptive behavior under (1) transient, (2) permanent, and (3) full-agent failures.}
  \label{fig:fault_handling}
\end{figure}
\subsection{RQ4: Latency and Scalability}
We evaluate \textit{End-to-End Latency}, \textit{Orchestration Time} (including \TaskConstellation creation and edits), and \textit{Total Task Execution Time} to characterize \ufo's efficiency and scalability. These metrics reflect the trade-off between reasoning overhead, arising from LLM-based planning, scheduling, and coordination, and execution gains obtained through distributed parallelism.

As shown in Figure~\ref{fig:time_comparison} (left), \ufo exhibits strong execution efficiency across all task difficulties. On average, the measured end-to-end latency is only \textbf{243.7 seconds}, notably shorter than the combined orchestration and task execution time of \textbf{352.5 seconds} (63.3 + 289.2). This corresponds to a \textbf{31\% reduction} in total completion time compared to a fully sequential workflow. Even for complex, cross-device tasks, \ufo maintains an average end-to-end latency of about \textbf{4 minutes}, demonstrating its ability to exploit distributed parallelism while preserving robust coordination and correctness. Compared to manual execution, which often takes tens of minutes or longer, \ufo delivers a substantial improvement in both speed and scalability. This efficiency gain stems from \ufo's concurrent design described in Section~\ref{sec:async}: multiple subtasks are executed in parallel across heterogeneous agents, while \cagent continues to refine the \TaskConstellation asynchronously in the background. Such overlap between orchestration and execution effectively hides reasoning latency and maximizes device utilization.

We also observe a clear correlation between task difficulty and latency: harder tasks incur longer orchestration times (84.2s for Hard vs. 55.1s for Easy) and longer end-to-end latency (325.2s vs. 195.9s), reflecting the increased complexity, number of subtasks, and cross-device coordination required. However, even for Hard tasks, the total end-to-end time remains within a few minutes, demonstrating \ufo's efficiency compared to manual execution of equivalent multi-device workflows.

Figure~\ref{fig:time_comparison} (right) further breaks down the average time spent in \TaskConstellation creation, editing, and individual task completion. Creation and editing are consistently fast (16–20s and 8–12s, respectively), indicating that \cagent is highly efficient at planning and adapting task DAGs in real time. Task completion, averaging around 48s overall, dominates the workflow, yet remains reasonable given the distributed nature and heterogeneity of devices. 

These results underscore several key insights: \textit{(i)} \ufo leverages cross-device parallelism effectively, enabling task execution to substantially overlap with reasoning and orchestration; \textit{(ii)} the orchestration engine scales gracefully with task complexity, keeping latency manageable even for multi-step, multi-host tasks; \textit{(iii)} the efficient creation and editing of the \TaskConstellation demonstrates that \cagent can dynamically adapt plans with minimal overhead, supporting robust and responsive execution in real-world scenarios. Collectively, these observations confirm that \ufo is both performant and scalable for heterogeneous, distributed multi-agent workloads.

\subsection{RQ5: Robustness and Fault Handling}
\label{subsec:rq5}

Finally, we evaluate \ufo's robustness and fault-tolerance capabilities by observing its behavior under three simulated device-agent failure scenarios. The test request is: \textit{``Run long\_job.sh concurrently on Linux 1--3 and report their running time on Notepad.''} 
Ideally, this request invokes three \texttt{LinuxAgent} instances to execute the script in parallel, followed by the \texttt{WindowsAgent} aggregating and recording the results in Notepad. We design three fault-injection scenarios to examine \ufo's adaptive responses:
\begin{itemize}
    \item \textbf{Scenario 1:} One \texttt{LinuxAgent} is intentionally disconnected but recovers shortly before the task completes.
    \item \textbf{Scenario 2:} The same \texttt{LinuxAgent} disconnects and does not recover for the remainder of the execution.
    \item \textbf{Scenario 3:} All \texttt{LinuxAgent} instances remain unavailable until the task ends.
\end{itemize}
These cases, illustrated in Figure~\ref{fig:fault_handling}, allow us to analyze how \ufo detects failures, adapts the \TaskConstellation accordingly, and maintains graceful degradation or recovery during runtime.

In \textbf{Scenario~1}, where \texttt{Linux~1} is temporarily disconnected, \ufo detects the failure event immediately. The corresponding subtask (\textit{Task~A}) is marked as failed, and upon receiving this signal, the \cagent proactively spawns a replacement task (\textit{Task~A'}) to retry execution. When \texttt{Linux~1} reconnects before the task deadline, the retry is successfully executed, and the results are correctly aggregated by the \texttt{WindowsAgent}.  This demonstrates \ufo's ability to automatically recover from transient device failures through adaptive re-planning without human intervention, showing that its fault-handling mechanism is both \textit{responsive} and \textit{self-healing}.

In \textbf{Scenario~2}, \texttt{Linux~1} remains disconnected throughout execution.  \ufo follows the same retry logic, creating a new \textit{Task~A'} and reassigning it to the same device after an adaptive delay.  However, as the device remains unavailable, the retry also fails. Instead of hanging indefinitely or aborting the entire workflow, \ufo recognizes this as a partial failure, continues executing the remaining subtasks on other available devices, and finally records the aggregated results, including explicit failure traces in the final Notepad report. This behavior shows that \ufo's orchestration framework can \textit{degrade gracefully} under partial failures, preserving useful progress and ensuring result completeness.

In \textbf{Scenario~3}, all \texttt{LinuxAgent} instances are disconnected, leading to consecutive retry failures across all subtasks. After exhausting retry attempts, the \cagent terminates the execution plan and reports the entire request as failed. 
The failure is propagated in a transparent manner, and \ufo refrains from producing hallucinated or incomplete results. 
This highlights the system's emphasis on \textit{fail-safe integrity}, where it prefers honest termination over speculative completion when recovery is infeasible.

Overall, these experiments demonstrate that \ufo exhibits a high degree of robustness in distributed environments. 
Its adaptive orchestration loop allows it to distinguish between transient and permanent failures, retry safely, and preserve progress when possible. More importantly, the system maintains end-to-end consistency even under multi-agent disruptions, validating the effectiveness of its hierarchical fault-tolerance design.

\subsection{Case Study: Cross-Device Orchestration in Action}
Lastly, we present three representative cases from \bench to demonstrate how \ufo effectively and efficiently orchestrates heterogeneous device agents to complete complex user requests across operating systems and application boundaries.

\subsubsection{Case 1: Collecting Logs and Writing Reports}
\begin{figure}[t]
  \centering
  \vspace{-1.em}
  \includegraphics[width=\textwidth]{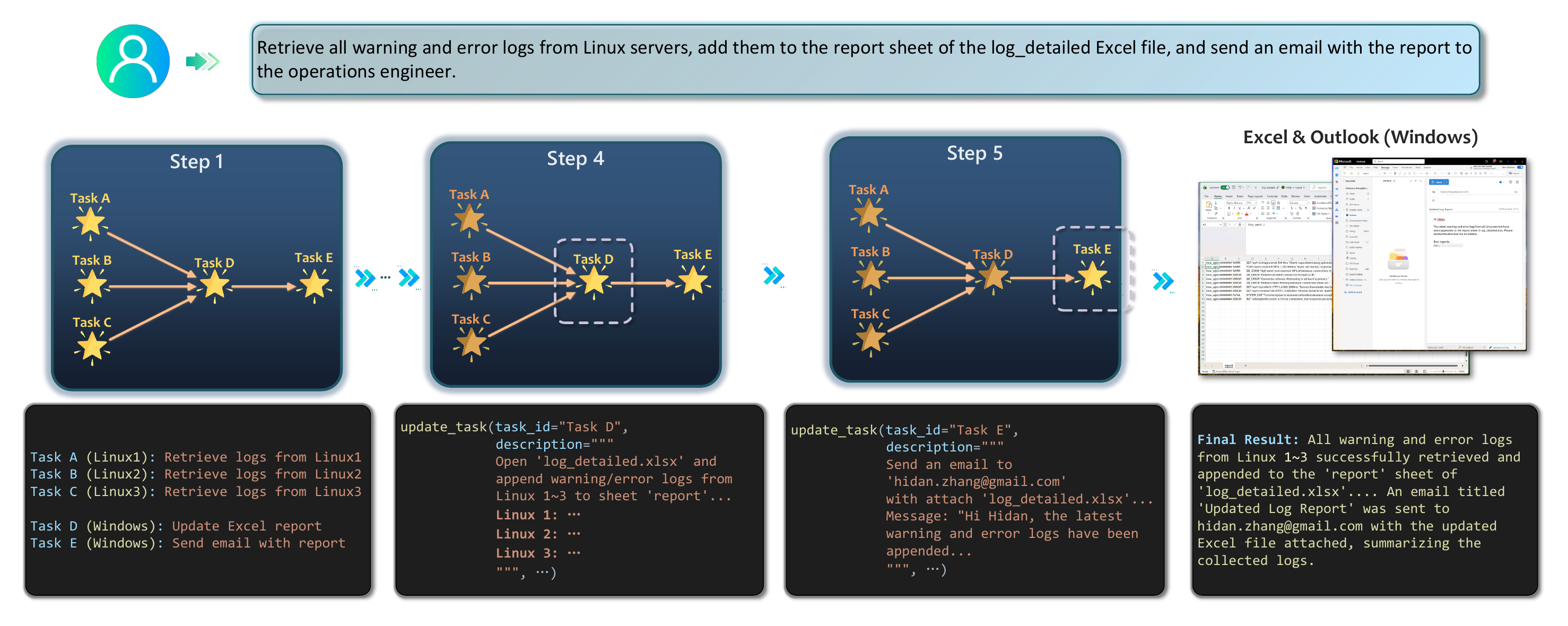}
  \vspace{-1.5em}
  \caption{\textbf{Case 1: Log collection and report generation.} 
  \ufo decomposes the high-level request into parallel log retrieval tasks on Linux agents and sequential report-generation tasks on Windows, demonstrating cross-device coordination and result aggregation.}
  \label{fig:case_log}
\end{figure}

As illustrated in Figure~\ref{fig:case_log}, this case requests \ufo to ``Retrieve all warning and error logs from Linux servers, add them to the `report' sheet of the \texttt{log\_detailed.xlsx} file, and send an email with the report to the operations engineer.'' While conceptually straightforward, this task traditionally requires tedious manual effort across multiple systems and interfaces.

\ufo autonomously decomposes the request into five \TaskStars: three for parallel log retrieval and two for final report generation and email dispatch. The first three \TaskStars are dispatched to distinct \textit{LinuxAgents}, each collecting logs asynchronously from its assigned host. Once completed, the results are automatically aggregated by the \cagent, which then spawns two sequential \TaskStars on the \textit{WindowsAgent} to insert the logs into Excel and compose an email summary.

This case exemplifies \ufo's strong parallelism and adaptive coordination capabilities. By overlapping asynchronous data collection with centralized result fusion, \ufo achieves both time efficiency and consistency across heterogeneous systems, significantly reducing human effort and turnaround time compared to manual workflows.

\subsubsection{Case 2: Task Allocation and Result Integration via Excel}
\begin{figure}[t]
  \centering
  \vspace{-1.em}
  \includegraphics[width=\textwidth]{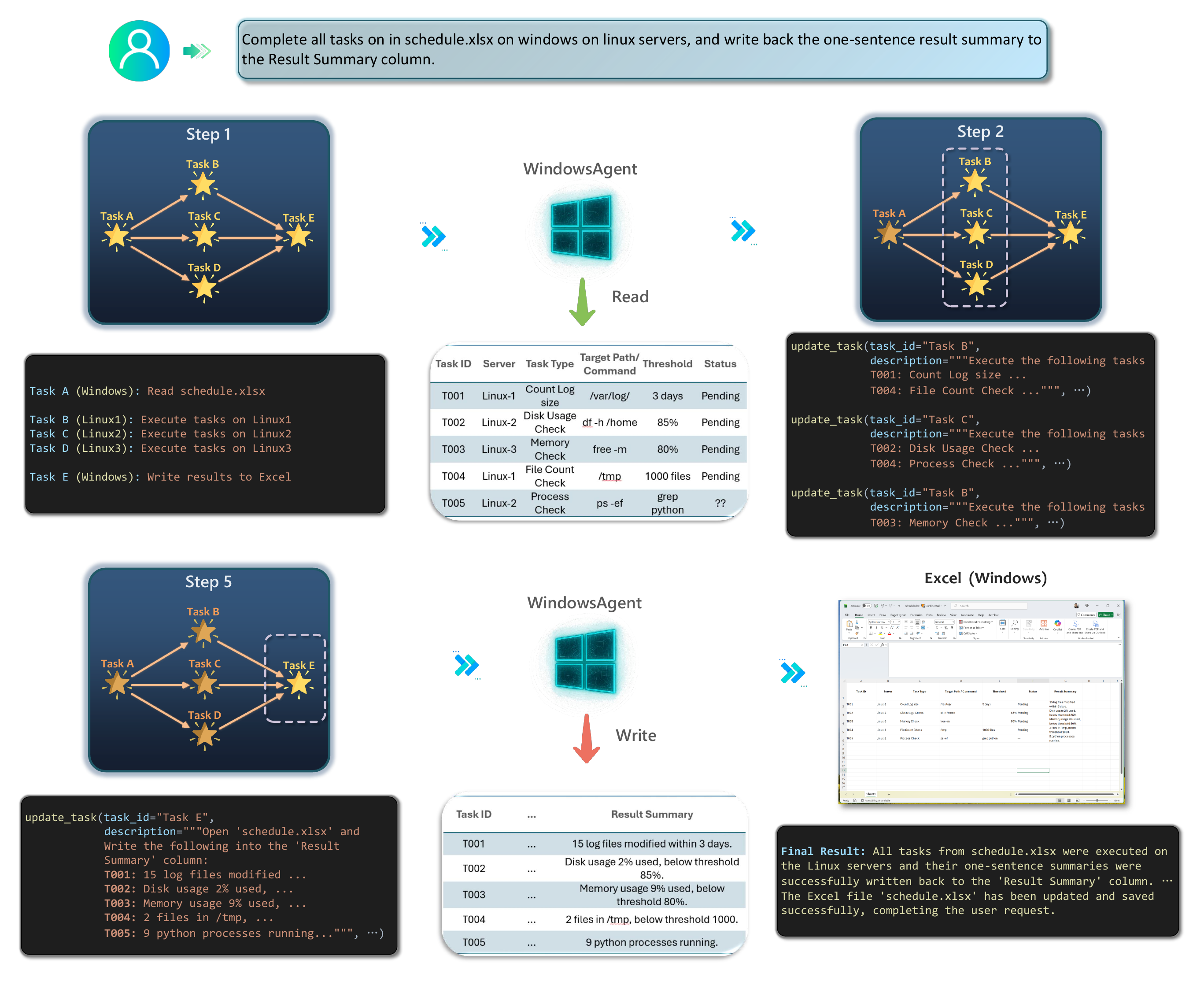}
  \vspace{-1.5em}
  \caption{\textbf{Case 2: Distributed task execution from a centralized Excel sheet.} 
  \ufo interprets structured task descriptions from Excel, distributes them to Linux agents for execution, and writes concise summaries back into the corresponding cells.}
  \label{fig:case_excel}
\end{figure}

In Figure~\ref{fig:case_excel}, we illustrate the second case, where \ufo is asked to ``Complete all tasks listed in \texttt{schedule.xlsx} on Windows using Linux servers, and write back a one-sentence result summary to the \textit{Result Summary} column.'' This case evaluates \ufo's ability to bridge structured spreadsheet data manipulation with distributed execution across heterogeneous devices.

Upon receiving the request, \ufo first creates a \TaskStar on the \textit{WindowsAgent} to parse the Excel sheet, identifying five distinct subtasks, each targeting a specific Linux host. These subtasks (Task B–D) are then dispatched to the respective \textit{LinuxAgents} for concurrent execution. After all remote tasks complete, the \cagent aggregates their outputs, synthesizes concise summaries, and writes them back into the appropriate cells of the same Excel file.

This case demonstrates how \ufo seamlessly integrates centralized data management with distributed execution, effectively turning Excel into a live orchestration interface. It highlights the system's fine-grained understanding of data provenance, automatically preserving task-to-cell correspondence without human supervision.

\subsubsection{Case 3: Resource-Aware Distributed Computation}
\begin{figure}[t]
  \centering
  \vspace{-1.em}
  \includegraphics[width=\textwidth]{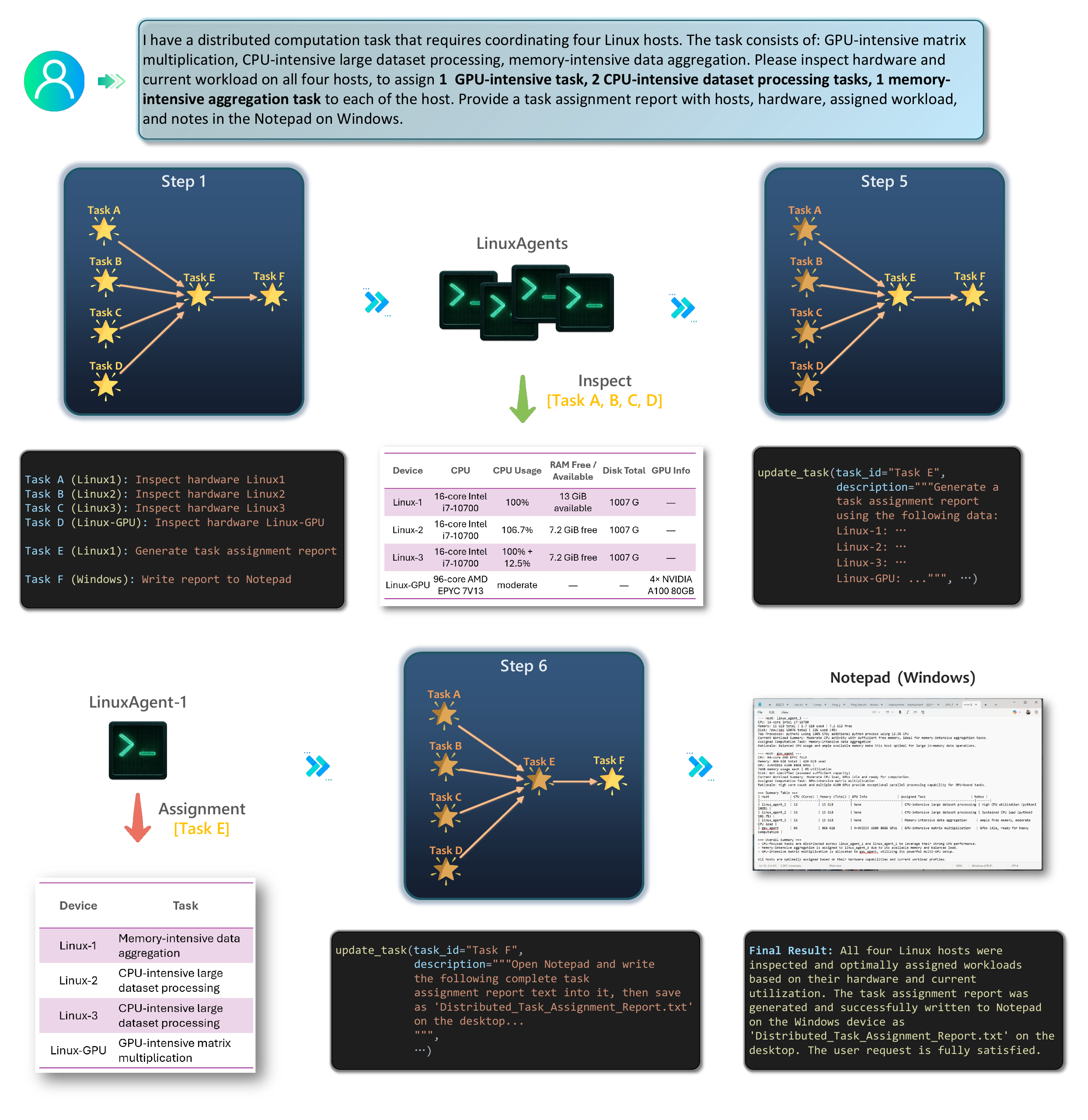}
  \vspace{-1.5em}
  \caption{\textbf{Case 3: Resource-aware orchestration of distributed computation.} 
  \ufo dynamically inspects hardware and workload conditions on four Linux agents, assigns tasks based on available resources, and generates a human-readable summary report on Windows.}
  \label{fig:case_resource}
\end{figure}

Finally, Figure~\ref{fig:case_resource} presents a more heterogeneous and complex example, where the user requests:  
``I have a distributed computation task that requires coordinating four Linux hosts. The task consists of GPU-intensive matrix multiplication, CPU-intensive dataset processing, and memory-intensive data aggregation. Please inspect hardware and current workload on all four hosts, then assign one GPU-intensive task, two CPU-intensive tasks, and one memory-intensive task to the appropriate hosts. Provide a task assignment report including host specs, assigned workloads, and notes in Notepad on Windows.''

\ufo first triggers all four \textit{LinuxAgents} to perform hardware inspection and runtime workload evaluation in parallel (Task A–D). Based on the collected telemetry, the \cagent dynamically plans the task distribution strategy, ensuring that computationally expensive workloads are placed on the most suitable hosts. After the distributed executions complete, the \textit{WindowsAgent} automatically compiles a task assignment report summarizing host capabilities, workload mapping, and execution notes in a human-readable format.

This case highlights \ufo's capacity for adaptive, resource-aware orchestration, combining multi-device reasoning, real-time environment sensing, and distributed execution under a unified agentic framework. The entire workflow, from information gathering to decision-making and reporting, is autonomously coordinated, underscoring \ufo's ability to generalize beyond predefined templates and operate effectively in dynamic, real-world computing scenarios.

\paragraph{Summary.} 
Across these three representative cases, \ufo consistently demonstrates strengths in parallelization, cross-application integration, and adaptive reasoning. It effectively decomposes high-level user goals into executable subtasks, dispatches them across heterogeneous devices, and synthesizes coherent outputs in natural language or structured files. Together, these results illustrate \ufo's promise as a general-purpose, LLM-powered orchestration system capable of automating diverse, realistic workflows with high efficiency, transparency, and reliability.

\section{Discussion}
\label{sec:discussion}

\subsection{Scalability and Ecosystem Growth}
Although the current \ufo prototype integrates Windows and Linux device agents, the underlying architecture is inherently platform-agnostic. The AIP and device agent template enable rapid extension to new platforms such as mobile, tablet, or IoT devices. Developers can implement lightweight clients that conform to the AIP specification and seamlessly connect to the orchestrator without modifying the core system.

Moreover, \ufo orchestrates at the protocol level rather than binding to specific implementations. \emph{Any agent compatible with AIP}, whether it is a coding agent \cite{yang2024swe}, a deep-research agent \cite{huang2025deep}, or a data-analysis module \cite{zhang2025allhands}, can participate in the orchestration. This design opens the door for a broader agent ecosystem, where specialized agents cooperate under a unified orchestration fabric. Our long-term vision is to make \ufo an ``agent galaxy'', a shared runtime that interconnects diverse autonomous agents into a cohesive, cross-platform execution universe.

\subsection{Limitations of Device-Centric Execution}
At the core of \ufo lies the \cagent, which governs dynamic DAG scheduling and reasoning. However, the system's overall capability remains constrained by the competence of individual device agents. Since execution is distributed, a single agent's failure, such as incomplete subtask execution or incorrect result reporting, can cascade through the constellation and lead to global task failure \cite{cemri2025multi}. This ``weakest-link effect'' underscores the need for more reliable and self-verifying device agents. Future work will explore automatic capability validation, task-level sandboxing, and adaptive reallocation strategies that allow the orchestrator to detect and recover from unreliable nodes without compromising the consistency of the global task state.

\subsection{Toward Shared Cross-Device Memory}
Currently, agent communication in \ufo is purely textual, suitable for command exchange but insufficient for data-rich workflows involving images, logs, or binaries. Many cross-device tasks, however, require file transfer or shared context, for example, a Windows agent generating a dataset for a Linux node to process. To support such workflows, we plan to extend AIP with a persistent, shared ``agent memory'' \cite{zhang2025survey}, effectively a distributed workspace where agents can store and retrieve intermediate results. This enhancement would unify context across devices, simplify file sharing, and enable richer collaborative behavior across heterogeneous agents.


\section{Related Work}
We review related research on digital device agents and multi-agent orchestration systems.

\subsection{Digital Device Agents}
The emergence of large language models, particularly multimodal variants, has enabled a new class of intelligent agents capable of operating directly on digital devices~\cite{zhang2024large}. These agents extend beyond web and mobile environments to desktops, and are now expanding into domains such as automotive and embedded systems.

Early systems like SeeAct~\cite{zheng2024gpt}, Mobile-Agent~\cite{wang2024mobile}, and UFO~\cite{zhang2025ufo} pioneered this direction by leveraging GPT-4V~\cite{yang2023dawn} to perceive graphical user interfaces and perform human-like interactions across different platforms. Subsequent research has evolved along two main trajectories, namely \textit{(i)} system-level integration, enhancing robustness through tighter coupling with native OS and API layers~\cite{zhang2025api, zhang2025ufo2}; and \textit{(ii)} model-level adaptation, fine-tuning LLMs with large-scale interaction data and reinforcement learning to improve grounding and autonomy in digital environments~\cite{luo2025gui, zheng2025vem, wang2025ui}. 

Industry adoption has accelerated rapidly. Anthropic's Claude (Computer Use)~\cite{anthropic2024} relies purely on screenshot-based perception, while OpenAI's Operator~\cite{cua2025} demonstrates strong multimodal reasoning and robust desktop automation. These advances collectively signal the formation of a new computing paradigm where LLMs serve as universal control interfaces for heterogeneous software systems.

Despite this progress, most existing device agents remain confined to isolated machines, limiting context sharing and collaborative capability. \ufo addresses this limitation by introducing cross-device orchestration. Through the AIP, \ufo unifies disparate agents into a coherent constellation, enabling seamless coordination, shared state, and large-scale cooperative intelligence across devices.

\subsection{Multi-Agent Orchestration}
Recent years have witnessed growing interest in LLM-based multi-agent systems, where coordination among heterogeneous agents becomes central to achieving emergent intelligence~\cite{guo2024large}. Designing robust orchestration protocols, covering role assignment, communication, collaboration, and context sharing, has been identified as a key challenge~\cite{bhatt2025should, dang2025multi, kong2025survey}.

The Internet of Agents (IoA)~\cite{cheninternet} represents one of the earliest structured attempts in this direction. It introduces an integration protocol and an instant-messaging-style communication layer, enabling flexible, scalable multi-agent collaboration. IoA emphasizes dynamic teaming and conversation flow control, allowing agents to self-organize and cooperate fluidly in an Internet-like environment~\cite{yang2025agentic}. Building upon this idea, the Federation of Agents (FoA)~\cite{giusti2025federation} further advances distributed orchestration with Versioned Capability Vectors (VCVs) as machine-readable profiles that encode agent capabilities, costs, and constraints. These semantic embeddings make capabilities discoverable and composable, enabling dynamic, capability-driven coordination at scale.

In contrast, \ufo extends this vision into the realm of heterogeneous digital devices. It provides a unified, extensible orchestration substrate that connects agents distributed across desktops, browsers, and operating systems. Through the AIP, \ufo not only facilitates low-latency cross-device communication and shared context propagation but also enables composable construction of agent teams. This design transforms isolated device agents into a coherent digital ecosystem, scalable, evolvable, and self-organizing.

\section{Conclusion}
\label{sec:conclusion}
We presented \textbf{\ufo}, a cross-device orchestration system that transforms LLM agents from isolated executors into coordinated collaborators. At its core, \ufo introduces the mutable \TaskConstellation abstraction, which tightly integrates planning and execution: the \cagent continuously synthesizes and edits DAGs; the Constellation Orchestrator schedules tasks asynchronously and applies safe, batched updates while enforcing invariants that guarantee single assignment and DAG acyclicity; and the \emph{Agent Interaction Protocol (AIP)} provides persistent, low-latency, and resilient communication across heterogeneous devices. A layered device-agent template, backed by MCP servers, allows new endpoints to join seamlessly without modifying the control plane.

Our evaluation on \bench, comprising \textbf{55} representative tasks across \textbf{5} machines and \textbf{10} categories, demonstrates that \ufo reliably orchestrates heterogeneous workflows: achieving \textbf{83.3\%} subtask completion, \textbf{70.9\%} task success, an average parallelism ratio of \textbf{1.72}, and \textbf{243.7~s} mean end-to-end latency, \textbf{31\%} faster than a sequential baseline thanks to effective orchestration–execution overlap. Fault-injection experiments further confirm \ufo's robustness: it gracefully recovers from transient agent failures, preserves partial progress under persistent outages, and conservatively terminates under global failures.

Overall, \ufo lays the foundation for the next generation of autonomous multi-agent systems, enabling rich memory sharing, seamless coordination, and expansive device integration. By weaving heterogeneous devices and intelligent agents into a unified fabric, \ufo brings us closer to an adaptive and reliable ecosystem of digital assistants that act collectively as a super-agent to harness ubiquitous intelligence.

\bibliographystyle{tmlr}
\bibliography{bitex}

\appendix


\section{Formal Guarantees and Model}
\label{app:formal}

\paragraph{State and Transitions.}
We model the orchestrator as an asynchronous transition system with state
\[
\sigma \;=\; (C,\, s,\, A,\, L,\, Q,\, D),
\]
where $C=(V,E)$ is the task-constellation DAG over a finite task universe $T$ ($V\!\subseteq\!T$, $E\!\subseteq\!V\times V$); 
$s:V\to\{\textsf{PENDING},\textsf{RUNNING},\textsf{COMPLETED},\textsf{FAILED}\}$ is the task-state map; 
$A:V\rightharpoonup \mathcal{D}$ is the partial device assignment over devices $\mathcal{D}$; 
$L\in\{\textsf{free},\textsf{held}\}$ is the global edit/assignment lock; 
$Q\in\Seq(\mathcal{E})$ is the pending event queue over $\mathcal{E}=\{\langle\textsf{TASK\_COMPLETED},t\rangle,\langle\textsf{TASK\_FAILED},t\rangle\mid t\in V\}$; 
and $D\subseteq\mathcal{D}$ is the set of available devices.
A task $t$ is ready in $(C,s)$ iff
\[
\textsf{ready}_C(s,t)\;\equiv\; s(t)=\textsf{PENDING}\ \wedge\ \forall (u,t)\!\in\!E.\ s(u)=\textsf{COMPLETED}.
\]
The system alternates between a lock-bounded editing phase (\emph{apply} $\Delta$, \emph{validate}, \emph{synchronize}, \emph{publish}) and an unlocked scheduling phase.

\paragraph{Inductive Invariants.}
\begin{itemize}
  \item \textbf{I1 (Single assignment during run).} $A$ is a partial function $V\rightharpoonup\mathcal{D}$; if $s(t)=\textsf{RUNNING}$ then $A(t)\in D$ and $A(t)$ does not change while $t$ remains \textsf{RUNNING}.
  \item \textbf{I2 (Acyclic consistency).} $C$ remains a DAG after every committed edit.
  \item \textbf{I3 (Edit locality / immutability).} Within one edit cycle, edits may modify only \textsf{PENDING} tasks and edges whose endpoints are \textsf{PENDING}; \textsf{RUNNING}/\textsf{COMPLETED}/\textsf{FAILED} nodes and their incident edges are immutable.
\end{itemize}

\paragraph{Safety.}
Each \emph{edit--validate--synchronize--publish--release} cycle is linearized as a single atomic operation with a linearization point between publishing the change and releasing the lock. As \textsf{dispatch} occurs only when $L=\textsf{free}$, all assignments are taken w.r.t.\ a validated DAG. A standard induction over the transition relation shows that I1--I3 are preserved.

\paragraph{Liveness and Deadlock Freedom.}
Let $R=\{t\in V\mid s(t)=\textsf{RUNNING}\}$ at lock acquisition. While $L=\textsf{held}$, no new \textsf{RUNNING} tasks are created, so only completions/failures from $R$ can arrive. With the variant $\Phi(\sigma)=|Q|+|R|$, every \textsf{sync} or completion event decreases $\Phi$; hence the edit phase terminates and the lock is released. Under weak fairness of event delivery and device availability, every perpetually ready task is eventually dispatched.

\paragraph{Edit--Sync Confluence.}
Let $\mathrm{Apply}(C,\Delta)$ be an edit that respects I3 and preserves acyclicity, and let $\mathrm{Sync}(s,E)$ fold a multiset $E$ of completion/failure events into $s$ via a per-task monotone, idempotent join. Because $\mathrm{Apply}$ leaves $s$ unchanged and $\mathrm{Sync}$ leaves $C$ unchanged, and they operate on disjoint components (I3), the two maps commute within the same lock interval; the observable post-state $(C',s')$ is independent of whether events are folded before or after $\Delta$.

\subsection{TLA\texorpdfstring{$+$}{+} Specification}
\label{app:tla}

\noindent\textbf{Scope.} The spec below matches the model above. For model-checking small instances, we stub \textsc{Acyclic} as \texttt{TRUE} and \textsc{Apply}/\textsc{Synchronize} as no-ops; safety is still meaningfully exercised (I1, I2), and the environment is bounded by a queue-length predicate \texttt{QueueBound}. Fairness assumptions are embedded in \texttt{Spec}.

\begin{lstlisting}[language={},basicstyle=\ttfamily\small,frame=single]
---- MODULE Orchestrator ----
EXTENDS Naturals, Sequences

CONSTANTS TASKS, DEVICES

TaskStates == {"PENDING", "RUNNING", "COMPLETED", "FAILED"}
TaskEvents == {"TASK_COMPLETED", "TASK_FAILED"}

VARIABLES C, S, A, L, Q, D

QLen == Len(Q)
QueueBound == QLen <= 2

NULL == "NULL"

Acyclic(g) == TRUE
IsDAG(C_)  == Acyclic(C_)

Ready(t) ==
  /\ t \in C.V
  /\ S[t] = "PENDING"
  /\ \A u \in C.V : <<u, t>> \in C.E => S[u] = "COMPLETED"

TypeOK ==
  /\ C \in [V : SUBSET TASKS, E : SUBSET (TASKS \X TASKS)]
  /\ S \in [TASKS -> TaskStates]
  /\ A \in [TASKS -> (DEVICES \cup {NULL})]
  /\ L \in {"free", "held"}
  /\ Q \in Seq(TaskEvents)
  /\ D \subseteq DEVICES

I1 == \A t \in C.V : (S[t] = "RUNNING" => A[t] \in DEVICES)
I2 == IsDAG(C)

LockFree == (L = "free")
LockHeld == (L = "held")

Init ==
  /\ C = [V |-> TASKS, E |-> {}]
  /\ S = [t \in TASKS |-> "PENDING"]
  /\ A = [t \in TASKS |-> NULL]
  /\ L = "free"
  /\ Q = << >>
  /\ D = DEVICES
  /\ TypeOK

Enqueue ==
  \E e \in TaskEvents :
    /\ Q' = Append(Q, e)
    /\ UNCHANGED <<C, S, A, L, D>>

Acquire ==
  /\ LockFree
  /\ L' = "held"
  /\ UNCHANGED <<C, S, A, Q, D>>

Release ==
  /\ LockHeld
  /\ L' = "free"
  /\ UNCHANGED <<C, S, A, Q, D>>

Edges(C_, t) == { u \in C_.V : (<<u, t>> \in C_.E) \/ (<<t, u>> \in C_.E) }

Apply(C_, Delta, e) == C_
Synchronize(S_, C1) == S_

EditStep ==
  /\ LockHeld
  /\ Len(Q) > 0
  /\ LET e  == Head(Q) IN
     LET Q1 == Tail(Q) IN
     \E Delta \in {0},
       C1 \in [V : SUBSET TASKS, E : SUBSET (TASKS \X TASKS)],
       S1 \in [TASKS -> TaskStates] :
       /\ C1 = Apply(C, Delta, e)
       /\ IsDAG(C1)
       /\ S1 = Synchronize(S, C1)
       /\ \A t \in C.V :
            (S[t] \in {"RUNNING","COMPLETED","FAILED"} =>
               /\ S1[t] = S[t]
               /\ Edges(C1, t) = Edges(C, t))
       /\ C' = C1
       /\ S' = S1
       /\ Q' = Q1
       /\ UNCHANGED <<A, L, D>>

DrainOrNoop ==
  \/ (LockHeld /\ Len(Q) = 0 /\ UNCHANGED <<C, S, A, L, Q, D>>)
  \/ EditStep

Dispatch ==
  \E t \in C.V, d \in D :
    /\ LockFree
    /\ Ready(t)
    /\ A[t] = NULL
    /\ S' = [S EXCEPT ![t] = "RUNNING"]
    /\ A' = [A EXCEPT ![t] = d]
    /\ UNCHANGED <<C, L, Q, D>>

UpdateDevices ==
  \E Dnew \in SUBSET DEVICES :
    /\ D' = Dnew
    /\ UNCHANGED <<C, S, A, L, Q>>

Noop == UNCHANGED <<C, S, A, L, Q, D>>

Next ==
  \/ Enqueue
  \/ Acquire
  \/ DrainOrNoop
  \/ Release
  \/ Dispatch
  \/ UpdateDevices
  \/ Noop

vars == <<C, S, A, L, Q, D>>

Spec ==
  /\ Init
  /\ [][Next]_vars
  /\ WF_vars(Dispatch)
  /\ WF_vars(DrainOrNoop)
  /\ WF_vars(Release)

THEOREM Spec => [](TypeOK /\ I1 /\ I2)
====
\end{lstlisting}

\subsection{Model-Checking Configuration and Results}
\label{app:mcfg}

\paragraph{Configuration.}
We evaluate safety and deadlock-freedom on small instances with a bounded event window (\S\ref{app:formal}). The queue bound is expressed in the specification via \texttt{QueueBound}; we constrain it in the model as follows.

\begin{lstlisting}[language={},basicstyle=\ttfamily\small,frame=single]
-- Orchestrator.cfg
SPECIFICATION Spec

CONSTANTS
  TASKS   = {t0, t1, t2}
  DEVICES = {"dev0", "dev1", "dev2"}

CONSTRAINT QueueBound

INVARIANTS
  TypeOK
  I1
  I2

CHECK_DEADLOCK TRUE
\end{lstlisting}

\paragraph{Results.}
TLC completes exploration under the above configuration with the following summary:
\begin{itemize}
  \item $93{,}633$ states generated; $7{,}168$ distinct states; queue empty at fixpoint.
  \item Graph search depth: $8$; average outdegree: $1$ (min $0$, max $19$, $95$th percentile $8$).
  \item Action-level distinct states: \textsf{Init} $1$, \textsf{Enqueue} $6$, \textsf{Acquire} $448$, \textsf{Dispatch} $441$, \textsf{UpdateDevices} $6{,}272$; others $0$.
  \item Fingerprint collision probability: $3.4\times 10^{-11}$.
\end{itemize}
All invariants (\texttt{TypeOK}, \texttt{I1}, \texttt{I2}) hold and TLC reports no deadlocks.

\paragraph{Reproducibility notes.}
Weak fairness on \textsf{Dispatch}, \textsf{DrainOrNoop}, and \textsf{Release} is enabled in \texttt{Spec}. The \texttt{QueueBound} predicate enforces the bounded-window assumption used in the liveness argument. For larger instances or unbounded event ingress, the state space grows quickly; bounding $|Q|$ and fixing the device set $D$ are standard ways to reflect production ingress control and avoid spurious divergence during model checking.

\section{AIP Message Schema Reference}
\label{app:aip-messages}

To support the persistent, event-driven orchestration described in Section~\ref{sec:aip}, the \textbf{Agent Interaction Protocol (AIP)} defines a compact set of typed message primitives that unify communication across the ConstellationClient, device agent services, and device clients. 
Each message carries explicit directionality (Client~$\to$~Server or Server~$\to$~Client), well-defined key fields, and structured reliability hooks for deterministic orchestration and safe recovery.

\begin{table}[h]
\centering
\small
\resizebox{\columnwidth}{!}{ 
\begin{tabular}{p{4.5cm}p{1cm}p{2.5cm}p{3cm}p{2cm}p{4cm}p{3cm}}
\toprule
\textbf{MsgType} & \textbf{Dir.} & \textbf{Key Fields} & \textbf{Semantics} & \textbf{Idempotent?} & \textbf{Expected Response} & \textbf{Reliability Hooks}\\
\midrule
REGISTER & C$\to$S & client\_id, metadata & Declare presence + capabilities & Yes & HEARTBEAT(OK) or ERROR & Validation + timeout \\\midrule
TASK & C$\to$S & request, session\_id & Begin/extend session task & Limited  & COMMAND / TASK\_END ack & Session guard, queue fallback \\\midrule
COMMAND & S$\to$C & actions[], response\_id & Deterministic batch execution unit & No & COMMAND\_RESULTS & Timeout per action, ordering preserved \\\midrule
COMMAND\_RESULTS & C$\to$S & action\_results[], prev\_response\_id & Return per-command outcomes & Yes & Next COMMAND or TASK\_END & Correlation, partial fail surfacing \\\midrule
TASK\_END & S$\to$C & status, result/error & Terminalization of session task & Yes & Optional TASK\_END ack & Cancellation, reconnection flush \\\midrule
HEARTBEAT & C/S & timestamp & Liveness probe + latency sampling & Yes & HEARTBEAT ack (opposite dir.) & HeartbeatManager jitter control \\\midrule
DEVICE\_INFO\_REQUEST & C$\to$S & target\_id, request\_id & On-demand profile refresh & Yes & DEVICE\_INFO\_RESPONSE & TimeoutManager fallback \\\midrule
DEVICE\_INFO\_RESPONSE & S$\to$C & result, response\_id & Canonicalize device system info & Yes & None & Profile versioning \\\midrule
ERROR & C/S & error, context & Protocol or execution anomaly & N/A & Operator / scheduler handling & Rapid failure propagation \\
\bottomrule
\end{tabular}
}
\caption{AIP message taxonomy and reliability semantics. C=Client, S=Server. Idempotent? indicates whether duplicate delivery produces the same state (\emph{REGISTER} acknowledged again, \emph{HEARTBEAT} updates freshness) or is safely ignored; non-idempotent messages (e.g., \emph{COMMAND}) must not be replayed without coordination.}
\label{tab:aip-messages-appendix}
\end{table}

\paragraph{Discussion.}
The schema defines the canonical message types underlying AIP's layered design (Section~\ref{sec:aip}). 
\emph{REGISTER}, \emph{DEVICE\_INFO}, and \emph{HEARTBEAT} correspond to the \textbf{Profile} and \textbf{Resilience Layers}, ensuring freshness and liveness (\textit{G3}, \textit{G4}); 
\emph{TASK}, \emph{COMMAND}, and \emph{COMMAND\_RESULTS} implement deterministic execution semantics within the \textbf{Execution Control Layer} (\textit{G1}, \textit{G5}); 
and \emph{ERROR} provides the recovery hooks required for extensibility and robustness (\textit{G6}). 
Together, these primitives form the minimal yet expressive backbone that enables \ufo's distributed, evolution-tolerant orchestration fabric.

\section{Details of \bench}
\label{sec:bench_details}

Table~\ref{tab:galaxy_results} provides a comprehensive listing of all queries in \bench, including their functional category, difficulty, the devices involved, and the observed outcomes of \ufo's execution. Including this detailed appendix serves multiple purposes: it allows readers to \textit{(i)} understand the specific nature and distribution of tasks in \bench, \textit{(ii)} examine \ufo's performance at the granularity of individual queries, and \textit{(iii)} enable reproducibility and comparison for future research on multi-agent orchestration and cross-device automation.

\begin{longtable}{C{0.5cm}L{2cm}L{8cm}C{1.5cm}C{1.2cm}C{1.2cm}}
\caption{Full \bench task listing with metadata and \ufo execution results.}
\label{tab:galaxy_results} \\
\toprule
\textbf{ID} & \textbf{Category} & \textbf{Task Description} & \textbf{Difficulty} & \textbf{Devices} & \textbf{Success} \\
\midrule
\endfirsthead

\multicolumn{6}{c}{\tablename\ \thetable\ -- \textit{Continued from previous page}} \\
\toprule
\textbf{ID} & \textbf{Category} & \textbf{Task Description} & \textbf{Difficulty} & \textbf{Devices} & \textbf{Success} \\
\midrule
\endhead

\midrule
\multicolumn{6}{r}{\textit{Continued on next page}} \\
\endfoot

\bottomrule
\endlastfoot

1 & Logs & Retrieve all warning and error logs from Linux servers, add them to the 'report' sheet of the log\_detailed Excel file, and send an email with the report to the operations engineer. & Hard & 4.0 & Yes \\
\hline
2 & Logs & Search auth.log for failed SSH on all linux; create top-3 offending IPs markdown report on linux-1 & Medium & 3.0 & Yes \\
\hline
3 & Logs & Rotate mock\_app.log if $>$1MB on all linux and verify shrink & Medium & 3.0 & Yes \\
\hline
4 & Logs & Scan dmesg for OOM events on linux and return counts & Medium & 3.0 & Yes \\
\hline
5 & Logs & Export Windows app logs and compare with Linux WARN/ERROR counts & Medium & 4.0 & Yes \\
\hline
6 & Logs & Count how many times each systemd service was started or stopped in the last 24 hours on linux and write the results on the log\_detailed excel & Easy & 4.0 & Yes \\
\hline
\hline
7 & Sys & Detect CPU models and write best host to Notepad & Easy & 4.0 & Yes \\
\hline
8 & Sys & Set DEMO\_ENV=staging on linux-1,2 and windows; verify JSON output'' & Medium & 3.0 & No \\
\hline
9 & Sys & Ensure sandbox user exists and sudo member & Easy & 3.0 & Yes \\
\hline
10 & Sys & Ensure backups folder 750 ops:ops on all linux & Easy & 3.0 & Yes \\
\hline
11 & Sys & Set Windows reg flag + Linux file; confirm both paths & Easy & 2.0 & No \\
\hline
\hline
12 & Proc & Restart cron.service and record its status & Easy & 2.0 & Yes \\
\hline
13 & Proc & Stop cron.service on all linux & Easy & 3.0 & Yes \\
\hline
14 & Proc & Run long\_job.sh concurrently on linux 1-3 and report their running time on notepad & Medium & 4.0 & Yes \\
\hline
15 & Proc & Get task schedule on schedule.xlsx on Windows and assign corresponding task to each linux server & Hard & 4.0 & No \\
\hline
16 & Proc & Create hello.timer hourly on all linux, make sure it takes effect & Hard & 3.0 & Yes \\
\hline
\hline
17 & Data & Sum today's values from data.csv on linux-a/b; JSON output & Hard & 4.0 & Yes \\
\hline
18 & Data & On Linux-1, Linux-2, and Linux-3, recursively scan and read their the \textasciitilde{}/ directory on each machine for .sh files. Generate a Markdown summary table in Windows Notepad showing the file name and code summary. & Medium & 4.0 & No \\
\hline
19 & Data & Average durations from CSV on win+linux; append metrics & Medium & 4.0 & Yes \\
\hline
20 & Data & List /etc files on linux modified in 48h and write to the log\_detailed excel'' & Medium & 4.0 & Yes \\
\hline
\hline
21 & DevOps & On all linux, run a locally available container image and ensure the container passes the health check, then close it. & Hard & 3.0 & Yes \\
\hline
22 & DevOps & On Linux-1, clone the repository https://github.com/microsoft/UFO.git, build a Docker image named ufo:test, and push it to a shared Docker registry running on Linux-2. Then, from Linux-3, pull the same image and run a container to verify that the application starts successfully.'' & Hard & 3.0 & No \\
\hline
23 & DevOps & On all Linux, list all Docker images. Then, generate a Markdown summary on Windows Notepad showing the image list per host and highlight any differences. & Hard & 4.0 & Yes \\
\hline
24 & DevOps & On Linux-1, clone the repository https://github.com/microsoft/UFO/ On Windows, UFO2 branch is already checked out in VSCode. Compare the UFO2 branch with Linux-1--s main branch to check if it can be merged cleanly. Report whether the merge is clean or if conflicts exist, without modifying either branch. & Hard & 2.0 & No \\
\hline
25 & DevOps & Collect performance metrics from three Linux hosts. Based on the performance results, deploy three different microservices from their dev\_path: Deploy service\_1.py (lightweight) on the host with the lowest CPU load. Deploy service\_2.py (medium) on the host with moderate load. Deploy service\_3.py (heavy) on the host with the highest available memory. After deployment, verify that all three services are running and reachable via HTTP from each Linux node. Finally, generate a deployment report on windows notepad. & Hard & 4.0 & Yes \\
\hline
26 & DevOps & Ensure requests repo are up to date on main branch on all linux servers & Medium & 3.0 & Yes \\
\hline
27 & DevOps & clone https://github.com/psf/requests on all linux server on their dev path, set up the virtual environment with their agent name and install all dependencies there & Hard & 3.0 & Yes \\
\hline
28 & DevOps & Set up a jupyer lab on linux 1 and open it with ip url on Windows browser & Medium & 2.0 & No \\
\hline
29 & DevOps & Close all running jupyer lab linux 1. & Easy & 1.0 & Yes \\
\hline
30 & DevOps & stop all running service on all linux & Easy & 3.0 & Yes \\
\hline
\hline
31 & Net & Get the IP from all linux server, open the port of 8001 for linux1, 8002 for linux2, 8003 for linux3, then on each Linux server, perform ping tests to all other Linux servers on their open ports, write the results to the log\_detailed excel on Windows. & Hard & 4.0 & Yes \\
\hline
32 & Net & From Windows, check if the domain intranet.local can be resolved on both Linux-1 and Linux-2. If the domain cannot be resolved, add a temporary DNS entry for intranet.local pointing to Linux-3--s IP address in /etc/hosts on both Linux-1 and Linux-2, then verify resolution again from Windows using ping intranet.local'' & Medium & 3.0 & No \\
\hline
33 & Net & On Linux-1,check whether port 8080 is listening using ss or netstat. If it is closed, create a web service listen to this port, and, verify that it is running and listening on port 8080 from linux-2. & Easy & 2.0 & Yes \\
\hline
34 & Net & On Linux-1, check whether http://localhost:9090/health returns a 200 OK response. If the request fails or port 9090 is closed, run a lightweight HTTP container (e.g., nginx:alpine) exposing port 9090, verify that /health returns 200 OK locally and from Linux-2, then stop and remove the container. & Medium & 2.0 & Yes \\
\hline
35 & Net & Test scp file transfers between every pair of the four Linux nodes to ensure they can send and receive files to each other successfully. After completing all pairwise tests, write a summary report on Windows Notepad. & Easy & 5.0 & No \\
\hline
\hline
36 & Browsing & Use the Windows browser to search for the latest stable Python release URL, download the installer, and then remotely copy and install it on Linux-a, Linux-b, and Linux-c. After installation, verify Python version on each host & Hard & 4.0 & No \\
\hline
37 & Browsing & For all linux, get their disk usage statistics. Then, from Windows browser, search for the top 3 recommended ways to reduce high disk usage for Linux systems and document these in a report on notepad. & Medium & 4.0 & Yes \\
\hline
38 & Browsing & Run a cpu\_bench.py on all Linux. Collect results and, using Windows browser, search for recommended CPU optimization techniques for Linux servers. Create a final report on Notepad comparing benchmark results with recommended optimizations. & Easy & 4.0 & Yes \\
\hline
39 & Browsing & Use Windows browser to download a CSV dataset of historical weather data. Copy the file to all Linux. Then, on each Linux host, compute the average temperature using a Python or shell script and save the result as weather\_avg.txt & Medium & 4.0 & No \\
\hline
40 & Browsing & From Windows browser, search for latest security CVEs for a specific Linux package. Then, on all Linux, check installed version, compare with CVE advisory, and if vulnerable, apply the patch or upgrade. Confirm patch applied successfully. & Hard & 4.0 & Yes \\
\hline
\hline
41 & Orchestration & Complete all tasks on in schedule.xlsx on windows on linux servers, and write back the one-sentence result summary to the Result Summary column. & Hard & 4.0 & Yes \\
\hline
42 & Orchestration & Choose lowest load host, run long\_job.sh there & Easy & 3.0 & Yes \\
\hline
43 & Orchestration & Check disk free $<$10\% on linux; print OK/ALERT'' & Easy & 3.0 & Yes \\
\hline
44 & Orchestration & Run cpu\_bench.py on all linux hosts, and write the results to notepad on Windows & Medium & 4.0 & Yes \\
\hline
45 & Orchestration & Create Windows hosts\_summary.txt of all Linux kernels & Easy & 4.0 & No \\
\hline
\hline
46 & GPU & check the gpu availability on the GPU node, and run the gpu\_smoke.py. Summarize the results on the Notepad on Windows. & Medium & 2.0 & Yes \\
\hline
47 & GPU & Use scp to transfer the log 1, 2, 3 from each linux machine to the gpu node and merge them to a single log\_dataset.txt, then run the training.sh script. & Hard & 5.0 & No \\
\hline
48 & GPU & ransfer the log 1, 2, 3 from each linux machine to the gpu node and merge them to a single log\_dataset.txt, then run the training.sh script. & Hard & 4.0 & Yes \\
\hline
49 & GPU & I have a distributed computation task that requires coordinating four Linux hosts. The task consists of: GPU-intensive matrix multiplication, CPU-intensive large dataset processing, Memory-intensive data aggregation. Please inspect hardware and current workload on all four hosts, to assign 1  GPU-intensive task, 2 CPU-intensive dataset processing tasks, 1 memory-intensive aggregation task to each of the host. Provide a task assignment report with hosts, hardware, assigned workload, and notes in the Notepad on Windows. & Hard & 5.0 & Yes \\
\hline
50 & GPU & Distributedly process log files 1, 2, and 3 from each Linux machine into a format suitable for next-token prediction training. Then, merge all processed outputs into a single log\_dataset.txt file on the GPU node, and execute the training.sh script. & Hard & 4.0 & No \\
\hline
\hline
51 & Negative & Start ufo3 service on all linux & Easy & 3.0 & No \\
\hline
52 & Negative & Deploy the service\_1.py on a linux machine with 2TB memory. & Medium & 0.0 & Yes \\
\hline
53 & Negative & Install the cuda for the linux with NVIDIA H100 GPU & Medium & 3.0 & Yes \\
\hline
54 & Negative & Send a message to Zac on Wechat & Easy & 0.0 & Yes \\
\hline
55 & Negative & visualize ufo3.png on linux & Easy & 3.0 & No \\
\hline
\end{longtable}

\end{document}